\documentclass[acmsmall]{acmart}
\renewcommand\footnotetextcopyrightpermission[1]{} % removes footnote with conference information in first column
\usepackage[false]{anonymous-acm}

\usepackage{amsmath}
\usepackage{amsfonts}
\usepackage{xspace}
\usepackage{listings}
\usepackage{mathpartir}
\usepackage{syntax}
\usepackage{subcaption}
\usepackage{hyphenat}
\usepackage{enumerate}
\usepackage{fancyvrb}
\usepackage{framed}
\usepackage{relsize}
\usepackage{bm}
\usepackage{tcolorbox}
\usepackage{textcomp}
\usepackage{mathpartir}
\usepackage{wrapfig}
\usepackage{mathtools} % reduce vertical spacing in align environments
\usepackage{tabularx}
\usepackage{caption}
\usepackage{array}
\usepackage[noend]{algorithm2e}
\usepackage{environ}
\usepackage{nopageno}
\usepackage{multicol}
\usepackage{lineno, blindtext}
\usepackage{etoolbox}
\usepackage{adjustbox}
\usepackage{changepage} % Required for the adjustwidth environment
\usepackage{tikz}
\usetikzlibrary{arrows, automata, positioning, decorations.pathmorphing}
\usepackage{forest}
\useforestlibrary{linguistics}

\newcommand{\ifempty}[3]{\ifstrequal{#1}{}{#2}{#3}}

\newcommand{\mathscale}[2]{\displaystyle\vcenter{\hbox{\scalebox{#1}{$#2$}}}}
\newcommand{\greencheck}{\mathscale{2}{\color{pinegreen}{\checkmark}}}

\NewEnviron{sentailsP}[7]{%
  \resizebox{#7}{!}{$
    \ifempty{#3}{}{\ensuremath{#3},~}
      \left[
      \hspace{-.8em}
      \begin{minipage}{#1\textwidth}
        \footnotesize
        \begin{algorithm}[H]
          #4
        \end{algorithm}
      \end{minipage}
      \hspace{-0.09\hsize},~#5~\mathscale{1.5}{\Vdash}_{#2}~#6
      \right]_S
    $}
  }

\NewEnviron{qentailsP}[8]{%
  \resizebox{#8}{!}{$
    \ifempty{#3}{}{\ensuremath{#3},~}
      \left[
      \hspace{-.8em}
      \begin{minipage}{#1\textwidth}
        \footnotesize
        \begin{algorithm}[H]
          #4
        \end{algorithm}
      \end{minipage}
      \hspace{-0.09\hsize},~#5,~#6~\mathscale{1.5}{\Vdash}_{#2}~#7
      \right]_{\Queue}
    $}
}

\NewEnviron{rentailsP}[8]{%
  \resizebox{#8}{!}{$
    \ifempty{#3}{}{\ensuremath{#3},~}
      \left[
      \hspace{-1em}
      \begin{minipage}{#1\textwidth}
        \footnotesize
        \begin{algorithm}[H]
          #4
        \end{algorithm}
      \end{minipage}
      \hspace*{-.025\hsize},~#5,~#6~\mathscale{1.5}{\Vdash}_{#2}~#7
      \right]_R
    $}
}

\NewEnviron{plet}[5]{
  \AlgoDontDisplayBlockMarkers\SetAlgoNoEnd
  \mathscale{#4}{
    \begin{minipage}{#5\textwidth}
      $\bm{#1}~{#2}\coloneq~$ \\
      \begin{algorithm}[H]
        #3
      \end{algorithm}
    \end{minipage}
  }
}

\newtheorem{theorem}{Theorem}

\ifx\noeditingmarks\undefined
  \newcommand{\pgwrapper}[3]{\begingroup \color{#1} #2: #3 \endgroup}
  \newcommand{\pgwrapperb}[1]{\textbf{#1}}
\else
   \newcommand{\pgwrapperb}[1]{}
   \newcommand{\pgwrapper}[2]{}
\fi

\newsavebox{\procbox}

\usepackage{listings,lstlangcoq,lstautogobble}
\lstdefinestyle{customcoq}{
  columns=flexible,
  mathescape=true,
  belowcaptionskip=1\baselineskip,
  breaklines=true,
  numbers=none,
  xleftmargin=\parindent,
  language=Coq,
  morekeywords={Variant, fun, Arguments, Type, cofix},
  emph={%
        ITree,CTree, Pid, Gt, Lt, Eq, proc, ur
  },
  emphstyle={\bfseries},
  showstringspaces=false,
  basicstyle=\small\ttfamily,
  keywordstyle=\bfseries,
  commentstyle=\itshape\color{red!40!black},
  identifierstyle=\itshape,
  stringstyle=\color{orange},
  escapeinside={<@}{@>}
}

\newcommand{\spaced}[1]{\ifempty{#1}{}{\ {#1}}}
\newcommand{\spacedr}[1]{\ifempty{#1}{}{{#1}\ }}
\newcommand{\dashed}[1]{\ifempty{#1}{\textcolor{gray}{\_}}{#1}}
\newcommand{\ticl}{\texttt{ticl}\xspace}
\newcommand{\Ticl}{\texttt{Ticl}\xspace}

\newcommand{\typed}[2]{\ensuremath{#1~\textcolor{gray}{\in~#2}}}

\newcommand{\tm}[1]{\ensuremath{\mathtt{#1}}} % terminals
\newcommand{\PROP}{\ensuremath{\mathbb{P}}}
\newcommand{\Type}{\ensuremath{\mathtt{Type}}}

\newcommand{\fin}[1]{\mathtt{fin}\spaced{#1}}
\newcommand{\unit}{\ensuremath{\mathtt{unit}}}
\newcommand{\ttt}{\mathtt{()}}
\definecolor{dgreen}{rgb}{0.0, 0.6, 0.0} % RGB values for a darker green
\definecolor{pinegreen}{rgb}{0.0, 0.47, 0.44}
\newcommand{\cbwd}{\textcolor{blue}{$\Leftarrow$}\xspace}
\newcommand{\ciff}{\textcolor{dgreen}{$\Leftrightarrow$}\xspace}
\newcommand{\coindeq}{\stackrel{\mathsf{coind}}{=}}
\newcommand{\ictree}[2]{\ensuremath{\mathtt{ictree}_{\ifempty{#1}{}{\textcolor{gray}{#1\ifempty{#2}{}{,~#2}}}}}}
\newcommand{\bbar}{\bm{|}~}

\newcommand{\ICTree}{\texttt{ICTree}\xspace}
\newcommand{\ICTrees}{\texttt{ICTrees}\xspace}

\newcommand{\sep}{~\mid~}
\newcommand{\mkw}[1]{\mbox{\sc{#1}}}
\newcommand{\upto}[1]{\ensuremath{\mkw{#1}^{UP}}}

\newcommand{\Tau}[1]{\ensuremath{\mathtt{Tau}\spaced{#1}}}
\newcommand{\Br}[2]{\ensuremath{\ifempty{#1}{\mathtt{Br}\spaced{#2}}{\mathtt{Br}\spaced{#1} \spaced{#2}}}}
\newcommand{\VisX}[3]{\ensuremath{\mathtt{Vis} \spaced{#1} \spaced{#2} \spaced{#3}}}
\newcommand{\Vis}[2]{\ensuremath{\mathtt{Vis} \spaced{#1} \spaced{#2}}}
\newcommand{\Ret}[1]{\ensuremath{\mathtt{Ret} \spaced{#1}}}
\newcommand{\stuck}{\ensuremath{\tm{\emptyset}}\xspace}

\newcommand{\sbisim}{\ensuremath{\sim}}

\newcommand{\R}{\ensuremath{\mathcal{R}}}

\newcommand{\bindc}{\ensuremath{>\!\!>\!\!=}}

\newcommand{\bind}[2]{\ensuremath{x \gets #1\tm{;\!;}~#2 ~x}\xspace}
\newcommand{\bindv}[3]{\ensuremath{#1 \gets #2\tm{;\!;}~#3}\xspace}

\newcommand{\wellfounded}[1]{\ensuremath{\mathtt{well\_founded}~#1}}
\newcommand{\iter}[2]{\ensuremath{\tm{iter\spaced{~#1}~\spaced{#2}}}\xspace}
\newcommand{\logE}[1]{\ensuremath{\mathtt{L}_{\textcolor{gray}{#1}}}}
\newcommand{\Log}[1]{\ensuremath{\mathtt{Log} \spaced{#1}}}
\newcommand{\llog}[1]{\ensuremath{\mathtt{log} \spaced{#1}}}
\newcommand{\option}[1]{\ensuremath{\tm{option}_{#1}}\xspace}
\newcommand{\Some}[1]{\ensuremath{\tm{Some}({#1})}\xspace}
\newcommand{\None}{\ensuremath{\tm{None}}\xspace}
\newcommand{\Map}[2]{\ensuremath{\tm{Map}_{\textcolor{gray}{\tm{#1, #2}}}}}
\newcommand{\Ctx}{\ensuremath{\mathcal{M}}}
\newcommand{\itern}{\tm{iter}\xspace}
\newcommand{\trigger}[1]{\ensuremath{\tm{trigger}~#1}\xspace}
\newcommand{\branch}{\tm{branch}\xspace}

\newcommand{\gfp}{\ensuremath{\tm{gfp}~}}

\newcommand{\sget}{\ensuremath{\tm{get}}}
\newcommand{\sput}[1]{\ensuremath{\tm{put}~#1}}
\newcommand{\cons}{\ensuremath{\tm{::}}}
\newcommand{\app}{\ensuremath{\tm{\ ++\ }}}

\newcommand{\sembra}[2]{\ensuremath{\llbracket \dashed{#2} \rrbracket_{\ifempty{#1}{}{_{\textcolor{gray}{\tm{#1}}}}}}}

\newcommand{\InstrM}[2]{\ensuremath{\mathtt{InstrM}_{\textcolor{gray}{{#1} ,{#2}}}}}
\newcommand{\instrv}[3]{\ensuremath{\mathtt{instr} \spaced{#1} \spaced{#2} \spaced{#3}}}
\newcommand{\stateT}[2]{\ensuremath{\mathtt{stateT}\spaced{\tm{#1}}\spaced{\tm{#2}}}}

\newcommand{\stimp}{\ensuremath{\tm{StImp}}\xspace}
\newcommand{\IAExp}{\ensuremath{\tm{AExp}}\xspace}
\newcommand{\IBExp}{\ensuremath{\tm{BExp}}\xspace}
\newcommand{\St}[2]{\ensuremath{\tm{state}_{\textcolor{gray}{#1}\ifempty{#2}{}{,#2}}}}
\newcommand{\IVar}[1]{\ensuremath{\tm{var}}~{#1}}
\newcommand{\IVal}[1]{\ensuremath{\tm{val}}~{#1}}
\newcommand{\IAdd}[2]{\ensuremath{#1 ~ + ~ #2}}
\newcommand{\ISub}[2]{\ensuremath{#1 ~ - ~ #2}}

\newcommand{\ILt}[2]{\ensuremath{#1 ~ < ~ #2}}

\newcommand{\IIf}[3]{\ensuremath{\text{\textbf{if}}~(#1)~\text{\textbf{then}}~ #2 ~\text{\textbf{else}}~ #3}}
\newcommand{\IWhile}[2]{\ensuremath{\text{\textbf{while}}~(#1)~\{#2\}}}
\newcommand{\IAssign}[2]{\ensuremath{#1~ \leftarrow ~ #2}}
\newcommand{\ISeq}[2]{\ensuremath{#1~;~#2}}

\newcommand{\ISkip}{\ensuremath{\tm{skip}}}

\newcommand{\meQ}[1]{\ensuremath{\tm{MeQ}\ifempty{#1}{}{_{#1}}}\xspace}
\newcommand{\QPop}{\ensuremath{\tm{pop}}\xspace}
\newcommand{\QPush}[1]{\ensuremath{\tm{push} \spaced{#1}}\xspace}
\newcommand{\QWhile}[1]{\ensuremath{\text{\textbf{while}}~(\tm{true})~\{#1\}}}
\newcommand{\QRet}[1]{\ensuremath{\tm{ret}~#1}}
\newcommand{\QBind}[2]{\ensuremath{#1 \bindc #2}}

\newcommand{\meS}[1]{\ensuremath{\tm{MeS}\ifempty{#1}{}{_{#1}}}\xspace}
\newcommand{\SWrite}[3]{\ensuremath{\tm{write} \spaced{#1} \spaced{#2} \spaced{#3}}\xspace}
\newcommand{\SRead}[2]{\ensuremath{\tm{read} \spaced{#1} \spaced{#2}}\xspace}
\newcommand{\SIf}[3]{\ensuremath{\text{\textbf{if}} ~#1~ \text{\textbf{then}}~#2~\text{\textbf{else}}~#3}\xspace}
\newcommand{\SRet}[1]{\ensuremath{\tm{ret}~#1}}
\newcommand{\SBind}[2]{\ensuremath{#1 \bindc #2}}

\newcommand{\meR}[1]{\ensuremath{\tm{MeR}\ifempty{#1}{}{_{#1}}}\xspace}
\newcommand{\RLoop}[2]{\ensuremath{\tm{loop}} \spaced{#1} \spaced{#2} \xspace}
\newcommand{\RBr}[2]{\ensuremath{#1~\oplus~#2}}
\newcommand{\RCall}[1]{\ensuremath{\tm{call} ~#1}\xspace}
\newcommand{\RRet}[1]{\ensuremath{\tm{ret}~#1}}
\newcommand{\RBind}[2]{\ensuremath{#1 \bindc #2}}

\newcommand{\World}[1]{\ensuremath{\mathcal{W}\ifempty{#1}{}{_{\textcolor{gray}{#1}}}}}
\newcommand{\Pure}{\ensuremath{\mathtt{Pure}}\xspace}
\newcommand{\Val}[1]{\ensuremath{\mathtt{Val} \spaced{#1}}\xspace}
\newcommand{\Obs}[2]{\ensuremath{\mathtt{Obs} \spaced{#1} \spaced{#2}}\xspace}
\newcommand{\Finish}[3]{\ensuremath{\mathtt{Finish} \spaced{#1} \spaced{#2} \spaced{#3}}\xspace}
\newcommand{\notdone}[1]{\ensuremath{\mathtt{not\_done} \spaced{#1}}}
\newcommand{\donewith}[2]{\ensuremath{\mathtt{done\_with} \spaced{#1} \spaced{#2}}}
\newcommand{\canstep}[2]{\ensuremath{\mathtt{can\_step} \spaced{#1} \spaced{#2}}}
\newcommand{\ktrans}[4]{\ensuremath{[\dashed{#1},\ \dashed{#2} ] \mapsto [ \dashed{#3},\ \dashed{#4} ]}}

\newcommand{\psix}{\ensuremath{\psi_\mathtt{X}}\xspace}
\newcommand{\Px}{\ensuremath{P_\mathtt{X}}\xspace}

\newcommand{\implL}{\ensuremath{\Rightarrow_L}}
\newcommand{\implR}{\ensuremath{\Rightarrow_R}}
\newcommand{\implLR}{\ensuremath{\Rightarrow_{LR}}}
\newcommand{\iffL}{\ensuremath{\Leftrightarrow_L}}
\newcommand{\iffR}{\ensuremath{\Leftrightarrow_R}}
\newcommand{\iffLR}{\ensuremath{\Leftrightarrow_{LR}}}

\newcommand{\now}[1]{\ensuremath{\mathtt{now}\spaced{#1}}\xspace}

\newcommand{\done}[1]{\ensuremath{\mathtt{done}\spaced{#1}}\xspace}
\newcommand{\doneq}[2]{\ensuremath{\mathtt{done_=}\spaced{#1}\spaced{#2}}\xspace}
\newcommand{\val}[1]{\ensuremath{\mathtt{val}\spaced{#1}}\xspace}
\newcommand{\finish}[1]{\ensuremath{\mathtt{finish}\spaced{#1}}\xspace}
\newcommand{\pure}{\ensuremath{\mathtt{pure}}\xspace}
\newcommand{\obs}[1]{\ensuremath{\mathtt{obs}\spaced{#1}}\xspace}
\newcommand{\inl}[1]{\ensuremath{\mathtt{inl}\spaced{#1}}\xspace}
\newcommand{\inr}[1]{\ensuremath{\mathtt{inr}\spaced{#1}}\xspace}

\newcommand{\AU}[2]{\ensuremath{\spacedr{#1} \tm{AU}      \spaced{#2}}\xspace}
\newcommand{\EU}[2]{\ensuremath{\spacedr{#1} \tm{EU}      \spaced{#2}}\xspace}
\newcommand{\AN}[2]{\ensuremath{\spacedr{#1} \tm{AN}      \spaced{#2}}\xspace}
\newcommand{\EN}[2]{\ensuremath{\spacedr{#1} \tm{EN}      \spaced{#2}}\xspace}
\newcommand{\AG}[1]{\ensuremath{\tm{AG}\spaced{#1}}\xspace}
\newcommand{\EG}[1]{\ensuremath{\tm{EG}\spaced{#1}}\xspace}

\newcommand{\AX}[1]{\ensuremath{\tm{AX} \spaced{#1}\xspace}}
\newcommand{\EX}[1]{\ensuremath{\tm{EX} \spaced{#1}\xspace}}
\newcommand{\AF}[1]{\ensuremath{\tm{AF} \spaced{#1}\xspace}}
\newcommand{\EF}[1]{\ensuremath{\tm{EF} \spaced{#1}\xspace}}

\newcommand{\entails}[3]{\ensuremath{\langle ~\dashed{#1}, \ \dashed{#2} \vDash ~\dashed{#3}~ \rangle}}
\newcommand{\entailsL}[3]{\ensuremath{\langle ~\dashed{#1}, \ \dashed{#2} \vDash_L ~\dashed{#3}~ \rangle}}
\newcommand{\entailsR}[3]{\ensuremath{\langle ~\dashed{#1}, \ \dashed{#2} \vDash_R ~\dashed{#3}~ \rangle}}
\newcommand{\entailsRR}[2]{\ensuremath{\langle~ \dashed{#1}, \ \dashed{#2} \vDash_R ~}}
\newcommand{\entailsLR}[3]{\ensuremath{\langle~ \dashed{#1}, \ \dashed{#2} \vDash_{LR} ~\dashed{#3} ~\rangle}}
\newcommand{\sentailsL}[4]{\ensuremath{[\dashed{#2}, \ \dashed{#3} \Vdash_L ~\dashed{#4} ]_{#1}}}
\newcommand{\sentailsR}[4]{\ensuremath{[\dashed{#2}, \ \dashed{#3} \Vdash_R ~\dashed{#4} ]_{#1}}}
\newcommand{\sentailsLR}[4]{\ensuremath{[\dashed{#2}, \ \dashed{#3} \Vdash_{LR} ~\dashed{#4} ]_{#1}}}
\newcommand{\sentailsRR}[2]{\ensuremath{[\dashed{#1}, \ \dashed{#2} \Vdash_R ~ }}

\newcommand{\qentailsL}[4]{\ensuremath{[\dashed{#1}, \ \dashed{#2},\ \dashed{#3} \Vdash_L ~\dashed{#4} ]_{\Queue}}}
\newcommand{\qentailsR}[4]{\ensuremath{[\dashed{#1}, \ \dashed{#2},\ \dashed{#3} \Vdash_R ~\dashed{#4} ]_{\Queue}}}
\newcommand{\qentailsLR}[4]{\ensuremath{[\dashed{#1}, \ \dashed{#2},\ \dashed{#3} \Vdash_{LR} ~\dashed{#4} ]_{\Queue}}}
\newcommand{\qentailsRR}[3]{\ensuremath{[\dashed{#1}, \ \dashed{#2},\ \dashed{#3} \Vdash_R ~}}

\newcommand{\mentailsLR}[4]{\ensuremath{[\dashed{#1}, \ \dashed{#2},\ \dashed{#3} \Vdash_{LR} ~\dashed{#4} ]_{\Lbl}}}

\newcommand{\rentailsL}[4]{\ensuremath{[\dashed{#1}, \ \dashed{#2},\ \dashed{#3} \Vdash_L ~\dashed{#4} ]_{R}}}
\newcommand{\rentailsR}[4]{\ensuremath{[\dashed{#1}, \ \dashed{#2},\ \dashed{#3} \Vdash_R ~\dashed{#4} ]_{R}}}
\newcommand{\rentailsLR}[4]{\ensuremath{[\dashed{#1}, \ \dashed{#2},\ \dashed{#3} \Vdash_{LR} ~\dashed{#4} ]_{R}}}

\newcommand{\Get}{\ensuremath{\tm{Get}}}
\newcommand{\Put}[1]{\ensuremath{\tm{Put~{#1}}}}
\newcommand{\svar}[1]{\ensuremath{\tm{var}~{#1}}}

\newcommand{\Queue}{\ensuremath{\mathcal{Q}}}

\newcommand{\MS}{\ensuremath{\mathcal{M}_\mathcal{S}}}
\newcommand{\Lbl}{\ensuremath{\mathcal{S}}}

\newcommand{\Pid}{\ensuremath{\tm{PID}_n}}
\newcommand{\Enet}{\ensuremath{E_{\tm{net}}}}
\newcommand{\hnet}{\ensuremath{h_{\tm{net}}}}
\newcommand{\Msg}{\ensuremath{\tm{Msg}_n}}
\newcommand{\Mailboxes}{\ensuremath{[\Msg]_n}}

\newcommand{\proc}[1]{\ensuremath{\tm{proc} \spaced{#1}}}

\usepackage{tabularray}

\NewEnviron{typsyntax}[2]{%
  \begin{flushleft}
    \begin{tabular}{l l}
      $\typed{#1}{#2}~=$
      &
        $\begin{aligned}[t] % Start aligned in math mode
          \BODY
        \end{aligned}$ % End aligned and math mode
    \end{tabular}
  \end{flushleft}
}

\NewEnviron{typcosyntax}[2]{%
  \begin{flushleft}
    \begin{tabular}{l l}
      $\typed{#1}{#2}~=$
      &
        $\begin{aligned}[t] % Start aligned in math mode
          \BODY
        \end{aligned}$ % End aligned and math mode
    \end{tabular}
  \end{flushleft}
}

\NewEnviron{typfunction}[2]{%
  \begin{flushleft}
    \begin{tabular}{l}
      \ensuremath{\mathtt{#1}~\textcolor{gray}{\in~#2}} \\[.5em]
      {\begin{minipage}{\textwidth}
          $\begin{aligned} % Start aligned in math mode
            \BODY
          \end{aligned}$
        \end{minipage}
      } % End aligned and math mode
    \end{tabular}
  \end{flushleft}
}

\newcommand{\typequation}[3]{
  \begin{typsyntax}{#1}{#2}
    #3 \\
  \end{typsyntax}
}

\newcommand{\typcoequation}[3]{
  \begin{typcosyntax}{#1}{#2}
    #3 \\
  \end{typcosyntax}
}

\newcommand{\sys}{\texttt{Ticl}\xspace}

\setcopyright{cc}
\setcctype{by}
\acmDOI{10.1145/3763091}
\acmYear{2025}
\acmJournal{PACMPL}
\acmVolume{9}
\acmNumber{OOPSLA2}
\acmArticle{313}
\acmMonth{10}
\received{2024-10-16}
\received[accepted]{2025-08-12}


\begin{document}
\SetAlgoNoLine
\SetInd{0.3em}{0.2em}
\SetStartEndCondition{ (}{)}{)}
\SetAlgoBlockMarkers{}{\}}%
\SetKwFor{While}{while}{ \{}{}
\SetKwIF{If}{ElseIf}{Else}{if}{ then}{elif}{else}{}%
\AlgoDisplayBlockMarkers

\title{Structural Temporal Logic for Mechanized Program Verification}

\authoranon{
\author{Eleftherios Ioannidis}
\orcid{0000-0003-2749-797X}
\affiliation{
  \institution{University of Pennsylvania}
  \city{Philadelphia}
  \state{PA}
  \country{United States}
}
\email{elefthei@cis.upenn.edu}

\author{Yannick Zakowski}
\orcid{0000-0003-4585-6470}
\affiliation{
  \institution{ENS Lyon, Inria}
  \city{Lyon}
  \country{France}
}
\email{yannick.zakowski@inria.fr}

\author{Steve Zdancewic}
\orcid{0000-0002-3516-1512}
\affiliation{
  \institution{University of Pennsylvania}
  \city{Philadelphia}
  \state{PA}
  \country{United States}
}
\email{stevez@cis.upenn.edu}

\author{Sebastian Angel}
\orcid{0000-0002-3798-5590}
\affiliation{
  \institution{University of Pennsylvania}
  \city{Philadelphia}
  \state{PA}
  \country{United States}
}
\email{sebastian.angel@cis.upenn.edu}
}

\begin{abstract}
  Mechanized verification of liveness properties for
  infinite programs with effects and nondeterminism is challenging.
  Existing temporal reasoning frameworks operate at the level of models such as traces and automata.
  Reasoning happens at a very low-level, requiring complex nested
  (co-)inductive proof techniques and familiarity with proof assistant
  mechanics (e.g., the guardedness checker).
  Further, reasoning at the level of models instead of
  program constructs creates a verification gap that loses the benefits of
  modularity and composition enjoyed by structural program logics such as Hoare
  Logic. To address this verification gap, and the lack of compositional proof techniques for temporal specifications,
  we propose \sys, a new structural temporal logic. Using \sys, we encode complex (co-)inductive proof techniques
  as structural lemmas and focus our reasoning on variants and
  invariants. We show that it is possible to perform compositional proofs of
  general temporal properties in a proof assistant, while working at a
  high level of abstraction.
  We demonstrate the benefits of \sys by giving mechanized proofs of safety and liveness properties for programs with
  scheduling, concurrent shared memory, and distributed consensus, demonstrating a low proof-to-code ratio.
\end{abstract}
\maketitle

\section{Introduction}

Mechanized program verification can guarantee that executable code
satisfies formal specifications categorized as either liveness or safety
properties. Liveness properties (``a good thing happens'') include
\emph{termination} and \emph{fairness}, as well as
\emph{always-eventually} properties. Liveness
properties appear in web servers (``the server
\emph{always-eventually} replies to requests''), operating systems
(``the memory allocator will \emph{eventually} return a memory page'',
``the scheduler is \emph{fair}'') and distributed protocols (``a
consensus is \emph{eventually} reached''). Despite their prevalence in
computer systems, liveness properties have been understudied compared
to safety properties (``a bad thing never happens''), for which
numerous general reasoning frameworks and verifications techniques
exist~\cite{appel2001indexed,jung2018iris,ahman2017dijkstra,maillard2019dijkstra,winterhalter2022partial,silver2021dijkstra}.

Arguably, the widespread success of mechanized safety verification has
been due to the development of program logics that are \emph{compositional}
and reason directly over the \emph{structure} of programs. An example is Hoare logic, with its basic construct, the Hoare triple $\{P\}\ c\ \{Q\}$, which specifies that if the precondition $P$ holds before executing the command $c$,
and $c$ terminates, then the postcondition $Q$ will hold afterward.
Hoare logic allows one to perform local reasoning by breaking down complex
programs into small components, and to verify individual parts without needing to understand the
whole. Then, using the \emph{sequence rule}, one can combine triples
$\{P\}~c_1~\{Q\}$ and $\{Q\}~c_2~\{R\}$ to get $\{P\}~c_1;c_2~\{R\}$, building bigger proofs from smaller subproofs.
Hoare rules are structural: they allow reasoning over standard program
constructs like assignment (\IAssign{x}{a}), conditionals
(\IIf{c}{a}{b}), and loops (\IWhile{c}{b}), hiding their semantic
interpretations.

Unfortunately, this picture could not be more different when it comes
to proving liveness properties. While there are very powerful logics
for reasoning about general concepts of progress and time, namely
\emph{temporal
logics}~\cite{pnueli1977temporal,emerson1982using,browne1988characterizing,
alur2002alternating, lamport1994temporal, kozen1984decision}, these
tend to focus on \emph{semantic models} of program
execution. In other words, instead of writing proofs about
standard program constructs as shown above, one first models programs
as automata or infinite traces and then reasons about these models instead~\cite{pnueli1977temporal,emerson1982using,sistla1987complementation,denicola1990action,alur2002alternating,
doczkal2016completeness,hawblitzel2015ironfleet}.
Mechanized reasoning in these semantic models is arduous, requiring nested induction and coinduction
techniques (Section~\ref{s:ticl:low-level}) and deep understanding of
complex mathematical concepts like the Knaster-Tarski lemma
(Appendix~\ref{appendix:ticl:coinduction}), and the proof assistant's
mechanics (e.g., the guardedness checker). Additionally, semantic proofs of liveness do not  compose with respect to
the \emph{sequence} and \emph{iteration} operators, causing proof scalability issues
for large programs.

{\bf Contributions:}
We introduce \emph{Temporal Interaction and Choice Logic}
(\ticl), a novel program logic inspired by \emph{Computation Tree
  Logic} (CTL)~\cite{emerson1982using} that is designed for modular,
mechanized verification of liveness and safety properties. \Ticl
extends CTL with program postconditions, similar to those in Hoare
logic. Using \ticl one can write and prove temporal specifications
(e.g., \emph{always, eventually, and always-eventually}) at a high-level of
abstraction. \Ticl proofs \emph{compose} with the \emph{sequence} and \emph{iteration} operators, addressing the
long-standing challenge of compositional verification of liveness properties. \Ticl has three goals:
\begin{enumerate}
\item Combine temporal specifications over finite and infinite traces
  in one proof system. This part is crucial for supporting
  composition, as \ticl needs ways to express postconditions that
  apply to the return values of terminating programs while also being
  able to handle programs that run forever.
\item Close the verification gap between executable programs and the
  formal models used in temporal logics (e.g., traces and transition systems).
  \Ticl achieves this using a new mathematical model of computation
  that we call \ICTrees. As a part of the Interaction Trees
  family~\cite{xia2019interaction, chappe2023choice}, \ICTrees encode
  programs in different programming
  languages~\cite{ZBY+21,LX+22,XZHH+20,chappe2025monadic,silver2023interaction,lee2023fair,SH+23}
  with support for program extraction and formal transition system semantics.
\item Develop a library of 50 high-level structural lemmas that proof
  engineers can readily apply to programs in order to prove liveness properties. These lemmas internalize more than 20,000 lines of
  complex (co-)inductive proofs, hiding that complexity from the user.
  To use these lemmas, proof engineers must define their programs
  using \ICTrees and then write temporal specifications
  as \ticl formulas (Section~\ref{s:ticl:usage}).
  These specifications are then proved structurally,
  in a manner similar to Hoare Logic proofs, and without the usual (co-)induction
  bureaucracy.
\end{enumerate}

We demonstrate that \ticl is sufficiently expressive to prove
meaningful safety and liveness specifications with a small
proof-to-program ratio. We use examples spanning
sequential, concurrent, and distributed programming: imperative
programs with heaps, a round-robin scheduler, concurrent programs with
shared memory, and a simple distributed consensus protocol. Our
development is formalized in the Rocq proof
assistant~\cite{rocq-refman} (formerly known as ``Coq''), relying
solely on the \emph{uniqueness of identity proofs} axiom (UIP or
\tm{eq\_rect\_eq}).  \Ticl is released under an open-source
license\textanon{\footnote{\href{https://github.com/vellvm/ticl}{https://github.com/vellvm/ticl}}}{
  and a link is included in the deanonymized version of this paper}.

{\bf Related Work:} Beyond LTL and CTL
(Section~\ref{s:ticl:discussion} offers a deeper comparison),
step-indexed logical relation frameworks like Iris
~\cite{appel2001indexed,jung2018iris} can prove safety but not
liveness properties. More recently, transfinite extensions to
step-indexing~\cite{spies2021transfinite} made it possible to prove
\emph{always} properties but not \emph{always-eventually}
properties. Certain liveness properties have been studied in a
\emph{syntactic}
setting~\cite{lee2023fair,dosualdo2021tada,liang2016program} but these
are limited in expressivity and do not provide a general framework for
arbitrary temporal specifications. For example, Fair Operational
Semantics~\cite{lee2023fair} are limited to binary
\emph{always-eventually} properties, specifically \emph{good} vs.
\emph{bad} events, and do not generalize to arbitrary liveness
properties. Many deductive verification frameworks for temporal
properties, for example Cyclist~\cite{tellez2017automatically},
CoqTLA~\cite{chajed2024coqtla} and the Maude
language~\cite{meseguer1992conditional} operate on the semantic level of
models, not on the syntactic level of code, missing the advantages of
structural program logics.

{\bf Limitations:} \ticl has extensive support for backwards reasoning
(systematically weakening a goal specification into smaller subgoals and proving
them), less support is included at this point for forward reasoning
(strengthening and combining known hypotheses to create new hypotheses). Some
support for forward reasoning is offered through custom tactics and inversion
lemmas we developed.
Still, as we
report in the feature table of Figure~\ref{f:ticl:ticl-table}, proving forward
reasoning principles for some of \ticl's constructions remains open question, which we
leave for future work.
\Ticl also inherits the same limitations of completeness (with respect to
specifying liveness properties) found in prior variants of temporal
logics~\cite{alpern1987recognizing,wolper83}.  We discuss this in more detail in
Section~\ref{s:ticl:liveness}.

\iffalse
%% Lef: Remove if short on space
{\bf Paper structure: } Section~\ref{s:ticl:definitions} defines
\ICTrees, their combinators, equivalence, Kripke semantics,
the syntax and semantics of \ticl formulas, and \ticl formula
equivalence.  Section~\ref{s:ticl:lemmas} presents the \ticl library
of structural, temporal logic lemmas over
\ICTrees. Section~\ref{s:ticl:usage} shows how to use \ticl to prove
temporal properties for the \stimp programming language.
Section~\ref{s:ticl:examples} walks through three examples of \ticl
safety and liveness proofs; a round-robin scheduler (always-eventually),
concurrent shared memory (always) and distributed consensus (eventually).
Section~\ref{s:ticl:discussion} compares \ticl with past related works
from temporal logics and program logics.
\fi

\section{Why are liveness properties so challenging to prove?}\label{s:ticl:low-level}
\begin{figure}[t!]
  \begin{center}
  \small
  \begin{minipage}{0.26\textwidth}
    \underline{Program (\tm{rr})} \\
    \footnotesize
    \begin{algorithm}[H]
      \While{true}{
        $p \gets \text{\textbf{pop}}()$\;
        $\text{\textbf{push}}(p)$
      }
    \end{algorithm}
  \end{minipage}
  \begin{minipage}{0.55\textwidth}
    $\forall~\bm{x},q,$\\[.5em]
    \hspace*{.25cm}\underline{State ($q_\tm{init}$)} \\[.3em]
    \hspace*{.5cm}$q \app [\bm{x}]$ \\[.5em]
    \hspace*{.25cm}\underline{Specification (\emph{always-eventually})} \\[.3em]
    \hspace*{.5cm}$\entails{\tm{rr}}{q_\tm{init}}{\AG{\AF{(\lambda~\tm{running}\Rightarrow \tm{running}=\bm{x})}}}$
  \end{minipage}
  \end{center}
  \caption{Example of a round-robin scheduler program (\tm{rr}). The
    initial state of the program (\tm{q_{init}}) consists of a
    designated thread \tm{x} appended to a list of other threads
    \tm{q}. The liveness specification, given in
    CTL~\cite{emerson1982using}, asserts that for all possible
    designated threads (\tm{x}) and for all initial lists of other
    threads (\tm{q}), thread \tm{x} will always (\tm{AG})-eventually
    (\tm{AF}) be popped from the queue (i.e., \tm{running = x}).}
\label{f:ticl:rotate}
\end{figure}

We illustrate the challenges of formally proving a basic liveness property
  for a small program.
Consider an operating system that maintains a queue of threads with some tasks, and
  a round-robin scheduler that processes each thread one after the other.
The program \tm{rr} in Figure~\ref{f:ticl:rotate} implements round-robin scheduling---a simple
  infinite loop removes a thread from the head of the queue and re-inserts
  it at the end. Our goal is to prove that a thread $\bm{x}$ will
  \emph{always-eventually} be scheduled (\tm{AG~AF}
  using CTL notation~\cite{emerson1982using}).

One approach to mathematical reasoning about infinite programs is to
  represent them as coinductive trees of events.  The infinite loop in
  Figure 1 unfolds to a coinductive stream of alternating $[\tm{pop},
  \tm{push}, \tm{pop}\ldots]$ events. Popping removes an
  element from the \emph{head} of the queue, while pushing 
  appends an element. Applying an infinite stream of alternating
  \tm{pop} and \tm{push} events to the initial queue state, $q_\tm{init}$, results in a coinductive tree of queue states,
  as shown in Figure~\ref{f:ticl:rotate-instr}.
  Each infinite trace depends on the length of $q$.

The goal property (``always-eventually $x$ is running'') is a nested inductive and coinductive predicate over
the coinductive tree in Figure~\ref{f:ticl:rotate-instr}.
Proving this property requires nested induction on the length of $q$, and
coinduction on each trace.  The proof is hard---working directly
with trees of traces and low-level induction/coinduction tactics is
neither \emph{modular} nor \emph{structural}. The trivial-looking example of
Figue~\ref{f:ticl:rotate} requires a non-trivial amount of
infrastructure to prove, most of which is not reusable for other
programs and specifications. With \ticl, proving the example from
Figure~\ref{f:ticl:rotate} is reduced to a simple application of the
\emph{invariance} rule for while loops (for a preview, see
Figure~\ref{f:ticl:meq-proof}).

\begin{figure}[t!]
  \vspace*{-.9cm}
\begin{tikzpicture}[
  grow = right, % Change direction to left-to-right
  sibling distance = 1.5em,
  level distance = 8em,
  edge from parent/.style = {draw, -latex}, % Arrow style
  every node/.style = {minimum size=1.8cm, align=center, fill=none, draw=none} % No fill, no box
  ]
  % Common parent node
  \node (parent) { $q \app [x]$ }
    % Chain 4: q_0 ... q_n -> [q_0 ... q_n, x] -> [...] -> [...]
    child { node { $[q_0,q_1, \dots, q_n, x]$ }
      child { node { $[q_1, \dots, q_n, x]$ }
        child { node { $[q_1,\dots, q_n, x, q_0]$ }
          child { node { $\dots$ }
          }
        }
      }
    }
    % Chain ...
    child { node { $\vdots$ } }
    % Chain 3: q_0, q_1 -> [q_0, q_1, x] -> [q_1, x] -> [q_1, x, q_0] -> ...
    child { node { $[q_0, q_1, x]$ }
      child { node { $[q_1, x]$ }
        child { node { $[q_1, x, q_0]$ }
          child { node { $\dots$ }
          }
        }
      }
    }
    % Chain 2: q_0 -> [q_0, x] -> [x] -> [x, q_0] -> ...
    child { node { $[q_0, x]$ }
      child { node { $[x]$ }
        child { node { $[x, q_0]$ }
          child { node { $\dots$ }
          }
        }
      }
    }
    % Chain 1: [] -> [x] -> [] -> [x] -> ...
    child { node { $[x]$ }
      child { node { $[]$ }
        child { node { $[x]$ }
          child { node { $\dots$ }
          }
        }
      }
    };
  \end{tikzpicture}
  \vspace*{-0.5cm}
  \caption{Coinductive traces for the \tm{rr} program (Figure~\ref{f:ticl:rotate}) and all possible initial states $q \app [x]$.}
  \label{f:ticl:rotate-instr}
\end{figure}

\section{Computational model: \ICTrees}\label{s:ictree}
In this section we introduce a model of computation that we call
\emph{Interaction and Choice Trees} (\ICTrees).
This denotational model, inspired by prior
works~\cite{xia2019interaction,chappe2023choice}, is expressive enough
to formalize programming languages with nondeterminism, nontermination,
and interaction (we discuss how it relates to prior works in detail in
Section~\ref{s:discussion:trees}).
Further, \ICTrees have standard combinators like sequencing (bind), iteration
and choice and support semantic interpretation~\cite{xia2019interaction}.
In Sections~\ref{s:ticl:usage},\ref{s:ticl:examples} we define the semantics of several programming languages using
\ICTrees, and write example programs in those languages.

\subsection{The \tm{ictree} computational model}\label{s:ticl:ctree-core}

The \ICTree coinductive datastructure represents (in-)finite,
nondeterministic, and effectful programs.
\ICTrees are defined in Figure~\ref{f:ticl:ctree-def} using four
kinds of nodes: visible event nodes (\tm{Vis}), silent nodes (\tm{Tau}),
nondeterministic choice nodes (\tm{Br}), and nodes returning a value
(\tm{Ret}).

\begin{figure}
  {\small
    \begin{typfunction}{\ictree{}{}}{(\Type \to \Type) \to \Type \to \Type}
    \ictree{E}{X} \coindeq &\quad \bbar \Ret{(\typed{x}{X})} && \bbar \VisX{(\typed{X}{\Type})}{(\typed{e}{E~X})}{(\typed{k}{X \to \ictree{E}{X}})} \\
    \quad                       &\quad \bbar \Tau{(\typed{t}{\ictree{E}{X}})} && \bbar \Br{(\typed{n}{\mathbb{N}})}{(\typed{k}{\fin{n} \to \ictree{E}{X}})}
  \end{typfunction}
  
  \typcoequation{\emptyset}{\ictree{E}{X}}{\Tau{\stuck}}

  \begin{typfunction}{\bindc~}{\ictree{E}{X} \to (X \to \ictree{E}{Y}) \to \ictree{E}{Y}}
    &&(\Ret{x})~&\bindc ~f   ~ = ~ f~x,                  && (\VisX{X}{e}{k})~&\bindc ~f ~ \coindeq ~ \VisX{X}{e}{(\lambda~(\typed{x}{X}) \Rightarrow (k~x) \bindc f)} \\
    &&(\Tau{t})~&\bindc ~f ~ \coindeq ~ \Tau{(t \bindc f)}, && (\Br{n}{k})~&\bindc ~ f  ~ \coindeq ~ \Br{n}{(\lambda~(\typed{i}{\fin{n}}) \Rightarrow (k~i) \bindc f)}
  \end{typfunction}

  \typequation{(x~\tm{\gets}~t~\text{;;}~k~x)}{\ictree{E}{Y}}{(\typed{t}{\ictree{E}{X}})~\bindc~\lambda~(\typed{x}{X}) \Rightarrow
    (\typed{k}{X \to \ictree{E}{Y}})~x}

  \begin{typfunction}{\itern}{(I \to \ictree{E}{I + R}) \to I \to \ictree{E}{R}}
    && \iter{step}{i} ~ \coindeq ~ (\tm{step~i}) \bindc \lambda~(\typed{lr}{I+R}) \Rightarrow
    \begin{cases}
    \Tau{(\iter{step}{i'})}, & lr = \inl{i'} \\
    \Ret{(r)},               & lr = \inr{r}
    \end{cases}
  \end{typfunction}

  \typequation{\tm{trigger}~ (\typed{e}{E~X})}{\ictree{E}{X}}{ \VisX{X}{e}{(\lambda~(\typed{x}{X}) \Rightarrow \Ret{x})}}
  \typequation{\tm{branch} ~ (\typed{n}{\mathbb{N}})}{\ictree{E}{\fin{n}}}{ \Br{n}{(\lambda~(\typed{i}{\fin{n}}) \Rightarrow \Ret{i})}}

  \begin{typfunction}{\oplus~}{\ictree{E}{X} \to \ictree{E}{X} \to \ictree{E}{X}}
    && \tm{l}~\oplus~\tm{r} ~=~ \Br{\_}{}~\biggl(\lambda~(\typed{i}{\fin{2}}) \Rightarrow
    \begin{cases}
        \tm{l}, & i = F_1 \\
        \tm{r}, & i = FS~F_1 \\
    \end{cases}\biggr)
  \end{typfunction}
}
\caption{Definition of the \ICTree datastructure and core \ICTree combinators.}
  \label{f:ticl:ctree-def}
\end{figure}

%% (Vis)
\tm{Vis} nodes store events (\typed{E}{Type \to Type}) representing
interactions of the program with the environment. An event
(\typed{e}{E~X}) is an action, expecting an environment response
($X$). For example, \tm{Pop} events emitted by the \textbf{pop}()
command in Figure~\ref{f:ticl:rotate} have type $E_\Queue~\mathbb{N}$
(Figure~\ref{f:ticl:meq-semantics}) and expect a natural number
response ($\mathbb{N}$) representing a thread ID. \tm{Vis} nodes have a child
node for every response---in the case of \tm{Pop} there is one child for
every possible natural number (Figure~\ref{f:ticl:ictrees}).

\begin{figure}
  \[
    \begin{array}[t]{ccc}
      \sembra{}{\tm{flip}} & \quad\quad\sembra{\Queue}{rr} & \instrv{h_\Queue}{\sembra{\Queue}{rr}}{[t_1,t_2]}  \\[.5em]
{\small
\begin{forest}
  % for tree={circle, , l sep=0.3cm, s sep = 5pt}
  [\Br{2}{}
    [\Ret{H}]
    [\Ret{T}, name=ret]
  ]
  \draw[dotted, draw opacity=0] (ret) -+(0, -2.3);  
\end{forest}}
      &
{\small
\begin{forest}
  % for tree={circle, , l sep=0.3cm, s sep = 5pt}
  [\Vis{\tm{Pop}}{}
    [\Vis{(\tm{Push}~0)}{}[\Tau{}, name=tau0]]
    [\Vis{(\tm{Push}~1)}{}[\Tau{}, name=tau1]]
    [\dots]
  ]
  \draw[dotted] (tau0) --+(0,-.6);
  \draw[dotted] (tau1) --+(0,-.6);
\end{forest}}
      &
{\small
\begin{forest}
  % for tree={circle, , l sep=0.3cm, s sep = 5pt}
  [\Vis{(\Log{t_1})}{}
  [\Tau{}
  [\Vis{(\Log{t_2})}{}, name=vis]]
  ]
  \draw[dotted] (vis) --+(0,-.6);
\end{forest}}
\end{array}
\]
\caption{Example \ICTrees denoting programs $\tm{flip}$, $\tm{rr}$, and the instrumentation of $\tm{rr}$ with two threads.}
\label{f:ticl:ictrees}
\end{figure}

%% (Br) ICTrees vs CTrees
\tm{Br} nodes represent finitary, nondeterministic choice.
For example, the program that flips a coin that can be either
heads or tails ($\tm{flip} ~\coloneq~H \oplus T$) denotes to the binary choice (\tm{Br} 2) node shown in Figure~\ref{f:ticl:ictrees} (left).
These nodes count as a ``step'' of computation that nondeterministically chooses
among $n$ possible continuations (see transition relation in
Section~\ref{s:ticl:ctree-kripke}). They are inspired by the
equivalent notion of so-called \textit{stepping} branching nodes
(\tm{BrS}) used by Choice Trees~\cite{chappe2023choice}.

%%% (Ret) and (Tau)
\tm{Ret} nodes capture the return value of a terminating computation (\Ret{} nodes have no children).
\tm{Tau} nodes are \emph{silent} steps, representing
internal computation. Their name originates from $\tau$ transitions in CCS~\cite{milner1980calculus}
which indicate an internal action that is separate from process communication.
\tm{Tau} nodes are necessary to model programs with control flow that 
might not terminate (e.g., program \tm{rr} in Figure~\ref{f:ticl:rotate}).
In the next section we introduce the \ICTrees transition relation, which transitively closes over \tm{Tau} nodes, making them unobservable to \ticl formulas (Figure~\ref{f:ticl:kripke-def}).
The idea is, we only care about program specifications with respect to their observable behavior. Hiding
\tm{Tau} nodes in the transition relation and equational theory---as we will see next---makes
the internal behavior of programs (e.g., unobservable control flow) irrelevant.
The \emph{stuck} (\tm{\emptyset}) \ICTree represents the deadlocked
program that cannot make progress, and is defined as an infinite
chain of \tm{Tau} nodes.

%%% ICTree combinators (bind, \oplus, trigger, branch)
\ICTrees are \textit{monads}, meaning the monadic composition operations bind (\bindc) and return (\Ret{}) are defined
(Figure~\ref{f:ticl:ctree-def}) and satisfy the monad laws (Figure~\ref{f:ticl:ctree-algebra}).
Such sequential composition of both finite and infinite programs is a key property of \ICTrees that makes them suitable models for \Ticl,
a logic supporting both finite and infinite specifications. Nondeterministic binary choice ($\oplus$) is defined by
matching on the nondeterminstic result of $\Br{2}{}$. Operations \trigger{}{} and \branch{}{} are wrappers
around visible event nodes (\Vis{}{}) and $n$-ary nondeterministic choice (\Br{}{}).

%%% ICTree iter
Looping programs---both finite and infinite---can be constructed through the \itern combinator in
Figure~\ref{f:ticl:ctree-def}, which accepts a \emph{stepping function}
(\typed{k}{I \to \ictree{E}{I+R}}) and seed ($\typed{i}{I}$). If the stepping function returns a value of type $I$
(iterator), the loop continues with a new iterator.
Otherwise, it terminates with a return value of type $R$ (result).
Note that \itern emits a \Tau{} node every loop repetition, indicating
the internal control flow.  For example, in Figure~\ref{f:ticl:ictrees} (center) we show how program \tm{rr} (encoded in \ICTrees from the code in Figure~\ref{f:ticl:rotate}) infinitely repeats after the \Tau{} node.

\begin{figure}[t!]
{\footnotesize
  \begin{mathpar}
    \inferrule*[right=SbRefl]{~}{t \sbisim t}
    \and
    \inferrule*[right=SbSym]{t \sbisim u}{u \sbisim t}
    \and
    \inferrule*[right=SbTrans]{t \sbisim u \and u \sbisim v}{t \sbisim v}
    \and
    \inferrule*[right=SbTau]{~}{\Tau{t}\sbisim t}
    \and
    \inferrule*[right=SbBind]{t \sbisim u \and (\forall x,~g~x\sbisim k~x)}{t\bindc g \sbisim u \bindc k}
    \and
    \inferrule*[right=SbBindL]{~}{\Ret{v} \bindc k \ \sbisim \ k~v}
    \and
    \inferrule*[right=SbBindR]{~}{\bind{t}{\mathtt{Ret}} \ \sbisim \ t}
    \and
    \inferrule*[right=SbBindAssoc]{~}{(t \bindc k) \bindc l \ \sbisim \  t \bindc (\lambda x \Rightarrow k~ x \bindc l)}
    \and
    {\mprset { fraction ={===}}
      \inferrule*[right=SbRet] {x = y} {\Ret{x} \sbisim \Ret{y}}
    }
    \and
    {\mprset { fraction ={===}}
      \inferrule*[right=SbVis]{\forall x,~h~x\sbisim k~x} {\Vis{e}{h} \sbisim \Vis{e}{k}}
    }
    \and
    {\mprset { fraction ={===}}
      \inferrule*[right=SbBr]{\forall x, ~h~x\sbisim k~x} % This is actually weaker than the LTS equiv in the repo
      {\Br{n}{h}\sbisim\Br{n}{k}}
    }
  \end{mathpar}
}
    \caption{Equational theory for \ICTree with respect to \emph{up-to-tau} equivalence relation ($\sbisim$). A double inference line indicates a coinductive rule.}
\label{f:ticl:ctree-algebra}
\end{figure}

% \begin{figure}
%   \begin{mathpar}
%     \interpstate{h}{(\Ret{r})}{w} \cong \Ret{(r, w)}
%     \and
%     \interpstate{h}{(\Tau{t})}{s} \cong \interpstate{h}{t}{s}
%     \and

%     \interpstate{h}{(\Vis{e}{k})}{s} \cong (\runStateT{(h~e)}{s}) \bindc
%     \lambda `(r, s) \Rightarrow \Tau{(\interpstate{h}{(k~x)}{s})}
%     \and
%     \interpstate{h}{(\Br{n}{k})}{s} \cong \Br{n}{(\lambda x \Rightarrow \Tau{(\interpstate{h}{(k~x)}{s})})}
%     \and
%     \interpstate{h}{(t \bindc k)}{s} \cong
%     (\interpstate{h}{t}{s}) \bindc \lambda `(r, s) \Rightarrow \interpstate{h}{k~x}{s}
%   \end{mathpar}
%   \caption{State interpreter equations for \ctrees.}
%   \label{f:ctree-instrument}
% \end{figure}

%% Up-to-tau equivalence
The addition of \Tau{} nodes makes programs with the same observable behaviors appear
different syntactically (e.g., a loop and its unfolding have the same behavior but are syntactically different).
This is undesirable because later when we introduce \ticl formulas 
we will want to say things like: a program with a finite loop terminates if and only if
the equivalent program by unrolling the loop terminates. 
To make such reasoning possible, we define a notion of \ICTree 
equivalence, called \emph{up-to-tau} equivalence, that \ICTrees inherits
from prior Interaction Tree works~\cite{xia2019interaction,chappe2023choice}.
Up-to-tau equivalence is invariant to (inductively) adding or removing taus on either side; it equates programs 
with the same observable behavior. For instance, it \textit{does not} equate the stuck \tm{\emptyset} tree with any terminating tree, such as \Ret{x}, but it does equate \tm{Ret\; x} with \Tau{(\Ret{x})}, since both terminate and return the same \tm{x}.   Up-to-tau equivalence ($\sbisim$) is defined coinductively in Figure~\ref{f:ticl:ctree-algebra}, along with some useful equations.
With this equational theory, programs can be simplified using the
monad laws, removing taus, unfolding loops, and more.

\begin{figure}
  {\small
  \typequation{\bm{\logE{W}}}{\Type \to \Type}{\bbar\typed{\Log{(\typed{w}{W})}}{\logE{W}~\unit}}
  \typequation{\bm{\tm{log}}~(\typed{w}{W})}{\ictree{\logE{W}}{\unit}}{\trigger{\Log{w}}}
  \typequation{\bm{\InstrM{S}{W}}}{\Type \to \Type}{\stateT{S}{\ictree{\logE{W}}{}}}

  \begin{typfunction}{\bm{\tm{instr}}}{ (E \leadsto \InstrM{S}{W}) \to \ictree{E}{} \leadsto \InstrM{S}{W}}
    &\quad && \instrv{h}{(\Ret{x})}{s} ~ = ~ \Ret{(x, s)},\\
    &\quad && \instrv{h}{(\Tau{t})}{s} ~ \coindeq ~ \Tau{(\instrv{h}{t}{s})} \\
    &\quad && \instrv{h}{(\VisX{X}{e}{k})}{s} ~ \coindeq ~ (h~e~s)~\bindc~(\lambda~`(\typed{x}{X}, \typed{s'}{S}) \Rightarrow \Tau{(\instrv{h}{(k~x)}{s'})}) \\
    &\quad && \instrv{h}{(\Br{n}{k})}{s} ~ \coindeq ~ \Br{n}{(\lambda~(\typed{i}{\fin{n}}) \Rightarrow \instrv{h}{(k~i)}{s})})\hspace*{4cm}
  \end{typfunction}
  }
  \caption{Instrumentation of an \ictree{E}{} produces the monad
    $\InstrM{S}{W}$, that ``remembers'' the temporal order of events $E$ via observations $\logE{W}$.}
  \label{f:ticl:ctree-instr}
\end{figure}

\subsection{\ICTree semantics and instrumentation}\label{s:ticl:instrumentation}
Up until this point we have treated \ICTrees as abstract syntax.
We will now assign to them semantic meaning.

The semantic meaning of events is given by a semantic handler
$h : E \leadsto M$, where $M$ is a monad compatible with the $\ICTree$
structure.  For example, the ``$\tm{read}~x$'' and ``$\tm{write}~x~v$'' events
could be interpreted as functions defined to operate in a shared state
monad, where the state type ($S$) is a map from variables to their
values. There, we would have $h_S(\tm{read}~x)$ defined by the
implementation
$\lambda~(\typed{s}{S})~\Rightarrow~\Ret{(s, s[x])}$. That is, the function that takes the current state $s$ and returns it, along with the value of $x$ in
that state. The operation $h_S(\tm{write}~x~v)$ would be defined by
the function
$\lambda~(\typed{s}{S})~\Rightarrow~ \Ret{((x \hookrightarrow v) \cup s, \ttt)}$. That is, the function
that takes the current state and returns the modified state along with
the unit value (signifying that write does not itself return a value).
In general, the monad $M$ of a handler must be compatible with the
$\itern$ and $\oplus$ constructs required by \ICTrees.  In the case above,
the types of the \tm{read} and \tm{write} operations are of the form:
$S \to \ictree{\tm{void}}{(S \times X)}$---i.e., they are functions from the
starting state to a (potentially diverging) \ictree{}{}.  This type is an
instance of the \textit{state monad transformer}:
$\stateT{S}{(\ictree{\tm{void}}{})}$.

% We revisit semantic
% interpretation for \ticl in Section~\ref{s:ticl:instrumentation}.

% In Section~\ref{s:ticl:ctree-core} we briefly touched on how an 
% \ictree{E}{} represents side-effects through
% \Vis{}{} events, and how an interpretation handler of type
% $h_S:~E \leadsto \stateT{S}{\ictree{\tm{void}}{}}$ evaluates events ($E$) to
% a state monad---a partial function over states $S$.
However, interpretation is insufficient for temporal reasoning---in
addition to the result of a program, formulas may specify the temporal
order of events.  This is analoguous to big-step vs. small-step
semantics: \ICTree interpretation
($\ictree{E}{} \leadsto \stateT{S}{\ictree{\tm{void}}{}}$) erases the
small-step event information in the tree's \Vis{}{} nodes, returning a
big-step reduction in the form of the partial function
\stateT{S}{\ictree{\tm{void}}{}}. At that point the provenance of
events has been erased.

We address the loss of provenance by introducing the notion of
\emph{instrumentation}.  Intuitively, instrumentation defines which
events and their accompanying (ghost) state are considered to be
relevant for the \ticl specifications.  Event instrumentation
interprets an event $(\typed{e}{E~X})$ over a slightly different state
monad ($\stateT{S}{\ictree{\logE{W}}{}}$) that we call the instrumentation
monad (\InstrM{S}{W} in Figure~\ref{f:ticl:ctree-instr}).  Our goal is
to interpret events $E$, leaving behind a trace of \emph{observation}
events of type \logE{W}. Observation events, or \emph{log events}, are
themselves uninterpreted events---the environment response is always of
type \unit. Observations are left behind by event interpretation and
signify a ``memory'' of an environment interaction, recorded as an
observation of type $W$. Log events can be erased without altering the
semantics of the program. Values of type $W$ encode auxilary (ghost)
state that can be queried by \ticl formulas. The specification author
is free to pick an arbitrary type $W$ to observe and reason about.

For example, program (\tm{rr}) from
Figure~\ref{f:ticl:rotate} denotes to the \ICTree in the middle of
Figure~\ref{f:ticl:ictrees}. The instrumentation handler for queues
($h_\Queue$ defined later in Figure~\ref{f:ticl:meq-semantics}) logs
the result of \tm{Pop} events, while it interprets away \tm{Push}
events.  As a result, instrumenting \tm{rr} results in the infinite
trace on the right of Figure~\ref{f:ticl:ictrees}, showing every value
popped from the head of the queue. To make salient the use of instrumentation handlers, we
provide several examples in Sections~\ref{s:ticl:stimp-semantics} and~\ref{s:ticl:ex-queue}--\ref{s:ticl:ex-election}.

\section{Temporal specifications: \Ticl}\label{s:ticl}
\Ticl is defined using a ternary entailment relation
$\entails{t}{w}{p}$, which can be read as ``program $t$
satisfies formula $p$, starting at a world $w$''. Our goal in this
section is to define the necessary components of \ticl entailment; $t$
is an \ICTree describing (in-)finite,
effectful, nondeterministic programs, $w$ captures the initial state
of the external \emph{world}, and $p$ is a \ticl formula.

\subsection{Kripke transition relation}\label{s:ticl:ctree-kripke}
Temporal logics are commonly defined over traces or transition
systems, stepping from one ``world'' to another. This will also be true 
of \ticl, so we review this concept and describe how transitions apply to \ICTrees.

\begin{wrapfigure}{R}{0.3\textwidth}
  \centering
  {\small
  \begin{tikzpicture}
    \draw (10,10) node(p) {\Pure};
    \draw (9,9) node(v) {\Val{x}};
    \draw (11,9) node(o) {\Obs{e}{v}};
    \draw (11,8) node(f) {\Finish{e}{v}{x}};

    \draw [->] (p) -- (v);
    \draw [->] (p) -- (o);
    \draw [->] (o) -- (f);

    \draw [->] (p) to[loop right] ();
    \draw [->] (o) to[loop right] ();
  \end{tikzpicture}
    }
  \caption{Transitions between Kripke worlds for \ICTrees.} 
  \label{f:ticl:world-po}
\end{wrapfigure}
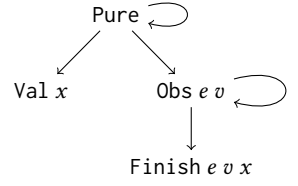
A Kripke \emph{world} (\World{E}), parametrized by an event type $E$, is a datatype that ``remembers'' the status of the program,
for instance, a past observation and/or its return value. 
A \Pure world indicates that no event has been observed yet. 
A world ``\Obs{e}{v}'' remembers the last observed event ($\typed{e}{E~X}$) and
the response it obtained from interacting with its environment ($\typed{v}{X}$). 
A world ``\Val{x}'' captures the return value
($x$) of a pure program that has terminated. A world ``\Finish{e}{v}{x}'' captures the
return value ($x$) of an effectful program that terminated, and the last event
($\typed{e}{E~X}$) and response ($\typed{v}{X}$). Worlds (\World{E}) are divided into
\tm{done} worlds (\tm{Val} and \tm{Finish}) and \notdone{} worlds (\tm{Pure} and \tm{Obs}),
indicating whether a program terminated or is still running. The predicate
\donewith{}{} (Figure~\ref{f:ticl:kripke-def})  enforces a postcondition ($\Px$) on a \tm{done} world and will
be used to define \ticl postconditions in Section~\ref{s:ticl:entails}.

\begin{figure}[t!]
  {\footnotesize
    \begin{mathpar}
      \typed{\World{E}}{\Type} ~=~ \Pure \sep \Obs{e}{v} \sep \Val{x} \sep \Finish{e}{v}{x} \\
      \notdone{\Pure}
      \and
      \notdone{(\Obs{e}{v})}
      \and
      \inferrule*{(\typed{\Px}{X \to \PROP}) \and \Px~x}{\donewith{\Px}{(\Val{x})}}
      \and
      \inferrule*{(\typed{\Px}{E~Y \to Y \to X \to \PROP}) \and Px~e~v~x}{\donewith{\Px}{(\Finish{e}{v}{x})}}
      \and
      \inferrule{\ktrans{t}{w}{t'}{w'}}{\ktrans{\Tau{t}}{w}{t'}{w'}}
      \and
      \inferrule{\notdone{w} \and 0 \leq i < n}{\ktrans{\Br{n}{k}}{w}{k~i}{w}}
      \and
      \inferrule{\notdone{w}}{\ktrans{\Vis{e}{k}}{w}{k\ v}{\Obs{e}{v}}}
      \and
      \ktrans{\Ret{x}}{\Pure}{\stuck}{\Val{x}}
      \and
      \ktrans{\Ret{x}}{\Obs{e}{v}}{\stuck}{\Finish{e}{v}{x}}
\end{mathpar}
}
\caption{Kripke transition relation for \ICTrees ($\mapsto$) and world predicates (\notdone{}, \donewith{}{}). $\Px$ is a type-theoretic \ticl postcondtion (described in Section~\ref{s:ticl:entails}).}
\label{f:ticl:kripke-def}
\end{figure}

Figure~\ref{f:ticl:kripke-def} defines the Kripke transition relation for 
  \ICTrees ($\ktrans{t}{w}{t'}{w'}$).
This is an irreflexive binary relation over pairs of \ictree{E}{} and 
  worlds (\World{E}), inductively defined over \Tau{} nodes.
Transitions only make sense in \notdone{} worlds since \done{} worlds 
  represent programs that have already terminated.
Within \notdone{} worlds, an \ICTree can either transition from 
  a \Pure world onto another \Pure world, it can
  observe an event and its result (\Obs{e}{v}),
  or it can terminate (\Val{x}).
An \ICTree program can transition from ``\Obs{e}{v}''
  onto another ``\Obs{e'}{v'}'' or it can terminate (\Finish{e}{v}{x}). 
We summarize these transitions in Figure~\ref{f:ticl:world-po}.

There are two goals informing our definition of the \ICTree transition
relation in Figure~\ref{f:ticl:kripke-def}: (1) respecting the
up-to-tau equivalence (Section~\ref{s:ticl:ctree-core}); and (2)
respecting the monad composition laws
(Figure~\ref{f:ticl:ctree-algebra}).  The first goal is achieved by
defining the transition relation inductively over \Tau{} nodes.  The
intuition is that \ticl observes the external behavior of programs,
and internal steps should not change the outcome of a \ticl
specification. Consequently, up-to-tau equivalent programs should
satisfy the same \ticl formulas. This is a restatement of the
well-known result that strong bisimulation preserves CTL
properties~\cite{EMERSON1990995}, which we are able to formally prove
in Section~\ref{s:ticl:uptotau-proper}.  The second goal---respecting
the laws of monadic composition---is achieved through the lemmas in
Figure~\ref{f:ticl:kripke-lemmas}. These lemmas describe how the
composition of two programs $t$ and $k$ ($\bind{t}{k}$) transition.
Either $t$ transitions to $t'$ and $\bind{t}{k}$ transitions to
$\bind{t'}{k}$, or $t$ terminates with return value $x$ (and possibly
an observation $e$) and then the continuation ($k~x$) transitions
($t'$). These lemmas allow us to break proofs of bind transitions into
smaller subproofs by case matching on $t$, a technique we use
thoroughly in our development.

\begin{adjustwidth}{0pt}{}
    \begin{figure}[t!]
      {\footnotesize
    \begin{mathpar}
      %% TrBindGoal
      \inferrule
      {\ktrans{t}{w}{t'}{w'} \and \notdone{w'}}
      {\ktrans{\bind{t}{k}}{w}{\bind{t'}{k}}{w'}}
      \and

      %% TrBindGoalVal
      \inferrule
      {\ktrans{t}{w}{\stuck}{\Val{x}} \and \ktrans{k~x}{w}{t'}{w'}}
      {\ktrans{\bind{t}{k}}{w}{t'}{w'}}
      \and

      %% TrBindGoalFinish
      \inferrule
      {\ktrans{t}{w}{\stuck}{\Finish{e}{v}{x}} \and \ktrans{k~x}{w}{t'}{w'}}
      {\ktrans{\bind{t}{k}}{w}{t'}{w'}}
    \end{mathpar}
    }
    \caption{Lemmas connecting Kripke transitions ($\mapsto$) to \ICTree composition.}
    \label{f:ticl:kripke-lemmas}
    \end{figure}
    \end{adjustwidth}

\subsection{Syntax of \ticl}\label{s:ticl:syntax}
A crucial question that \ticl must answer is how to handle both
infinite and terminating program specifications. Temporal logics like
LTL and CTL assume infinite traces, whereas finite LTL assumes finite
traces~\cite{pnueli1977temporal,emerson1986sometimes,de2013linear}.
Our goal is a specification language that works for both finite and
infinite programs \emph{compositionally}. For instance, an ``always''
proof should be broken up into a finite ``until'' prefix and an
infinite ``always'' suffix.

The \ticl syntax (Figure~\ref{f:ticl:ticl-syntax}) is inspired by
CTL~\cite{emerson1982using} using the same path-quantified temporal
operators, with some notable differences. There are two syntactic
categories in \ticl: \emph{prefix} formulas ($\varphi$) that represents
predicates on the prefix of a tree (or on infinite trees), and
\emph{suffix} ($\psix$) formulas that represent postconditions on
terminating trees. Suffix formulas (\psix) reference prefix formulas ($\varphi$) on the
left-hand side argument of their binary temporal operators
($\AN{}{}, \AU{}{}, \EN{}{}, \EU{}{}$). This is reasonable, as the
formula on the left must be satisfied before the one on the right.
Due to their appearance on the left side of temporal operators we also
refer to prefix formulas as \emph{left} ($L$) formulas and to suffix
formulas as \emph{right} ($R$). We assign meaning to formulas with the
two ternary entailment relations $\vDash_L$ and $\vDash_L$ in
Definition~\ref{f:ticl:entails-def}, and overload the notation
$\vDash_{LR}$ to indicate we refer to both $\vDash_L$ and $\vDash_R$.

The dual syntax is novel compared to LTL, CTL and
TLA~\cite{yu1999model}. To motivate the dual syntax, consider the
alternative---what if we only chose one syntactic class of \ticl formulas,
either suffix $(\psix)$ formulas or prefix formulas $(\varphi)$.
If we only have suffix formulas then every program (including infinite
programs) must have a postcondition. The only reasonable postcondition
for an infinite program is $\bot$, a choice made by partial correctness
program logics prohibiting sound proofs of liveness. If we
only have prefix formulas we lose program postconditions, and by
extension, sequential proof composition. At the risk of jumping ahead,
\entailsR{t}{w}{\AF{\AX{\done{\R}}}} means that program $t$ will
eventually terminate satisfying postcondition $\R$. As we will see
shortly in Section~\ref{s:ticl:lemmas}, a goal
\entailsL{\bind{t}{k}}{w}{\AF{\varphi}} can be broken into two subgoals
\entailsR{t}{w}{\AF{\AX{\R}}} and
$\forall~x~w, \R~x~w \to \entailsL{k~x}{w}{\AF{\varphi}}$, where postcondition
$\R$ specifies the codomain of $t$ and the domain of its continuation
($k$).  We therefore need formulas recognizing
infinite programs (prefixes) and formulas for finite programs
with postconditions (suffixes), and a way to compose them.

The syntax and semantics of CTL~\cite{emerson1986sometimes} coincide
with prefix formulas.  The syntax and semantics of suffix formulas
closely resemble those of finite LTL~\cite{de2013linear}. Moreover,
prefix and suffix formulas have different structural lemmas with
respect to sequential composition and iteration.  For example, if $t$
can run forever, so can $\bind{t}{k}$, for any $k$.  However, if $t$
terminates with postcondition $\R$ and $\bind{t}{k}$ runs forever, it
must be because the continuation $k$ runs forever starting at $\R$. We
revisit this type of lemmas in Section~\ref{s:ticl:ticl-bind}.

\renewcommand{\syntleft}{\normalfont\itshape}
\renewcommand{\syntright}{}
\renewcommand{\ulitleft}{\normalfont}
\renewcommand{\ulitright}{}

\begin{figure}[t!]
  {\small
\begin{adjustwidth}{0pt}{}
  \begin{minipage}[t]{0.31\textwidth}
    \begin{grammar}
    <$\varphi$, $\varphi'$> ::=
    $\now{(\typed{P}{\World{E} \to \PROP})}$
    \alt $\AN{\varphi}{\varphi'}$
    \alt $\EN{\varphi}{\varphi'}$
    \alt $\AU{\varphi}{\varphi'}$
    \alt $\EU{\varphi}{\varphi'}$
    \alt $\AG{\varphi}$
    \alt $\EG{\varphi}$
    \alt $\varphi \land \varphi'$
    \alt $\varphi \lor \varphi'$
    \end{grammar}
  \end{minipage}
  \hspace*{0.05cm}
  \begin{minipage}[t]{0.33\textwidth}
    \begin{grammar}
    <$\psix$, $\psix'$> ::=
    $\done{(\typed{\Px}{X \to \World{E} \to \PROP})}$
    \alt $\AN{\varphi}{\psix}$
    \alt $\EN{\varphi}{\psix}$
    \alt $\AU{\varphi}{\psix}$
    \alt $\EU{\varphi}{\psix}$
    \alt $\psix \land \psix'$
    \alt $\psix \lor \psix'$
    \end{grammar}
  \end{minipage}
  \hspace*{0.3cm}
  \begin{minipage}[t]{0.15\textwidth}
    \vspace*{-.8cm}
    \begin{align*}
      &\top &&\ = \ \now{(\lambda\ \_ . \top)} \\
      &\bot &&\ = \ \now{(\lambda\ \_ . \bot)} \\
      &\Top       &&\ = \ \done{(\lambda\ \_\ \_ . \top)} \\
      &\Bot       &&\ = \ \done{(\lambda\ \_\ \_ . \bot)} \\
      &\AX{}{~p} &&\ =\ \AN{\top}{p} \\
      &\EX{}{~p} &&\ =\ \EN{\top}{p} \\
      &\AF{}{~p} &&\ =\ \AU{\top}{p} \\
      &\EF{}{~p} &&\ =\ \EU{\top}{p}
    \end{align*}
  \end{minipage}

  \vspace*{0.1cm}
  \begin{adjustwidth}{0pt}{}
  \begin{minipage}[t]{0.40\textwidth}
    \begin{alignat*}{2}
    & \pure      && =\ \now{(\lambda\ w.\ w = Pure)} \\
    & \obs p     && =\ \now{(\lambda\ w.\ w = \Obs{e}{v}\ \land\ p\ e\ v)} \\
    & \doneq x w && =\ \done{(\lambda\ x'\ w'.\ w = w' \land x = x')} \\
    \end{alignat*}
  \end{minipage}
  \hspace*{.3cm}
  \begin{minipage}[t]{0.45\textwidth}
    \begin{alignat*}{2}
      & \val p     && =\ \done{(\lambda\ x\ w.\ w = \Val{x} \land p\ x)} \\
      & \finish p  && =\ \done{(\lambda\ x\ w.\ w = \Finish{e}{v}{x}\ \land\ p\ x\ e\ v)} \\
    \end{alignat*}
  \end{minipage}
  \end{adjustwidth}
\end{adjustwidth}
}
\vspace*{-.4cm}
  \caption{Syntax of \ticl prefix formulas ($\varphi$), suffix formulas ($\psix$), and useful syntactic notations.}
  \label{f:ticl:ticl-syntax}
\end{figure}

A notational difference of \ticl with CTL is the \emph{next} operators
$\tm{AN}$ and $\tm{EN}$ are binary, unlike the $\tm{AX}$ and $\tm{EX}$
operators of CTL which are unary. We reclaim their unary versions using
syntactic notations (see Figure~\ref{f:ticl:ticl-syntax}). We
elaborate on the comparison of \ticl with CTL in
Section~\ref{s:ticl:ctl-comparison}.

\subsection{Semantics of \ticl formulas}\label{s:ticl:entails}
\Ticl is defined using binary and unary operators. The meaning of
\ticl temporal operators is indicated by their two letters.
For the first letter, ``A'' stands for \emph{all paths} and ``E'' stands
for \emph{exists a path}; these are the same path quantifiers from
CTL~\cite{emerson1982using}. For the second letter, ``N'' stands for
``Next''. For example, in the binary operator $\AN{\varphi}{\varphi'}$, {the formula $\varphi$ must hold now, and $\varphi'$ must hold in every
possible single step}. 
In contrast, in $\EN{\varphi}{\varphi'}$ (``exists next''), 
  {the formula $\varphi$ must hold now and there exists a single step that satisfies $\varphi'$}. 
``U'' stands for ``Until''.
For example, in $\AU{\varphi}{\psix}$, {$\varphi$ must hold in all paths until
eventually $\psix$ holds, then $\varphi$ does not have to hold any
longer}. The ``G'' stands for ``Globally'', as in the formula under
this operator must hold \emph{forever}.
For example, $\EG{\varphi}$. Using syntactic notations we
define ``F'' as ``Finally'', a unary version of the inductive ``U''
that has no left-hand requirement.  In CTL ``X'' stands for ``neXt''
but in \ticl ``X'' is simply a unary version of the binary next
``N'', with no left-hand formula. Finally, the base formulas
\tm{now} and \tm{done} are shallow predicates of the
metalanguage and apply to the current world (\World{E}) each
time.

\begin{figure}[t!]
  {\footnotesize
\begin{alignat*}{2}
  &\typed{\canstep{t}{w}}{\PROP}\ &&= \ \exists~t',w',~ \ktrans{t}{w}{t'}{w'} \\
  &\typed{\tm{anc}~P~Q~t~w}{\PROP} \ &&=\ P~t~w~\land~ \canstep{t}{w} ~\land~
                            \forall~t',w',~\ktrans{t}{w}{t'}{w'} \to Q~t'~w' \\
  &\typed{\tm{enc}~P~Q~t~w}{\PROP} \ &&=\ P~t~w~\land~
                                        \exists~t',w',~\ktrans{t}{w}{t'}{w'}~\land~ Q~t'~w' \\
  &\typed{\tm{agc}~P~Q~t~w}{\PROP} \ &&=\ \gfp (\tm{anc}~P)~Q~t~w \\
  &\typed{\tm{egc}~P~Q~t~w}{\PROP} \ &&=\ \gfp (\tm{enc}~P)~Q~t~w
\end{alignat*}
\begin{mathpar}
  \inferrule{Q~t~w}{\tm{auc}~P~Q~t~w}
  \and
    \inferrule{\tm{anc}~P~(\tm{auc}~P~Q)~t~w}{\tm{auc}~P~Q~t~w}
  \and
  \inferrule{Q~t~w}{\tm{euc}~P~Q~t~w}
  \and
    \inferrule{\tm{enc}~P~(\tm{euc}~P~Q)~t~w}{\tm{euc}~P~Q~t~w}
\end{mathpar}
}
\caption{\emph{Next} (\tm{anc} and \tm{enc}), \emph{globally} (\tm{agc} and \tm{egc}) and
  \emph{until} (\tm{auc} and \tm{euc}) higher-order predicates.}
\label{f:ticl:ticl-shallow}
\end{figure}

Before jumping into the semantics of \ticl formulas ($\vDash_{LR}$) we
must first define the \emph{shallow} predicates of Figure~\ref{f:ticl:ticl-shallow}.
Definitions~\tm{anc}, \tm{enc}, \tm{agc}, \tm{egc}, \tm{auc}, and \tm{euc} 
  are \emph{higher-order predicates} in type theory:
they take predicates of type $\ictree{E}{X} \to \World{E} \to \PROP$ as
arguments and transport them under their modal operator to get
``future'' predicates of the same type. For example, \tm{anc} means
``forall-next'', \tm{enc} means ``exists-next'', \tm{agc} means
``forall-globally'', \tm{auc} means ``forall-until'', etc.

The basic predicates are \tm{anc} and \tm{enc}, connecting predicates $P, Q$ to
the \ICTree transition relation in Figure~\ref{f:ticl:kripke-def}. All other
predicates are (co-)inductively defined in terms of \tm{anc} and \tm{enc} in Figure~\ref{f:ticl:ticl-shallow}.
One difference of our definition compared to temporal logics such as CTL is
the restriction $\tm{can\_step}$ on \emph{forall-next} (\tm{anc}).
Predicate $\tm{can\_step}$ asserts the existence of at least one transition and is crucial to
prove the soundness of \ticl. Because the transition relation is not left-total,
and allows for stuck states ($\stuck$), omitting \canstep{}{} allows
\entailsL{\stuck}{w}{\AX{\bot}} to be provable in one step---by
introducing the hypothesis $\ktrans{\stuck}{w}{t'}{w'}$ and concluding the proof by contradiction.
Predicate $\tm{can\_step}$ prohibits vacuously proving statements by asserting that \stuck can transition. %\sa{What's a stuck transition?}

Path induction for \emph{until} operators
(Figure~\ref{f:ticl:ticl-entails}) is implemented by the inductive,
higher-order predicates $\tm{auc}$ and $\tm{euc}$
(Figure~\ref{f:ticl:ticl-shallow}). There are two cases. The base case asserts $Q$ holds, and the
inductive case asserts $P$ holds now, while $\tm{auc}~P~Q$ (or $\tm{euc}~P~Q$)  holds next.
Path coinduction (\emph{always}) is implemented by the $\tm{agc}$ and $\tm{egc}$ greatest fixpoints.
The $\gfp$ operator in Figure~\ref{f:ticl:ticl-shallow} and the associated machinery for completing
coinductive proofs, namely up-to-principles~\cite{pous2016coinduction}, are presented in detail
in Appendix~\ref{appendix:ticl:coinduction}.

\begin{figure}[t!]
  {\footnotesize
  \begin{equation*}
    \begin{aligned}
    &\begin{rcases}
      \typed{\sembra{L}{\varphi}}{\forall~X,~\ictree{E}{X} \to \World{E} \to \PROP}, \  &
      \typed{\sembra{R}{\psix}}{\ictree{E}{X} \to \World{E} \to \PROP} \quad \\
    \end{rcases}\text{ denotations to shallow predicates} \\   
    &\begin{rcases}
      \sembra{L}{\now{P}}        = \lambda~\_~w.~\notdone{w}~ \land ~ P\ w, \ &
      \sembra{R}{\done{\Px}}     = \lambda~\_~w.~\donewith{Px}{w} \quad \\
    \end{rcases}\text{ Base case predicates} \\ 
    &\begin{rcases}
      \sembra{L}{\AN{\varphi}{\varphi'}}     = \tm{anc}~\sembra{L}{\varphi}~~\sembra{L}{\varphi'}, &
      \sembra{L}{\EN{\varphi}{\varphi'}}     = \tm{enc}~\sembra{L}{\varphi}~~\sembra{L}{\varphi'}  \quad \\
      \sembra{R}{\AN{\varphi}{\psix}}  = \tm{anc}~\sembra{L}{\varphi}~~\sembra{R}{\psix}, &
      \sembra{R}{\EN{\varphi}{\psix}}  = \tm{enc}~\sembra{L}{\varphi}~~\sembra{R}{\psix} \quad \\
    \end{rcases}\text{ \emph{Next} operators} \\
    &\begin{rcases}
      \sembra{L}{\AU{\varphi}{\varphi'}}     = \tm{auc}~\sembra{L}{\varphi}~~\sembra{L}{\varphi'}, &
      \sembra{L}{\EU{\varphi}{\varphi'}}     = \tm{euc}~\sembra{L}{\varphi}~~\sembra{L}{\varphi'}  \quad \\      
      \sembra{R}{\AU{\varphi}{\psix}}  = \tm{auc}~\sembra{L}{\varphi}~~\sembra{R}{\psix}, &
      \sembra{R}{\EU{\varphi}{\psix}}  = \tm{euc}~\sembra{L}{\varphi}~~\sembra{R}{\psix} \quad \\
    \end{rcases}\text{ \emph{Until} operators (inductive)} \\
    &\begin{rcases} 
      \sembra{L}{\AG{\varphi}}                 = \tm{agc}~\sembra{L}{\varphi}, &
      \hspace*{.98cm} \sembra{L}{\EG{\varphi}} = \tm{enc}~\sembra{L}{\varphi} \quad \\
    \end{rcases}\text{ \emph{Globally} operators (coinductive)} \\
    &\begin{rcases}
      \sembra{LR}{p \land q}     = \lambda~t~w.~\sembra{LR}{p}~t~w \land \sembra{LR}{q}~t~w, \ &
      \sembra{LR}{p \lor q}      = \lambda~t~w.~\sembra{LR}{p}~t~w \lor \sembra{LR}{q}~t~w \quad \\
    \end{rcases}\text{ Propositional operators}       
    \end{aligned}
  \end{equation*}
}
\caption{\Ticl formula denotations (\sembra{LR}{\textcolor{gray}{\_}}) defined by induction on $\varphi$ and $\psix$.}
\label{f:ticl:ticl-entails}
\end{figure}

The semantics of \ticl formulas---denoted to type theory---are defined by induction on
formulas $\varphi$ and $\psix$ in Figure~\ref{f:ticl:ticl-entails}. The \ticl entailment relations then become simple predicate
applications to an \ICTree $t$ and world $w$ as stated in Definition~\ref{def:ticl:ticl-entails}.
\begin{definition}[Ticl Entailment]\label{def:ticl:ticl-entails}
  {\small
    \begin{alignat*}{4}
      &\entailsL{t}{w}{\varphi}       && = \sembra{L}{\varphi}~t~w,\quad&
      &\entailsR{t}{w}{\psix}   && = \sembra{R}{\psix}~t~w
    \end{alignat*}}
  \label{f:ticl:entails-def}
\end{definition}

\subsection{\Ticl formula equivalence}\label{s:ticl:equivalence}
\Ticl entailments \entailsLR{t}{w}{p} are type-theoretic propositions (\PROP) that form a complete lattice
$(\PROP, \to)$. Consequently, denotations of \ticl formulas (\typed{\sembra{LR}{p}}{\ictree{E}{X}\to\World{E}\to\PROP})
which are type-theory predicates, also form a complete lattice, with respect to the pointwise implications
$\implL$ and $\implR$ in Definition~\ref{def:ticl:ticl-equiv} (shown below).
Taking an implication in both directions introduces an equivalence relation on \ticl formulas $(\iffLR)$.
Two \ticl formulas $p, q$ are (semantically) equivalent ($p \iffLR q$) when for all
trees ($t$) and worlds ($w$), $\entailsLR{t}{w}{p}$ if and only if $\entailsLR{t}{w}{q}$.

%% Partial order and equivalence on formulas
\begin{definition}[Partial order and equivalence]\label{def:ticl:ticl-equiv}
{\small
\begin{alignat*}{4}
  \varphi \implL \varphi'         & = &  \forall\ t, w,\ \entailsL{t}{w}{\varphi} \to \entailsL{t}{w}{\varphi'} & \qquad &
  \varphi \iffL \varphi'          & = && \ \varphi \implL \varphi'\ \text{ and }\ \varphi' \implL \varphi  \\
  \psix \implR \psix' & = &  \ \forall\ t, w,\ \entailsR{t}{w}{\psix} \to \entailsR{t}{w}{\psix'} & \qquad &
  \psix \iffR \psix'  & = && \ \psix \implR \psix'\ \text{ and }\ \psix' \implR \psix
\end{alignat*}
}
\end{definition}

Now that we have a notion of formula equivalence, building a library
of useful (in-)equalities enables fluent proof manipulation. For example, if
$p \iffLR q$ and the goal is $\entailsLR{t}{w}{p}$, we can rewrite it
to $\entailsLR{t}{w}{q}$ instead. Similarly under \ticl operators:
\entailsLR{t}{w}{\AX{p}} is equivalent to \entailsLR{t}{w}{\AX{q}},
\entailsLR{t}{w}{r \land q} to \entailsLR{t}{w}{r \land q}, and so on.
Some useful (in-)equalities in the \ticl library are shown
in Figure~\ref{f:ticl:ticl-algebra}. We elide the boolean algebra laws for space.

\begin{figure}[t]
  {\small\centering
      \begin{minipage}[t]{0.4\textwidth}
        \begin{alignat*}{4}
      & \AN{p}{q}              &&  \implLR\ \EN{p}{q}        & (\AN{}{}\text{-weaken}) & \\
      & \AU{p}{q}              &&  \implLR\ \EU{p}{q}        & (\AU{}{}\text{-weaken}) & \\
      & \AG{\varphi}                 && \implL\ \EG{\varphi}              & (\AG{}{}\text{-weaken}) & \\
      & \AN{p}{q}              && \implLR\ \AU{p}{q}          & (\AN{}{}\text{-until}) & \\
      & \EN{p}{q}              && \implLR\ \EU{p}{q}          & (\EN{}{}\text{-until}) & \\
      & \AG{\varphi}                 && \implL\ \varphi                   & (\AG{}{}\text{-M}) & \\
      & \EG{\varphi}                 && \implL\ \varphi                   & (\EG{}{}\text{-M}) & \\
      & \EG{(\varphi \land \varphi')}      && \implL\ \EG{\varphi} \land EG{\varphi'} & (\EG{}{}\text{-and}) & \\
      & \AG{\varphi} \lor AG{\varphi'}     && \implL\ \AG{(\varphi \lor \varphi')}      & (\AG{}{}\text{-or}) & \\
      & \EG{\varphi} \lor EG{\varphi'}     && \implL\ \EG{(\varphi \lor \varphi')}      & (\EG{}{}\text{-or}) &
    \end{alignat*}
  \end{minipage}
  \hspace*{0.25cm}
  \begin{minipage}[t]{0.55\textwidth}
    \begin{alignat*}{4}
      & \AU{p}{q}            &&  \iffLR\ q \lor (\AN{p}{\AU{p}{q}}) & \quad (\AU{}{}\text{-unfold}) & \\
      & \EU{p}{q}            &&  \iffLR\ q \lor (\EN{p}{\EU{p}{q}}) & \hfill (\EU{}{}\text{-unfold}) &  \\
      & \AG{\varphi}               &&  \iffL\ \AN{\varphi}{\AG{\varphi}} & \quad (\AG{}\text{-unfold}) & \\
      & \EG{\varphi}               &&  \iffL\ \EN{\varphi}{\EG{\varphi}} & \quad (\EG{}\text{-unfold}) & \\
      & \AU{p}{q}            &&  \iffLR\ \AU{p}{\AU{p}{q}}      & \hfill (\AU{}{}\text{-idem}) & \\
      & \EU{p}{q}            &&  \iffLR\ \EU{p}{\EU{p}{q}}      & \hfill (\EU{}{}\text{-idem}) & \\
      & \EG{\EG{\varphi}}          &&  \iffL\ \EG{\varphi}                        & \hfill (\EG{}{}\text{-idem}) & \\
      & \AG{\AG{\varphi}}          &&  \iffL\ \AG{\varphi}                        & \hfill (\AG{}{}\text{-idem}) & \\
      & \AG{(\varphi \land \varphi')}    &&  \iffL\ \AG{\varphi} \land AG{\varphi}            & \hfill (\AG{}{}\text{-and}) &
    \end{alignat*}
  \end{minipage}
}
\caption{Representative \ticl formula implications and equivalences.}
\label{f:ticl:ticl-algebra}
\end{figure}

\subsection{\ICTree equivalence under \Ticl entailment}\label{s:ticl:uptotau-proper}
Let us take stock of what we have achieved so far in this section and what our remaining goals are.
We defined two ternary entailment relations over \ICTrees and worlds, \entailsL{t}{w}{\varphi} and \entailsR{t}{w}{\psix},
giving meaning to \ticl formulas. We established a rewriting system over \ticl formulas ($\varphi \Leftrightarrow_L \varphi'$ and
$\psix \Leftrightarrow_R \psix'$) and proved equations useful for simplifying formulas under entailment.
We switch our attention back to programs ($t$) and remind ourselves of up-to-guard equivalence of programs
in Section~\ref{s:ticl:ctree-core}. To prove a specification \entailsLR{t}{w}{p} sometimes it is
convenient to simplify the formula ($p$), but sometimes it is convenient to simplify the program $t$
by substituting it with an equivalent program $u$ (where $t \sbisim u$).

For example, for the monadic bind simplification from
Section~\ref{s:ticl:ctree-core} ($\bind{\Ret{x}}{k} \sbisim k~x$) it
seems intuitive that the following goals are also equivalent
$\entailsLR{\bind{\Ret{x}}{k}}{w}{p} \Leftrightarrow \entailsLR{k~x}{w}{p}$. This
intuition is correct and we were able to mechanize the well-known
proof that (strong) bisimulation preserves temporal
properties~\cite{EMERSON1990995}. At its core, the proof relies on the
following lemma relating Kripke transitions and program equivalence.
The transition relation $\ktrans{t}{w}{t'}{w'}$ itself is \textit{not}
up-to-tau invariant---it is easy to find a counter-example with a \Tau{}
node in the middle of the tree. However, Lemma \mkw{ExEquiv} (shown
below) is a weaker version of up-to-tau invariance for transitions
that \textit{is} provable:

\begin{lemma}[ExEquiv] \ \\
\( \forall s,\; t,\; w,\; w',\ (s \sim t) \to (\ktrans{s}{w}{s'}{w'}) \to
  \exists\ t',\ \ktrans{t}{w}{t'}{w'} \land\; (s' \sim t') \)
\end{lemma}

By using lemma \mkw{ExEquiv} and by induction on the structure of \ticl
formulas, we are able to prove that rewriting with up-to-tau equivalence ($\sbisim$) under
\ticl entailment ($\vDash_{L,R}$) is correct, namely:

\begin{theorem}[Up-to-tau equivalence preserves \ticl formulas] \ \\
\(  \forall t,\; u,\; w,\; p,\ t \sbisim u \to \entailsLR{t}{w}{p} \to \entailsLR{u}{w}{p} \)
\end{theorem}

The property of \emph{up-to-tau invariance}
enables equational reasoning (Figure~\ref{f:ticl:ctree-algebra}) in
conjunction with structural proof techniques (next
Section~\ref{s:ticl:lemmas}) resulting in a remarkably flexible proof system.

\section{Structural lemmas for \ICTree}\label{s:ticl:lemmas}
The equational theories of \ticl formulas (\iffLR) and \ICTrees (\sbisim)
allow us to simplify a specification such as \entailsL{t}{w}{\AG{\varphi}}, but are insufficient to fully prove it.
The inequalities of Figure~\ref{f:ticl:ticl-algebra}
unfold the ``always'' operator ($\AG{\varphi} \iffL\ \AN{\varphi}{\AG{\varphi}}$) but there will always
be an \AG{\varphi} proof obligation left over. In this section we give structural lemmas
connecting \ICTree composition ($\bindc, \oplus$) and iteration (\itern) to \ticl
operators (\AU{}{}, \EU{}{}, \EG{} and \EG{}) allowing us to fully prove \ticl specifications.

The lemmas in this section \emph{internalize} low-level (co-)inductive
proofs to simple structural lemmas over \ICTrees. As \ICTrees form a
denotational basis for many programming languages, the lemmas in this
section form a logical basis for many temporal logics defined over
those languages.  In the next section (Section~\ref{s:ticl:usage}) we
will see how to use \ticl to define a new programming language, its
denotation to \ICTrees and its structural lemmas, with only a few
lines of definitions.  The table in Figure~\ref{f:ticl:ticl-table}
shows the cartesian product of \ICTree structures and \ticl
temporal operators. We have identified and proved backward-reasoning
lemmas (\cbwd) for sequential composition and iteration (\bindc,
\itern) and bidirectional lemmas (\ciff) for all \ICTree
nodes and nondeterministic choice ($\oplus$). We conjecture there are
useful inversion lemmas for \bindc$~$ and \itern which we
leave for future work.

The proof rules, collected in Figure~\ref{f:ticl:ticl-table},
correspond to lemmas stated with regards to the entailment relation
(Definition~\ref{def:ticl:ticl-entails}). All rules are proven---that
is, each rule of our logic is sound---and the collection of all proofs
corresponds to a statement of soundness for \ticl. In our Rocq
development a syntactic representation of entailment is provided, to
facilitate automation.  We prove that syntactic entailment implies
semantic entailment by induction.

\begin{figure}[t!]
  {\small
  \begin{center}
\begin{tabular}{|c|*{6}{c|}|*{4}{c|}}
  \hline
  & \multicolumn{6}{c|}{Prefix ($\varphi$)} & \multicolumn{4}{c|}{Suffix (\psix)} \\
  \hline
  & AN & EN & AU & EU & AG & EG & AN & EN & AU & EU \\
  \hline
  \Ret{} & \ciff & \ciff & \ciff & \ciff & \ciff & \ciff & \ciff & \ciff & \ciff & \ciff \\
  \hline
  \Br{}{} & \ciff & \ciff & \ciff & \ciff & \ciff & \ciff & \ciff & \ciff & \ciff & \ciff \\
  \hline
  \Vis{}{} & \ciff & \ciff & \ciff & \ciff & \ciff & \ciff & \ciff & \ciff & \ciff & \ciff \\
  \hline
  \stuck & \ciff & \ciff & \ciff & \ciff & \ciff & \ciff & \ciff & \ciff & \ciff & \ciff \\
  \hline
  \bindc & \cbwd & \cbwd & \cbwd & \cbwd & \cbwd & \cbwd & \cbwd & \cbwd & \cbwd & \cbwd \\
  \hline
  \itern & \cbwd & \cbwd & \cbwd & \cbwd & \cbwd & \cbwd & \cbwd & \cbwd & \cbwd & \cbwd \\
  \hline
\end{tabular}
\end{center}
}
\caption{Library of structural lemmas for \ICTree combinators and \ticl operators. Backwards-reasoning
  lemmas are indicated by \cbwd and bidirectional lemmas by \ciff.}
\label{f:ticl:ticl-table}
\end{figure}

\subsection{Sequential composition}\label{s:ticl:ticl-bind}
\begin{figure}[t!]
  {\footnotesize
    \begin{adjustwidth}{0pt}{}
  \begin{mathpar}
    %%% ctll_bind_l
    \mprset{vskip=.05cm}
    \inferrule*[Right=BindL]
    {\entailsL{t}{w}{\varphi}}
    {\entailsL{\bind{t}{k}}{w}{\varphi}}
    \vspace*{-.5cm}
    \\\\
    \hspace*{-.8cm}
    %%% aul_br
    \inferrule*[Right=BrAU$_L$]
    {\entailsL{t \oplus u}{w}{\varphi} \\\\ \entailsL{t}{w}{\AU{\varphi}{\varphi'}} \and \entailsL{u}{w}{\AU{\varphi}{\varphi'}}}
    {\entailsL{t \oplus u}{w}{\AU{\varphi}{\varphi'}}}\and
    \hspace*{-.2cm}
    %%% eul_br
    \inferrule*[Right=BrEU$_L$]
    {\entailsL{t \oplus u}{w}{\varphi} \\\\ \entailsL{t}{w}{\EU{\varphi}{\varphi'}}~ \lor~ \entailsL{u}{w}{\EU{\varphi}{\varphi'}}}
    {\entailsL{t \oplus u}{w}{\EU{\varphi}{\varphi'}}}
    \and
    \hspace*{-1.2cm}
   %%% aul_bind_r
    \inferrule*[Right=BindAU$_L$]
    {\entailsR{t}{w}{\AU{\varphi}{\AX{\done{\R_Y}}}}
      \\\\
      \forall~y,w,~\R_Y~y~w \to \entailsL{k~y}{w}{\AU{\varphi}{\varphi'}}}
    {\entailsL{\bind{t}{k}}{w}{\AU{\varphi}{\varphi'}}}
    \and
    %%%
    \hspace*{-.2cm}
    %%% aul_bind_r_eq
    \inferrule*[Right=BindAU$_{L=}$]
    {\entailsR{t}{w}{\AU{\varphi}{\AX{\doneq{y}{w'}}}}
      \\\\
      \entailsL{k~y}{w'}{\AU{\varphi}{\varphi'}}}
    {\entailsL{\bind{t}{k}}{w}{\AU{\varphi}{\varphi'}}}
    \vspace*{-.2cm}
    \\\\
    %%%
    \hspace*{-1cm}
    %%% aur_bind_r
    \inferrule*[Right=BindAU$_R$]
    {\entailsR{t}{w}{\AU{\varphi}{\AX{\done{\R_Y}}}}
      \\\\
      \forall~y,w,~\R_Y~y~w \to \entailsR{k~y}{w}{\AU{\varphi}{\psix'}}}
    {\entailsR{\bind{t}{k}}{w}{\AU{\varphi}{\psix'}}}
    \and
    \hspace*{-.2cm}
    %%% ag_bind_r
    \inferrule*[Right=BindAG]
    {\entailsR{t}{w}{\AU{\varphi}{\AX{\done{\R_Y}}}}
      \\\\
      \forall~y,w,~\R_Y~y~w \to \entailsL{k~y}{w}{\AG{\varphi}}}
    {\entailsL{\bind{t}{k}}{w}{\AG{\varphi}}}
  \end{mathpar}
  \end{adjustwidth}
}
\caption{Representative \ICTree structural lemmas for nondeterminism and sequential composition.}
  \label{f:ticl:ticl-bind}
\end{figure}

In Section~\ref{s:ticl:ctree-core} we define the sequential composition of \ICTrees (\Ret{},\bindc) and in Section~\ref{s:ticl:syntax} we motivate \ticl postconditions (\psix), asserting the existence
of compositional liveness lemmas. Figure~\ref{f:ticl:ticl-bind} shows some of those lemmas.
The goal is to distribute temporal specifications over the
sequential $(\bindc)$ and parallel $(\oplus)$ composition of programs. As a
result, we get modular subproofs for general liveness properties, analogous to the sequence rule for safety properties in Hoare logic.

For example, if \bind{t}{k} is a terminal application with the ability
to print to standard output, and the goal is to prove that it will
\emph{eventually} print $\entailsL{\bind{t}{k}}{w}{\AF{\obs{\tm{PRINTS}}}}$, there are
two cases to consider:

\begin{enumerate}
\item Either $t$ prints to the terminal, use the \mkw{BindL} lemma (Figure~\ref{f:ticl:ticl-bind})
  to prove it and ignore the continuation ($k$).

\item Or the continuation $k$ prints, use the \mkw{BindAU$_L$} lemma
  to show $t$ always terminates with postcondition $\R_Y$. Then for all possible return values
  (\typed{y}{Y}) and worlds (\typed{w'}{\World{E}})
  in the postcondition ($\R_Y~y~w'$), we must show the continuation ($k~y$)
  eventually prints to the terminal \entailsL{k~y}{w'}{\AF{\obs{\tm{PRINTS}}}}.
\end{enumerate}

\noindent While structural lemmas are proven for
both the universal ($AN, AU, AG$) and existential ($EN, EU, EG$) \ticl
operators, we focus our exposition on universal quantifiers, noting
that the same lemmas apply to their existential versions.

For deterministic programs, the convenience lemma \mkw{BindAU$_{L=}$}
assumes that a linear path can be traversed in finite steps---remember the
syntactic notation $\doneq{y}{w'}$ introduced in
Figure~\ref{f:ticl:ticl-syntax} uses equality to value $y$ and world
$w'$ as the postcondition.  This simplifying assumption lifts the need
to manually specify postconditions of deterministic
programs.\footnote{In practice, $y$ and $w'$ are replaced with
  existential variables in Rocq proofs, delaying their instantiation
  until program $t$ returns, largely automating the use of this rule.}

\subsection{Iteration}\label{s:ticl:ticl-iter}
The iteration \ICTree combinator (\itern), defined in Section~\ref{s:ticl:ctree-core}, encodes both finite and infinite loops.
In this section we prove lemmas that show loop \emph{termination}, \emph{liveness}, and \emph{invariance}
(Figure~\ref{f:ticl:ticl-iter}) by using loop variant and invariant relations over the loop body.

The loop \emph{termination} rule (\mkw{IterAU$_R$}) proves a loop
terminates with postcondition $\psix$. It requires specifying
a loop invariant relation ($\R$), and a binary \emph{well-founded}
relation called the loop variant ($R_v$)---well-founded relations have no infinite
chains, ensuring the loop terminates in finite steps.
There are two obligations, the inductive step, and base case of the underlying
induction.
\begin{enumerate}
\item If the loop body's return value $lr = \inl{i'}$, the loop continues.  The loop invariant $\R$
  must be satisfied before and after the loop body, much like in Hoare
  Logic. The next iteration (represented by iterator $i'$ and world
  $w'$) must be ``smaller'' according to the loop variant
  ($\R_v~(i',w')~(i, w)$).
\item If $lr = \inr{r}$ the loop terminates concluding the proof. All
  that is left is to show that the loop returns a state satisfying the loop
  postcondition $\entailsR{\Ret{r}}{w}{\AN{\varphi}{\psix}}$.
\end{enumerate}

\noindent The loop \emph{liveness} rule (\mkw{IterAU$_L$}) is slightly different
than the termination rule: it expects formula $\varphi'$ to be eventually
satisfied, even if the loop keeps running afterwards---possibly
forever. Note that this is the only meaning of liveness in temporal logics
like LTL and CTL. \Ticl differentiates between those two very
different cases with the two rules (\mkw{IterAU$_R$} and
\mkw{IterAU$_L$}) in Figure~\ref{f:ticl:ticl-iter}---the differences are
further highlighted in Section~\ref{s:ticl:ctl-comparison}. Similar to
loop \emph{termination}, the \emph{liveness rule} expects two
relations---the loop invariant ($\R$) and the loop variant
($\R_v$)---and produces two proof obligations, except now we
get to chose which one is satisfied in each iteration ($k~i$)
\begin{enumerate}
\item Either the loop body satisfies the liveness property $\entailsL{\tm{k~i}}{w}{\AU{\varphi}{\varphi'}}$.
\item Or, the loop continues ($lr = \inl{i'}$) and the new iterator ($i'$) and world ($w'$) satisfy
  the invariant ($\R$) and are ``smaller'' with respect to the well-founded variant ($\R_v$).
\end{enumerate}

Working directly with well-founded relations in Rocq can be difficult,
so we define simplified versions of rules \mkw{IterAU$_{R}$} and
\mkw{IterAU$_L$} expecting a \emph{ranking function}
(\mkw{IterAU$_{L,\mathbb{N}}$}).  A ranking function maps iterators and worlds
to the natural numbers (\typed{f}{I \to \World{E} \to \mathbb{N}}), such that
successive pairs of iterator and world are strictly monotonically
decreasing. Finding suitable ranking functions for complex loops can
be challenging. In Example~\ref{s:ticl:ex-election} we demonstrate a new
lemma we call \emph{liveness split}, that reduces liveness proofs to
smaller liveness proofs, with smaller ranking functions. Recent,
orthogonal work on automatic inference of ranking functions~\cite{yao2024mostly} also works well with \ticl, as \ticl's
  iteration rules can use such inferred ranking functions to produce formal proofs of liveness.

\def \MathparLineskip {\lineskip=.3cm}
\begin{figure}[t!]
  \begin{adjustwidth}{0pt}{}
  {\footnotesize
    \begin{mathpar}
      \hspace*{-2cm}
    %%% aul_iter_r
    {\mprset{vskip=.1cm}
    \inferrule*[Right=IterAU$_L$,flushleft]
    {\R~i~w \\ \wellfounded{\R_v} \\\\
    {\begin{aligned}
      \forall&~i,w,~\R~i~w \to \\
      & \entailsL{\tm{k~i}}{w}{\AU{\varphi}{\varphi'}}~\lor \\
      & \entailsRR{\tm{k~i}}{w} \varphi~\tm{AU}~\tm{AX}~\tm{done}~(\tm{\lambda}~lr~w'~ \Rightarrow \\
      & \qquad \exists~i',~lr = \inl{i'} ~ \land~ \R~i'~w'~ \\
      & \qquad \land~ \R_v~(i',w')~(i, w)) ~ \rangle\\
    \end{aligned}}}
    {\entailsL{\iter{k}{i}}{w}{\AU{\varphi}{\varphi'}}}}\and
  \hspace*{.3cm}
    %%% aul_iter_r_nat
    {\mprset{vskip=.1cm}
      \inferrule*[Right=IterAU$_{L,\mathbb{N}}$,flushleft]
      {\R~i~w \\\\
    {\begin{aligned}
      \forall&~i,~w,~\R~i~w \to \\
      & \entailsL{\tm{k~i}}{w}{\AU{\varphi}{\varphi'}}~\lor \\
      & \entailsRR{\tm{k~i}}{w} \varphi~\tm{AU}~\tm{AX}~\tm{done}~(\tm{\lambda}~lr~w'~ \Rightarrow \\
      & \qquad \exists~i',~lr = \inl{i'} ~ \land~ \R~i'~w'~ \\
      & \qquad \land~ f~i'~w' < f~i~w ) ~ \rangle\\
    \end{aligned}}}
{\entailsL{\iter{k}{i}}{w}{\AU{\varphi}{\varphi'}}}}
\\\\
\hspace*{-1.2cm}
  %%% aur_iter_r
  {\mprset{vskip=.1cm}
    \inferrule*[Right=IterAU$_R$,flushleft]
    {\R~i~w \\ \wellfounded{\R_v} \\\\
    {{\begin{aligned}
      \forall&~i,~w,~\R~i~w \to \\
      & \entailsRR{k~i}{w} \varphi~\tm{AU}~\tm{AX}~\tm{done}~(\tm{\lambda}~lr~w'~ \Rightarrow \\
      & \qquad {\begin{cases*}
        \ \R~i'~w'\land\R_v~(i', w')~ (i, w),             & \text{if }lr = \inl{i'} \\
        \ \entailsR{\Ret{r}}{w'}{\AN{\varphi}{\psix}}, & \text{if }lr = \inr{r} \\
      \end{cases*}}\\
      & )~\rangle
      \end{aligned}}}}
    {\entailsR{\iter{k}{i}}{w}{\AU{\varphi}{\psix}}}}\and
  \hspace*{0cm}
    %%% ag_iter_r
    {\mprset{vskip=.1cm}
    \inferrule*[Right=IterAG,flushleft]
    {\R~i~w \\\\
    {{\begin{aligned}
      \forall&~i,~w,~\R~i~w \to \\
      & \entailsL{\iter{k}{i}}{w}{\varphi}~\land \\
      & \entailsRR{\tm{k~i}}{w} \tm{AX}(\varphi~\tm{AU}~\tm{AX}~\tm{done}~(\tm{\lambda}~lr~w'~ \Rightarrow \\
      & \qquad \exists~i',~lr = \inl{i'} ~ \land~ \R~i'~w')) ~ \rangle\\
      \end{aligned}}}}
    {\entailsL{\iter{k}{i}}{w}{\AG{\varphi}}}}
  \end{mathpar}
}
\end{adjustwidth}
  \caption{Representative \ticl iteration lemmas for operators \AU{}{} and \AG{} and \ICTrees.}
  \label{f:ticl:ticl-iter}
\end{figure}

\Ticl addresses \emph{nonterminating} loops with the \emph{invariance}
rule (\mkw{IterAG}) in Figure~\ref{f:ticl:ticl-iter}. This rule can
prove both \emph{always} and \emph{always-eventually} properties by
specifying a suitable loop invariant $\R$ and two proof obligations:
\begin{enumerate}
\item The inner formula $\varphi$ must hold for the whole loop
  $\entailsL{\iter{k}{i}}{w}{\varphi}$. For example, if $\varphi$ is an \emph{eventually}
  property the loop might run multiple times before satisfying the base formula.
\item The loop body ($k~i$) must take at least one step and eventually
  terminate, satisfying $\varphi$ in every step. Stepping once is required
  for the loop to be productive, preventing unsoundness issues with
  cyclic proofs~\cite{pous2016coinduction}. In addition, the loop body
  ($k~i$) must always continue ($lr = \inl{i'}$), and the new iterator
  $i'$ and world $w'$ must satisfy the loop invariant
  ($\R$).\footnote{It is not always the case the loop body will
    terminate. For example, there may be nested infinite
    loops. Unfolding the outer \itern loop reveals its definition in
    terms of bind (\bindc). The \mkw{BindL} rule in
    Figure~\ref{f:ticl:ticl-bind} eliminates the outer loop,
    allowing us to apply the \emph{invariance} rule on the inner
    loop.}
\end{enumerate}

\noindent The \emph{invariance} rule (\mkw{IterAG}) is significant
because it discharges a coinductive (infinite) proof to two finite
subproofs. This single rule encapsulates all of \ticl's coinduction
techniques, and has proven sufficiently general to
complete every \emph{always} and \emph{always-eventually} proof in our
evaluation.

\begin{figure}[t!]
  {\small
    \begin{typsyntax}{\IAExp}{Type}
      & \hspace*{.5em} \bbar \IVar{(\typed{s}{string})} \quad \bbar \IVal{(\typed{n}{\mathbb{N}})} \\      
      & \hspace*{.5em} \bbar \IAdd{(\typed{x}{\IAExp})}{(\typed{y}{\IAExp})} \quad \bbar \ISub{(\typed{x}{\IAExp})}{(\typed{y}{\IAExp})} 
    \end{typsyntax}
  
    \begin{typsyntax}{\IBExp}{Type}
      \hspace*{.5em} \bbar \ILt{(\typed{x}{\IAExp})}{(\typed{y}{\IAExp})}
    \end{typsyntax}

    \begin{typsyntax}{\stimp}{Type}
      & \bbar \IAssign{(\typed{s}{string})}{(\typed{y}{\IAExp})}
      \quad \bbar \IIf{\typed{c}{\IBExp}}{(\typed{x}{\stimp})}{(\typed{y}{\stimp})} \\
      & \bbar \ISeq{(\typed{l}{\stimp})}{(\typed{r}{\stimp})}
      \quad \bbar \IWhile{\typed{c}{\IBExp}}{\typed{t}{\stimp}}
      \quad \bbar \ISkip
    \end{typsyntax}
  }
  \caption{Syntax of a small imperative language \stimp with mutable state and nondeterminism.}
  \label{f:ticl:imp-syntax}
\end{figure}
\begin{figure}[t!]
  {\small
    \typequation{\St{\Ctx}{}}{\Type \to \Type}
    { \bbar (\typed{\Get}{\St{\Ctx}{\Ctx}}) \quad \bbar (\typed{\Put{(\typed{m}{\Ctx})}}{\St{\Ctx}{\unit}})}

    \typequation{\sget}{\ictree{\St{\Ctx}{}}{\Ctx}}{  \trigger \Get}
    \typequation{\sput (\typed{m}{\Ctx})}{\ictree{\St{\Ctx}{}}{\unit}}{  \trigger (\Put{m})}
    % Handler
    \begin{typfunction}{h_{\Ctx}}{\St{\Ctx}{} \leadsto \InstrM{\Ctx}{\Ctx}{}}
      & h_{\Ctx}~(\typed{\Get}{\St{\Ctx}{\Ctx}})~(\typed{m}{\Ctx}) ~=~ \Ret{(m,m)} \hspace*{6cm} \\
      & h_{\Ctx}~(\typed{\Put{m'}}{\St{\Ctx}{\unit}})~(\typed{\_}{\Ctx}) ~=~ \log{m'}~\tm{;\!;}~\Ret{(\ttt, m')}
    \end{typfunction}
    %% AExp denotation
    \begin{typfunction}{\sembra{A}{\textcolor{gray}{\_}}}{\IAExp \to \ictree{\St{\Ctx}{}}{\mathbb{N}}}
      &  \sembra{A}{\IVar{s}} ~=~ \sget \bindc (\lambda~m \Rightarrow~ \Ret{m[s]}),\quad
      && \sembra{A}{\IAdd{x}{y}} ~=~ \bindv{a}{\sembra{A}{x}}{\bindv{b}{\sembra{A}{y}}{\Ret{(a + b)}}} \hspace*{.4cm} \\
      &  \sembra{A}{\IVal{n}} ~=~ \Ret{n}, \quad 
      && \sembra{A}{\ISub{x}{y}} ~=~ \bindv{a}{\sembra{A}{x}}{\bindv{b}{\sembra{A}{y}}{\Ret{(a - b)}}} \\
    \end{typfunction}
    %% BExp denotation
    \begin{typfunction}{\sembra{B}{\textcolor{gray}{\_}}}{\IBExp \to \ictree{\St{\Ctx}{}}{\mathbb{B}}}
      &\sembra{B}{\ILt{x}{y}} ~=~ \bindv{a}{\sembra{B}{x}}{\bindv{b}{\sembra{B}{y}}{\Ret{(a < b)}}} \hspace*{6.2cm} \\
    \end{typfunction}
    %% StImp denotation
    \begin{typfunction}{\sembra{S}{\textcolor{gray}{\_}}}{\stimp \to \ictree{\St{\Ctx}{}}{\unit}}
      &\sembra{S}{\IAssign{s}{x}} ~=~ \bindv{a}{\sembra{S}{x}}{ \bindv{m}{\sget}{\sput{((s \hookrightarrow a) \cup m)}}},
      \quad\sembra{S}{\ISeq{t}{u}} ~=~ \sembra{S}{t}\tm{;\!;}~\sembra{S}{u} \\
      &\sembra{S}{\IIf{c}{t}{u}} ~=~ \sembra{B}{c} \bindc \biggr( \lambda~(\typed{cv}{\mathbb{B}}) \Rightarrow
        \begin{cases}
          \sembra{S}{t}, & \text{if }cv \\
          \sembra{S}{u}, & \text{otherwise}
        \end{cases}\biggr),
      \quad \sembra{S}{\ISkip} ~=~ \Ret{\ttt}, \\
      &\sembra{S}{\IWhile{c}{t}} ~=~ \iter{\Biggr(\lambda~\ttt \Rightarrow \sembra{B}{c} \bindc \biggr( \lambda~cv \Rightarrow
          \begin{cases}
            \sembra{S}{t}\tm{;\!;}~\Ret{(\inl{\ttt})}, & \text{if }cv \\
            \Ret{(\inr{\ttt})}, & \text{otherwise}
          \end{cases}\biggl)\Biggl)}{\ttt}
    \end{typfunction}
    \vspace*{0.1cm}
    \typequation{\sentailsLR{A}{(\typed{x}{\IAExp})}{(\typed{m}{\Ctx})}{p}}{\PROP}
    { \entailsLR{\instrv{h_{\Ctx}}{\sembra{A}{x}}{m}}{\Obs{(\Log{m})}{\ttt}}{p}}

    \typequation{\sentailsLR{B}{(\typed{c}{\IBExp})}{(\typed{m}{\Ctx})}{p}}{\PROP}
    { \entailsLR{\instrv{h_{\Ctx}}{\sembra{B}{c}}{m}}{\Obs{(\Log{m})}{\ttt}}{p}}

    \typequation{\sentailsLR{S}{(\typed{t}{\stimp})}{(\typed{m}{\Ctx})}{p}}{\PROP}
    { \entailsLR{\instrv{h_{\Ctx}}{\sembra{S}{t}}{m}}{\Obs{(\Log{m})}{\ttt}}{\varphi}}    
  }
  \caption{Instrumentation and entailment of \stimp programs by denotation to \ictree{\St{\Ctx}{}}{}.}
  \label{f:ticl:imp-semantics}
\end{figure}

\section{Using \Ticl and \ICTrees}\label{s:ticl:usage}
This section demonstrates how to use \ticl and \ICTrees to define
new programming languages and temporal proof systems. Starting
with a simple imperative language called
\stimp, we give a step-by-step recipe to get started with \ticl liveness proofs:
\begin{enumerate}
\item Define the syntax and denotational semantics of \stimp
  using \ICTrees
  (Figures~\ref{f:ticl:imp-syntax} and \ref{f:ticl:imp-semantics}).
\item Define an instrumentation handler from \stimp events (\St{\Ctx}{}) to
  an instrumentation monad (\InstrM{\Ctx}{\Ctx}), chosing an appropriate
  ghost-state to observe ($h_{\Ctx}$ in Figure~\ref{f:ticl:imp-semantics}).
\item Using the \ticl library of structural lemmas (Section~\ref{s:ticl:lemmas}),
  prove high-level structural rules for \stimp (Figure~\ref{f:ticl:imp-lemmas}).
\end{enumerate}
\Ticl works at a high-level of abstraction---all of the above fits in
~500 lines of Rocq proofs and definitions---a small number even
compared to program logics for safety properties. Steps (1) and (2)
above are the same as those required for working with Interaction
Trees~\cite{xia2019interaction}. The proofs of the language-specific structural rules in step (3) are
high-level, syntactic, and fairly repetitive.

\subsection{Instrumentation of \stimp}\label{s:ticl:stimp-semantics}
\def \MathparLineskip {\lineskip=0.2cm}
\begin{figure}
  {\footnotesize
    \begin{mathpar}
      {
        \hspace*{-1.3cm}
        \mprset{vskip=.1cm}
        \inferrule*[Right=If$_S$AU$_L$,flushleft]
        {\sentailsR{B}{c}{m}{\AX{\doneq{(b,m)}{}}} \\\\
          {
            \quad\land \begin{cases}
              \sentailsR{S}{t}{m}{\AU{\varphi}{\varphi'}}, &\text{if } b \\
              \sentailsR{S}{u}{m}{\AU{\varphi}{\varphi'}}, &\text{otherwise}
            \end{cases}
          }
        }
        {\sentailsL{S}{\IIf{c}{t}{u}}{m}{\AU{\varphi}{\varphi'}}}
      }
      \and
      {
        \hspace*{.4cm}
        \mprset{vskip=.1cm}
        \inferrule*[Right=While$_S$AG]
        {
          {\begin{aligned}
            \R&~m \to \forall~m,~\R~m \to \\
               & \sentailsL{S}{\IWhile{c}{t}}{m}{\varphi}~\land \\
               & \sentailsR{B}{c}{m}{\AX{\doneq{(\tm{true},m)}{}}}~\land \\
               & \sentailsRR{t}{m}
                  \tm{AX}(\varphi~\tm{AU}~\tm{AX}~\tm{done}~\R)]
            \end{aligned}}}
        {\sentailsL{S}{\IWhile{c}{t}}{m}{\AG{\varphi}}}
      }
      \\\\
      {
        \hspace*{-1.3cm}
        \mprset{vskip=.1cm}
        \inferrule*[Right=While$_S$AU$_L$]
        {
          {\begin{aligned}
            \R&~m \to \forall~m,~\R~m \to \sentailsR{B}{c}{m}{\AX{\doneq{(b,m)}{}}} \\
            & \quad \land \begin{cases}
              \sentailsL{S}{t}{m}{\AU{\varphi}{\varphi'}}~\lor~& \quad \\
              \sentailsRR{t}{m}
              \varphi~\tm{AU}~\tm{AX}~\tm{done}~(\tm{\lambda}~m' \Rightarrow & \quad \\
              \quad \R~m'~\land~f~m' < f~m)~]_S, & \text{if }b \\
              \sentailsL{S}{\ISkip}{m}{\varphi'}, & \text{otherwise}
            \end{cases}
          \end{aligned}
        }
      }
      {\sentailsL{S}{\IWhile{c}{t}}{m}{\AU{\varphi}{\varphi'}}}
    }
    \and
    {
      \hspace*{.4cm}
      \mprset{vskip=.1cm}
      \inferrule*[Right=Seq$_S$AU$_L$]
      {\sentailsR{S}{t}{m}{\AU{\varphi}{\doneq{m'}{}}} \\\\
        \sentailsL{S}{u}{m'}{\AU{\varphi}{\varphi'}}}
      {\sentailsL{S}{\ISeq{t}{u}}{m}{\AU{\varphi}{\varphi'}}}
    }
    \end{mathpar}
  }
  \caption{Representative structural lemmas for language \stimp and \ticl operators $\AU{}{},\AG{}$.}
  \label{f:ticl:imp-lemmas}
\end{figure}

In Section~\ref{s:ticl:instrumentation} we defined
\emph{instrumentation}, a mechanism to evaluate events and record
proof relevant ghost-state. In this section, we demonstrate the process of giving semantics
to \stimp programs in two stages: (1) denoting \stimp syntax to an \ICTree, and (2) instrumenting the \ICTree
to an appropriate instrumentation monad.

For the first stage, the denotation brackets ($\sembra{S}{}$) in Figure~\ref{f:ticl:imp-semantics}
translate the syntax of \stimp (Figure~\ref{f:ticl:imp-syntax}) to an
$\ictree{\St{\Ctx}{}}{}$. The shared state $\Ctx$ is a map from
\tm{string} indices to natural number ($\mathbb{N}$) values. Low-level operations on maps and
their lemmas are assumed; $m_1 \cup m_2$ is
map union, $s \hookrightarrow x$ is the singleton map with key $s$ and value
$x$, $m[s]$ is the partial ``get'' that returns an option $\Some{v}$,
such that $v$ is the value associated with key $s$, or $\None$ if key $s$
does not exist in $m$. An \stimp \IVar{s} expression retrieves the value
of variable $s$ from shared state ($m[s]$), while an assignment (\IAssign{s}{x}) statement
updates the shared memory with the new value $((s \hookrightarrow a) \cup m)$.

The second stage defines the \stimp instrumentation handler
($h_{\Ctx}$) in Figure~\ref{f:ticl:imp-semantics}. This handler gives
a semantic meaning to $\sget$ and $\sput{}$ events, while also
specifying the \emph{ghost-state} to be observed by the proof
system. We chose to only instrument $\sput{m}$ events---by calling
$\log{m}$---to ensure that the instrumentation monad only remembers
events \emph{overwriting} the state, while events reading from the state
are evaluated and erased. Our choice of handler affects what
properties can be proved in \ticl later. For example, we cannot prove
``an empty heap is read'' using a handler that erases reads
($h_{\Ctx}$). Handlers are a flexible mechanism for observing program behavior;
we will see more examples of that in Section~\ref{s:ticl:examples}.

Putting the two stages together, $\instrv{h_{\Ctx}}{}{}$ ``applies''
the instrumentation handler over the tree
($\typed{\sembra{S}{t}}{\ictree{\St{\Ctx}{}}{}}$) with initial state
$s$. The end-to-end entailment relation \sentailsLR{S}{t}{s}{p},
connects an \stimp program ($t$) to a \ticl specification ($p$).

\subsection{\stimp structural lemmas}\label{s:ticl:stimp-lemmas}
\begin{figure}[t!]
  \begin{adjustwidth}{0pt}{}
    \begin{mathpar}
    \inferrule*[Right=$v\stackrel{\mathclap{\tiny\mbox{?}}}{>}N$,leftskip=.5cm,rightskip=-.5cm]{
      \inferrule*[Right=Seq$_S$AU$_L$]{
          \inferrule*[Right=While$_S$AU$_L$,rightskip=-.3cm]{
          \inferrule*[Right=${m[c]\stackrel{\mathclap{\tiny\mbox{?}}}{=}0}$]{
            \inferrule*[Right=${m[c]\stackrel{\mathclap{\tiny\mbox{?}}}{=}1}$]{
              \mathscale{0.7}{
                \begin{aligned}
                  v>&~N, ~ m[c] \leq v,~ m[r]+m[c] = v,~m[c] = 1, \\[-.4em]
                      & \quad\quad\sentailsR{B}{0 < c}{m}{\AX{\doneq{(\tm{true},m)}{}}}~\land~\greencheck \\
                      & \quad\quad\sentailsL{S}{c \gets c-1;~r \gets r+1}{m}{\AF{(\svar{r \geq N})}}
                \end{aligned}
              } \and
              \mathscale{0.7}{
                \begin{aligned}
                  m[c]&> 1, ~\sentailsR{B}{0 < c}{m}{\AX{\doneq{(\tm{true},m)}{}}}~\land~\greencheck \\
                  & \sentailsRR{c \gets c-1;~r \gets r+1}{m}\tm{AF}~\tm{AX}~\tm{done}~(\tm{\lambda}~m' \Rightarrow \\
                  & \quad\quad m'[c] \leq v\land~ m'[r]+m'[c] = v \\
                  & \quad\quad \land~v - m'[r] < v - m[r])~]t_S \\
                \end{aligned}
              }
            }
            {
              \mathscale{0.7}{
                \begin{aligned}
                  v>&~N,~ m[c] \leq v,~ m[r]+m[c] = v,\\
                  m[&c] = 0, \\
                    & \sentailsR{B}{0 < c}{m}{\AX{\doneq{(\tm{false},m)}{}}}~\land~\greencheck \\
                    & \sentailsL{S}{\ISkip}{m}{\svar{r \geq N}}
                \end{aligned}
              } \and
              \mathscale{0.7}{
                \begin{aligned}
                  m[c]&> 0,~ \sentailsR{B}{0 < c}{m}{\AX{\doneq{(\tm{true},m)}{}}}~\land \\
                  & \sentailsL{S}{c \gets c-1;~r \gets r+1}{m}{\AF{(\svar{r \geq N})}}~\lor~ \\
                  & \sentailsRR{c \gets c-1;~r \gets r+1}{m}\tm{AF}~\tm{AX}~\tm{done}~(\tm{\lambda}~m' \Rightarrow \\
                  & \quad\quad m'[c] \leq v\land~ m'[r]+m'[c] = v \\
                  & \quad\quad \land~v - m'[r] < v - m[r])~]_S \\
                \end{aligned}
              }
            }
          }
          {
            \mathscale{0.7}{
              \begin{aligned}
                &v>N,~\R~(r \hookrightarrow 0) \cup (c \hookrightarrow v)~\greencheck \\[2em]
                &\bm{\R~m} \coloneq m[c] \leq v \land m[r] + m[c] = v \\
                &\bm{f~m} \coloneq v - m[r]
              \end{aligned}
            } \and {
              \hspace*{-.5cm}
              \mathscale{0.7}{
                \begin{aligned}
                  \forall~m,& ~\exists~b,~\R~m \to \sentailsR{B}{0 < c}{m}{\AX{\doneq{(b,m)}{}}} \\
                      & \quad \land \begin{cases}
                        \sentailsL{S}{c \gets c-1;~r \gets r+1}{m}{\AF{(\svar{r \geq N})}}& \hspace*{-.3cm}\lor \\
                        \sentailsRR{c \gets c-1;~r \gets r+1}{m}\tm{AF}~\tm{AX}~\tm{done}~(\tm{\lambda}~m'&\hspace*{-.3cm}\Rightarrow \\
                        \quad\quad \R~m'~\land~f~m' < f~m)~]_S, & \hspace*{-.5cm}\text{if }b \\
                        \sentailsL{S}{\ISkip}{m}{\svar{r \geq N}}, & \hspace*{-.5cm}\text{otherwise}
                      \end{cases}
                \end{aligned}
              }
            }
          }
        }
        {
          \mathscale{0.8}{
            \inferrule*[Left=$v>N${,},vskip=.2cm]{
              s' = (r \hookrightarrow 0) \cup (c \hookrightarrow v)~\greencheck
            }
            {
              \sentailsL{S}{r \gets 0}{(c \hookrightarrow v)}{\AF{\AX{\doneq{s'}{}}}}
            }
          } \and
          {
            \begin{sentailsP}{.23}{L}{}
              {
                \While{$0 < c$}{
                  $c \gets c - 1$ \;
                  $r \gets r + 1$ \;
                }
              }
              {s'}
              {\AF{(\svar{r \geq N})}}
              {4.5cm}
            \end{sentailsP}
          }
        }
      }
      {
        {
          \mathscale{0.7}{v \leq N},
          \begin{sentailsP}{.23}{L}{}
            {
              $r \gets 0$ \;
              \While{$0 < c$}{
                $c \gets c - 1$ \;
                $r \gets r + 1$ \;
              }
            }
            {(\tm{c} \hookrightarrow \tm{v})}
            {\svar{c \leq N}}
            {4.5cm}
          \end{sentailsP}
          \mathscale{.7}{\greencheck}
        }\and
        {
          \mathscale{0.7}{v > N},
          \begin{sentailsP}{.23}{L}{}
            {
              $r \gets 0$ \;
              \While{$0 < c$}{
                $c \gets c - 1$ \;
                $r \gets r + 1$ \;
              }
            }
            {(\tm{c} \hookrightarrow \tm{v})}
            {\AF{(\svar{r \geq N})}}
            {5cm}
          \end{sentailsP}
        }
      }
    }
    {
      \begin{sentailsP}{.23}{L}{\forall~v,N}
        {
          $r \gets 0$ \;
          \While{$0 < c$}{
            $c \gets c - 1$ \;
            $r \gets r + 1$ \;
          }
        }
        {(\tm{c} \hookrightarrow \tm{v})}
        {\svar{c \leq N} \lor \AF{(\svar{r \geq N})}}
        {7.2cm}
        \end{sentailsP}
    }
  \end{mathpar}
\end{adjustwidth}
\caption{Example structural liveness proof for a simple \stimp loop program using \ticl.}
\label{f:ticl:imp-proof}
\end{figure}

Equipped with the end-to-end \ticl entailments over \stimp programs,
we proceed to ``lift'' the \ICTree rules in
Figures~\ref{f:ticl:ticl-bind} and \ref{f:ticl:ticl-iter} to the level of
\stimp program structures. Representative \stimp lemmas---focusing on
operators \AU{}{} and \AG{}---are shown in
Figure~\ref{f:ticl:imp-lemmas}. The full array of program structures
and temporal opertors is proven in our development.

Structural rules for \stimp (Figure~\ref{f:ticl:imp-lemmas})---much like
structural rules of \ICTrees---are backwards reasoning, meaning the goal
is in the bottom and proof obligations are given on the top of the
inference line. The obligations generated are ``smaller'' than
the goal they apply to, either targeting a subprogram of the original
program, or a subformula of the original formula. For example, in the
\emph{invariance} rule \mkw{WhileAG}, the first proof obligation
$\varphi$ is a subformula of $\AG{\varphi}$, while the other two proof obligations
refer to the loop conditinal and loop body. 

\subsection{Example: structural proof of liveness for \stimp}\label{s:ticl:stimp-liveness}

Now we demonstrate a \ticl structural proof in practice, by proving liveness of a program
from the T2 CTL benchmark suite~\cite{brockschmidt2016t2}.  The
program in Figure~\ref{f:ticl:imp-proof} (bottom) is a simple
\bm{$\tm{while}$} loop. The goal specification has been generalized with quantifiers
($\forall~v,N$) which are known to be challenging for model checking
systems~\cite{cimatti2022verification,
  pnueli2000liveness,farzan2016proving,brockschmidt2016t2}, but not \Ticl and the Rocq
proof assistant. We want to prove that either the initial value of
variable $c$ is $\leq$ a constant
($\typed{N}{\mathbb{N}}$), or eventually the value of $\tm{r}$ becomes
$\geq$ to $N$. Base formulas (for example \svar{n >
  0}) dereference variable $n$ from the current state (notation for $\exists~v, m[n] = \Some{v}$ and $v > 0$).

The proof begins by case analysis on value $v$. If $v \leq N$, we
have proved the left-side of the ``or'' in our goal. Here, $v$ is the
initial value of variable $c$.  Otherwise if $v > N$, use the sequence rule
(\mkw{Seq$_S$AU$_L$} in Figure~\ref{f:ticl:imp-lemmas}) to update the
state ($s'$) with the new assignment
($\IAssign{r}{0}$).

At this point, the \bm{$\tm{while}$} loop starts at the new state ($s'$) and our goal is
to prove the ``eventually'' property $\AF{(\svar{r \geq N})}$. The 
liveness lemma (\mkw{While$_S$AU$_L$} in Figure~\ref{f:ticl:imp-lemmas}) applies.
For its loop invariant ($\R$),
notice the sum of values in $r$ and $c$ remains constant throughout the loop
($m[r] + m[c] = v$)\footnote{Existantials are omitted in natural number
  propositions, so $m[r] + m[c] = v$ is shorthand notation for
  $\exists~v_r,~m[r]=\Some{v_r}$ and $\exists~v_c,~m[c]=\Some{v_c}$ such that
  $v_r+v_c=v$.}. For its ranking function ($f$), notice that variable $c$ is
at its greatest at value $v$, progressively decreasing every iteration
$(f~m \coloneq v - m[c])$. The remaining proof is straightforward.
\begin{enumerate}
\item Prove the initial loop state satisfies the loop invariant ($\R~(r \hookrightarrow 0) \cup (c \hookrightarrow v)$).
\item For each state ($m$) satisfying the invariant $\R~m$, the loop body must terminate and satisfy the loop invariant and the ranking function condition.
\item Taking three cases on the value of $c$ concludes the proof in Figure~\ref{f:ticl:imp-proof}:
  \begin{itemize}
  \item[-] If $m[c]=0$ then $m[r] = v$, so $\svar{r \geq N}$ is true.
  \item[-] If $m[c]=1$ then $m[r] = v - 1$, the loop body adds $1$ to $r$, so again, $\svar{r \geq N}$.
  \item[-] If $m[c]>1$ then it is easy to prove the loop invariant and variant are satisfied at termination of the loop body.
  \end{itemize}
\end{enumerate}

\section{Motivating examples}\label{s:ticl:examples}
We evaluated \ticl by structurally verifying several examples from the
T2 CTL benchmark suite~\cite{brockschmidt2016t2} (like the example in
Figure~\ref{f:ticl:imp-proof}) and three use cases inspired from computer systems.
In this section, we prove liveness and safety properties for a round-robin
scheduler, a secure concurrent shared memory system, and a
distributed consensus protocol.

In Section~\ref{s:ticl:usage} we defined the deep-embedding\footnote{Deep-embedding
  refers to the treatment of variables as datatypes and the explicit
  handling of substitution. Shallow-embedding languages avoid using a
  datatype for their abstract syntax altogether, in favor of programming in the
  proof assistant's metalanguage. Mixed-embedding languages
  have an abstract syntax, but use the proof assistant's variable and
  substitution mechanisms~\cite{chlipala2021skipping}.} imperative
language \stimp and its \ticl theory. In this section we define three mixed embedding
languages (\meQ{}, \meS{}, \meR{})~\cite{chlipala2021skipping} and a
shallow-embedding language (\ictree{\Enet}{}).  Our goal is to demonstrate the
flexibility of \ticl over different programming language techniques,
events, and temporal specifications.

\subsection{Round-robin scheduler}\label{s:ticl:ex-queue}
The syntax of the \meQ{} language for the round-robin scheduler
(\tm{rr}) from Figure~\ref{f:ticl:rotate} is given in
Figure~\ref{f:ticl:meq-syntax}. A shared queue ($\Queue$) with elements $T$
maintains the order of threads in the scheduler. Language \meQ{}
interfaces with the queue through the $\QPop$ and $\QPush{}$ instructions
and has sequential composition (\bindc) and an infinite loop program structures (\QWhile{t}).
This simple language is sufficient to prove that a thread $x$ will
\emph{always-eventually} get scheduled, the nested temporal property
we saw early on in Figure~\ref{f:ticl:rotate}. Nested temporal properties
like always-eventually pose a challenge for deductive verification,
not only to prove but to formally state.

\begin{figure}[t!]
  \begin{adjustwidth}{0pt}{}
    {\small
      \typequation{\Queue}{\Type}{\tm{list}~T}
      \typequation{E_{\Queue}}{\Type \to \Type}{
        ~\bbar \typed{\tm{Push}~(\typed{x}{T})}{E_{\Queue}~\unit}
        \quad \bbar \typed{\tm{Pop}}{E_{\Queue}~T}
      }
      \vspace*{.1cm}
      \begin{typsyntax}{\meQ{}}{\Type \to \Type}
        &\bbar \typed{\QPop}{\meQ{T}} \quad \bbar \typed{\QPush{(\typed{x}{T})}}{\meQ{\unit}} \\
        &\bbar \typed{\QWhile{\typed{p}{\meQ{X}}}}{\meQ{\unit}} \\
        &\bbar \QBind{(\typed{a}{\meQ{A}})}{(\typed{k}{A \to \meQ{B}})}
        \quad \bbar \typed{\QRet{(\typed{a}{A})}}{\meQ{A}}
      \end{typsyntax}
    }
  \end{adjustwidth}
  \caption{Language \meQ{} for a round-robin scheduler with a mutable queue ($\Queue$).}
  \label{f:ticl:meq-syntax}
\end{figure}

\begin{figure}[t!]
  \begin{adjustwidth}{0pt}{}
    {\small
      \vspace*{.1cm}
      \begin{typfunction}{h_{\Queue}}{E_{\Queue} \leadsto \InstrM{\Queue}{T}}
        & h_{\Queue}~(\typed{\tm{Push}\ n}{E_{\Queue}~\unit})~(\typed{q}{\Queue}) ~=~ \Ret{(\ttt, q \app [n])} \hspace*{6cm}\\
        & h_{\Queue}~(\typed{\tm{Pop}}{E_{\Queue}~\mathbb{N}})~(\typed{h \cons ts}{\Queue})
        ~=~ \llog{h}~\tm{;\!;}~\Ret(h, ts) \\
        & h_{\Queue}~(\typed{\tm{Pop}}{E_{\Queue}~\mathbb{N}})~(\typed{[]}{\Queue}) ~=~ \stuck
      \end{typfunction}
      \vspace*{.1cm}
      %% MeQ denotation
      \begin{typfunction}{\sembra{Q}{\textcolor{gray}{\_}}}{\meQ{A} \to \ictree{E_{\Queue}}{A}}
        &\sembra{Q}{\QPop} ~=~ \trigger{\tm{Pop}},
        \quad \sembra{Q}{\QPush{x}} ~=~ \trigger{(\tm{Push}~x)}, \hspace*{5cm} \\
        &\sembra{Q}{\QWhile{t}} ~=~ \iter{\left(\lambda~\ttt \Rightarrow \sembra{Q}{t} \bindc \left( \lambda~\_ \Rightarrow \Ret{(\inl{\ttt})}\right) \right)}{\ttt}, \\
        & \sembra{Q}{\QBind{x}{k}} ~=~ \sembra{Q}{x} \bindc (\lambda~a \Rightarrow \sembra{Q}{k~a}),
        \quad \sembra{Q}{\QRet{x}} ~=~ \Ret{x},
      \end{typfunction}
      \vspace*{0.1cm}
      \typequation{\qentailsLR{(\typed{t}{\meQ{A}})}{(\typed{q}{\Queue})}{(\typed{w}{\World{\logE{T}}})}{p}}{\PROP}
       {\entailsLR{\instrv{h_{\Queue}}{\sembra{Q}{t}}{q}}{w}{p}}
    }
  \end{adjustwidth}
  \caption{Denotation \sembra{Q}{\textcolor{gray}{\_}} and \ticl entailment for queue language \meQ{}.}
  \label{f:ticl:meq-semantics}
\end{figure}

Following the steps in Section~\ref{s:ticl:usage}, the denotation of
\meQ{} programs (\ictree{E_\Queue}{}) is shown in
Figure~\ref{f:ticl:meq-semantics}.  The queue instrumentation handler
($h_{\Queue}$) keeps track of popped elements, but not pushed
elements. The scheduler performs both actions in sequence, so
provenance information is not lost through this choice; still,
different target properties might require defining a different
handler.

The end-to-end \ticl entailment relation for \meQ{}
\qentailsLR{t}{q}{w}{p} is a \emph{quaternary} relation---in contrast to
previous ternary entailments we have seen. Its arguments are the
program $t$, queue $q$, current world $w$, and \ticl formula
$p$. In the last example (Figure~\ref{f:ticl:imp-semantics}) we used a ternary
relation, because the ghost-state coincided with the program
state (\Ctx). However, in Figure~\ref{f:ticl:meq-semantics} the
ghost-state represents elements popped ($T$) while the program state
represents the queue ($\Queue$), which necessitates keeping track of both.

\def \MathparLineskip {\lineskip=0.2cm}
\begin{figure}
  {\footnotesize
    \begin{mathpar}
      {
        \hspace*{-1.3cm}
        \mprset{vskip=.1cm}
        \mathscale{0.9}{
          \inferrule*[Right=While$_{\Queue}$AU$_L$]
          {
            {
              \begin{aligned}
                \R&~q~w\quad \to \forall~q~w,~\R~q~w \to  \\
                  & \qentailsL{t}{q}{w}{\AU{\varphi}{\varphi'}}~\lor \\
                  & \qentailsRR{t}{q}{w}\varphi~\tm{AU}~\tm{AX}~\tm{done}~(\tm{\lambda}~q'~w' \Rightarrow \\
                  & \quad\quad \R~q'~w'~\land~f~q' < f~q)~]_{\Queue}
              \end{aligned}
            }
          }
          {\qentailsL{\QWhile{t}}{q}{w}{\AU{\varphi}{\varphi'}}}
        }
    }
    \and
    {
      \hspace*{.4cm}
      \mprset{vskip=.1cm}
      \mathscale{0.9}{
        \inferrule*[Right=While$_{\Queue}$AG]{
          {
            \begin{aligned}
              \R&~q~w \to \forall~q~w,~\R~q~w \to  \\
                & \qentailsL{\QWhile{t}}{q}{w}{\varphi}~\land \\
                & \qentailsR{t}{q}{w}{\AX{(\AU{\varphi}{\AX{\done{\R}}})}}
            \end{aligned}
          }
        }
        {\qentailsL{\QWhile{t}}{q}{w}{\AG{\varphi}}}
      }
    }
  \end{mathpar}
  }
  \caption{Liveness and invariance loop lemmas for queue language \meQ{}.}
  \label{f:ticl:meq-lemmas}
\end{figure}

At this point we can prove the \emph{liveness} and
  \emph{invariance} lemmas in Figure~\ref{f:ticl:meq-lemmas}, which
  we will need to complete the \emph{always-eventually} proof in Figure~\ref{f:ticl:meq-proof}.
These proofs are short and reuse the \itern lemmas
from Figure~\ref{f:ticl:ticl-iter}. Comparing with the \stimp loop lemmas
in Figure~\ref{f:ticl:imp-lemmas}, the \meQ{} loop lemmas are
simpler; \meQ{} only supports infinite loops so there
is no case analysis on the loop conditional.

Finally, we proceed backwards (from bottom to top) through the
always-eventually proof in Figure~\ref{f:ticl:meq-proof}.
Start by using the \emph{invariance} rule \mkw{While$_{\Queue}$AG} with loop
invariant ($\R$): world $w$ must be not done, the queue will never be empty
($\exists~h,ts,~q=h::ts$), and either the head element is the target thread
($h = x$), or not ($h \neq x$), in which case $x$ must appear at some position $i$ in the
queue's tail ($\tm{find}~x~ts = \Some{i}$). Applying the invariance rule leaves three proof obligations:
\begin{enumerate}
\item The loop invariant must be initially satisfied ($\R~(q\app [x])~w$).
\item The loop body steps, then eventually terminates, respecting the
  invariant $\R$.
\item The loop must satisfy the inner \emph{eventually} property
  $(\AF{\obs{(\lambda~hd \Rightarrow ~hd=\tm{x})}})$.
\end{enumerate}

The first subproof is easy to prove. The second is also
straightforward by case analysis on the head of the queue ($h = x$).
The third subproof (inner eventually) requires the \emph{liveness}
lemma (\mkw{While$_{\Queue}$AU$_L$}) from
Figure~\ref{f:ticl:meq-lemmas}.  Since this is the same loop as
before, we reuse the loop invariant ($\R$) from the
\emph{invariance} rule. The ranking function $f$ is given simply by
the index of $x$ in the queue. The queue always contains $x$ (by
\R), meaning $f$ is total. We conclude by low-level
reasoning over lists and using the loop invariants.  For further
details the reader can refer to our Rocq development; the syntax and semantics of \meQ{},
(Figure~\ref{f:ticl:meq-syntax}), \ticl structural lemmas
(Figure~\ref{f:ticl:meq-lemmas}) and the \emph{always-eventually} proof in
Figure~\ref{f:ticl:meq-proof} span 137 lines of Rocq definitions
and 362 lines of proofs.

\begin{figure}[t!]
  \begin{adjustwidth}{0pt}{}
    \begin{mathpar}
    \inferrule*[Right=While$_{\Queue}$AG,leftskip=1.6cm]{
      \inferrule*[Right=${h\stackrel{\mathclap{\tiny\mbox{?}}}{=}x}$]{
        \inferrule*[Right=And]{
          \inferrule*[Right=While$_{\Queue}$AU$_L$]{
            \mathscale{0.65}{
              \begin{aligned}
                &\forall~x,w,h,ts,i,~\notdone{w},~h \neq x,~\tm{find}~x~ts = \Some{i},\\
                &\R'~(h::ts)~w~\greencheck \\[1em]
                &\bm{\R'~q'~w'} \coloneq \notdone{w'}~\land \\
                &\quad \exists~h',ts',~q' = h' :: ts'~\land \\
                &\quad (h' = x~\lor~(h' \neq x~\land~\exists~i',\tm{find}~x~ts' = \Some{i'})) \\
                &\bm{f~q} \coloneq \tm{find}~x~q
              \end{aligned}\hspace*{-1cm}
            } \and
              \mathscale{0.7}{
                \begin{aligned}
                  \forall&~q,w,~\R'~q~w \to \greencheck \\
                     & \qentailsL{p \gets \QPop();~\QPush{p}}{q}{w}{\AF{\obs{(\lambda~hd \Rightarrow ~hd=\tm{x})}}}~\lor \\
                     & \qentailsRR{p \gets \QPop();~\QPush{p}}{q}{w}~\tm{AF}~\tm{AX}~\tm{done}~(\tm{\lambda}~q'~w' \Rightarrow \\
                     & \quad\quad \R'~q'~w'~\land~f~q' < f~q)~]_{\Queue}
                \end{aligned}\hspace*{-.5cm}
              }
          }
          {
            \mathscale{0.69}{
              \begin{aligned}
                \forall~&x,w,h,ts,i,~\notdone{w},~h \neq x,~\tm{find}~x~ts = \Some{i}, \\
                & \qentailsL{\QWhile{p \gets \QPop();~\QPush{p}}}{(h :: ts)}{w}{\AF{\obs{(\lambda~hd \Rightarrow ~hd=\tm{x})}}}
              \end{aligned}
            }
          }
        }
        {
          \mathscale{0.7}{
            \begin{aligned}
              &\forall~x,w,ts,\\
              & \notdone{w} \to ~\greencheck\\
              & [ \QWhile{p \gets \QPop();~\QPush{p}}, \\
              & \quad\quad q,w~\Vdash_L~\AF{\obs{(\lambda~hd \Rightarrow ~hd=\tm{x})}} ]_{\Queue} \\
              & \land~\qentailsRR{p \gets \QPop();~\QPush{p}}{(x::ts)}{w} \\
              & \quad\quad\tm{AX}(\AF{\obs{(\lambda~hd \Rightarrow ~hd=\tm{x})}})~\tm{AU}~\tm{AX}~\tm{done}~\R)]_{\Queue}
            \end{aligned}
          } \and
          \mathscale{0.7}{
            \begin{aligned}
              &\forall~x,w,h,ts,i,\\
              &\notdone{w} \to h \neq x \to \tm{find}~x~ts = \Some{i} \to\\
              & [ \QWhile{p \gets \QPop();~\QPush{p}}, \\
              & \quad\quad q,w~\Vdash_L~\AF{\obs{(\lambda~hd \Rightarrow ~hd=\tm{x})}} ]_{\Queue} \\
              & \land~\qentailsRR{p \gets \QPop();~\QPush{p}}{(h::ts)}{w} \\
              & \quad\quad\tm{AX}(\AF{\obs{(\lambda~hd \Rightarrow ~hd=\tm{x})}})~\tm{AU}~\tm{AX}~\tm{done}~\R)]_{\Queue}~\greencheck
            \end{aligned}
          }
        }
      }
      {
        \mathscale{0.7}{
          \begin{aligned}
            &\forall~q,x,w,~\R~(q \app [x])~w~\greencheck \\[1em]
            &\bm{\R~q~w} \coloneq \notdone{w}~\land \\
            &\quad \exists~h,ts,~q = h :: ts~\land\\
            &\quad (h = x~\lor~(h \neq x~\land \\
            &\quad\quad \exists~i,\tm{find}~x~ts = \Some{i}))
          \end{aligned}
        }
        \and
        \hspace*{1cm}
        \mathscale{0.7}{
          \begin{aligned}
            &\forall~x,q,w~,~\R~q~w \to \\
            & [ \QWhile{p \gets \QPop();~\QPush{p}}, \\
            & \quad\quad q,w~\Vdash_L~\AF{\obs{(\lambda~hd \Rightarrow ~hd=\tm{x})}} ]_{\Queue} \\
            & \land~\qentailsRR{p \gets \QPop();~\QPush{p}}{q}{w} \\
            & \quad\quad\tm{AX}(\AF{\obs{(\lambda~hd \Rightarrow ~hd=\tm{x})}})~\tm{AU}~\tm{AX}~\tm{done}~\R)]_{\Queue}
          \end{aligned}
        }
      }
    }
    {
      \begin{qentailsP}{.29}{L}{\forall~\tm{q, x}}
    {
      \While{\tm{true}}{
        $p \gets \QPop();~\QPush{p}$
      }
    }
    {\tm{(q \app [x])}}
    {\Pure}
    {\AG{\AF{\obs{(\lambda~hd \Rightarrow ~hd=\tm{x})}}}}
    {9.5cm}
  \end{qentailsP}
    }
  \end{mathpar}
\end{adjustwidth}
\caption{Structural always-eventually proof for round-robin in \meQ{} using \ticl.}
\label{f:ticl:meq-proof}
\end{figure}

\subsection{Secure concurrent shared memory}\label{s:ticl:ex-sec}
For the next example let us switch gears and prove confidentiality of concurrent reads
and writes over a shared memory with security labels---a safety property. The proposed system is inspired by Mandatory
Access Control (MAC). Language \meS{} in Figure~\ref{f:ticl:mes-syntax}
uses a mutable heap (\MS), where each cell is \emph{tagged} with an information-flow
security label, either \emph{low} security ($L$) or high security
($H$). Labels form a preorder with accessibility
relation ($\leq$)---the smallest reflexive, transitive relation such that
$L \leq H$ holds.

Tagged memory is accessed by instructions \SRead{l_i}{x} and
\SWrite{l_i}{x}{y}, where $l_i$ is the permission level of the
instruction, $x$ is the address, and $y$ is the value to write. The
goal is to prove every read instruction (\SRead{l_i}{x}) accesses a
memory cell with a security level that is $\leq$ to
its permission level ($l_m \leq l_i$).  We prove this \emph{always} property for two
interleaved processes at different security levels, \emph{alice} and
\emph{bob}. The nondeterministic interleaving of \emph{alice} and \emph{bob} is a superset of all
concurrent traces---by proving safety in the interleaving we guarantee
safety in all concurrent executions.

The instrumentation handler ($h_S$), the denotational semantics of
the process language (\meS{}) and scheduler language (\meR{}), the structural
lemmas for \meR{}, and the complete safety proof can be found in
Appendix~\ref{appendix:ticl:mes}. The definition of the languages
(Figure~\ref{f:ticl:mes-syntax}), \ticl structural lemmas
(Figure~\ref{f:ticl:mes-lemmas}), and the safety proof
(Figure~\ref{f:ticl:mes-proof}) required 174 lines of Rocq definitions
and 242 lines of proofs.

\begin{figure}[t!]
  \begin{adjustwidth}{0pt}{}
    {\small
      \typequation{\Lbl}{\Type}{ \bbar L \quad \bbar H}
      \typequation{\MS}{\Type}{ \Map{\mathbb{N}}{(\mathbb{N} * \Lbl)}}
      \typequation{E_{\Lbl}}{\Type \to \Type}{
        & ~\bbar \typed{\tm{Read}~(\typed{l}{\Lbl})~(\typed{x}{\mathbb{N}})}{E_{\Lbl}~\option{\mathbb{N}}} \\
        & ~\bbar \typed{\tm{Write}~(\typed{l}{\Lbl})~(\typed{x}{\mathbb{N}})~(\typed{v}{\mathbb{N}})}{E_{\Lbl}~\unit}
      }
      \begin{typsyntax}{\meS{}}{\Type \to \Type}
        & \bbar \typed{\SRead{(\typed{l}{\Lbl})}{(\typed{n}{\mathbb{N}})}}{\meS{\option{\mathbb{N}}}} \\
        & \bbar \typed{\SWrite{(\typed{l}{\Lbl})}{(\typed{n}{\mathbb{N}})}{(\typed{v}{\mathbb{N}})}}{\meS{\ttt}} \\
        & \bbar \typed{\SIf{(\typed{c}{\mathbb{B}})}{(\typed{t}{\meS{A}})}{(\typed{u}{\meS{A}})}}{\meS{A}} \\
        & \bbar \typed{(\SBind{\typed{a}{\meS{A}}}{(\typed{k}{A \to \meS{B}})})}{\meS{B}}
        \quad \bbar \typed{\SRet{(\typed{a}{A})}}{\meS{A}}
      \end{typsyntax}
      \begin{typsyntax}{\meR{}}{\Type \to \Type}
        & \bbar \typed{\RLoop{(\typed{k}{X \to \meR{X}})}{(\typed{x}{X})}}{\meR{\unit}} \\
        & \bbar \typed{\RBr{(\typed{l}{\meR{A}})}{(\typed{r}{\meR{A}})}}{\meR{A}} 
        \quad \bbar \typed{\RCall{(\typed{p}{\meS{A}})}}{\meR{A}} \\
        & \bbar \typed{(\RBind{\typed{a}{\meR{A}}}{(\typed{k}{A \to \meR{B}})})}{\meR{B}}
        \quad \bbar \typed{\RRet{(\typed{a}{A})}}{\meR{A}} \\
      \end{typsyntax}
    }
  \end{adjustwidth}
  \caption{Process language \meS{} has read-write access to a security
    labelled heap. Scheduler language \meR{} has infinite loops
    ($\tm{loop}$), process calls ($\tm{call}$) and nondeterministic
    choice ($\oplus$).}
  \label{f:ticl:mes-syntax}
\end{figure}

\subsection{Distributed Consensus}\label{s:ticl:ex-election}
For the last example, we present a different approach to modeling
systems, and a new liveness composition lemma. Instead of defining a
programming language syntax, following the steps in
Sections~\ref{s:ticl:usage},~\ref{s:ticl:ex-queue}, and \ref{s:ticl:ex-sec},
we model a distributed, message-passing system in the
metalanguage of the Rocq proof assistant using \ICTrees directly. This
shallow-embedding provides a shortcut to the interesting part of the
protocol proof, reducing the syntax and denotation overhead.

\begin{lrbox}{\procbox}%
\begin{lstlisting}[style=customcoq,basicstyle=\footnotesize\ttfamily]
proc $\typed{(\typed{pid}{\Pid})}{\ictree{\hnet}{\unit}}\coloneq$
   $m$ <- recv $pid$;
   match $m$ with
   | C $candidate$ =>
     match compare $candidate$ $pid$ with
       | Gt => send $pid$ (C $candidate$)
       | Lt => Ret tt
       | Eq => send $pid$ (E $pid$)
     end
   | E $leader$ => send $pid$ (E $leader$)
   end.
  \end{lstlisting}
\end{lrbox}

\begin{wrapfigure}{R}{0.35\textwidth}
  \centering
  \scalebox{0.85}{\usebox{\procbox}}
  \caption{Leader election process.}
  \label{f:ticl:election-proc}
\end{wrapfigure}

The goal of the protocol is \emph{leader election}; processes must
reach consensus on which process will be the \emph{leader}. Leader
election is a common component of many distributed protocols like
Paxos~\cite{lamport2001paxos}. We are interested in the
liveness property ``eventually a leader is elected''.  For simplicity
we assume there are no network, process, or Byzantine
failures.  Modeling failures by using \ICTree's nondeterminism
is entirely possible, but doing so is beyond the scope of this paper.

Processes (Figure~\ref{f:ticl:election-proc}) perform message-passing
events (\tm{send} and \tm{recv}) defined in Figure~\ref{f:ticl:election-syntax}.
The messages are delivered in a
unidirectional ring (\emph{uniring}) configuration in a clockwise
manner, as shown in Figure~\ref{f:ticl:election-draw}. Process
scheduling is also in a uniring, following the same pattern.  Each
process sends and receives one of two kinds of messages: proposing a
candidate PID $(C_i)$ and announcing a leader ($E_i$). The formal
definition of messages and mailboxes are in
Appendix~\ref{appendix:ticl:election}. There are two distinct phases
in this leader election protocol:
\begin{enumerate}
\item \emph{Aggregating candidate nominations}: initially, every process
  ($pid$) self-nominates to be the leader.  Processes receive a
  candidate message ($C_{candidate}$). If the candidate PID received is greater
  than the process' own PID ($pid < candidate$), the message is propagated. If
  the candidate PID is less, the message is dropped.
\item \emph{Announcing the leader}: if a process ($pid$) receives their own
  candidacy message back ($C_{pid}$), they announce themselves the elected
  leader ($E_{pid}$). A process that receives an election announcement
  ($E_{leader}$) propagates it. The protocol diverges in the end, sending the leader
  announcement in cycles forever. The process with the
  highest PID will always be the leader.
\end{enumerate}

\noindent So far in \ticl we demonstrated uses of modular liveness rules
following the structure of programs (e.g., loops).
In the protocol in Figure~\ref{f:ticl:election-draw} the
``modules''---the logical parts of the problem we identified as basic
building blocks---are the phases of the protocol, not the scheduler
loop. Using the \emph{liveness} lemma (\mkw{IterAU$_L$}) would require
establishing an invariant and variant that apply to both
phases of the protocol, which is quite challenging and not very
modular. The following theorem illustrates a new liveness composition rule we
call \emph{liveness split}:

\begin{theorem}[Liveness split] \ \\
  {\footnotesize
  \begin{mathpar}
    %%% aul_iter_split
    {\mprset{vskip=.1cm}
      \hspace*{-1.5cm}
      \inferrule*[Right=SplitAU$_{L,\mathbb{N}}$]{
        \hspace*{-2.9cm}
        \R~i~w \\
        \hspace*{1cm}
    {(\forall~i~w, ~\R_I~i~w \to \entailsL{\iter{k}{i}}{w}{\AU{\varphi}{\varphi'}})} \\\\
    {\begin{aligned}
      \forall~i,&w,~\R~i~w \to \\
          & \entailsRR{\tm{k~i}}{w} \varphi~\tm{AU}~\tm{AX}~\tm{done}~(\tm{\lambda}~lr~w'~ \Rightarrow
            ~\exists~i',~lr = \inl{i'} ~ \land~ \R_I~i'~w'~ \rangle \quad \lor \\
          & \entailsRR{\tm{k~i}}{w} \varphi~\tm{AU}~\tm{AX}~\tm{done}~(\tm{\lambda}~lr~w'~ \Rightarrow
            ~\exists~i',~lr = \inl{i'} ~ \land~ \R~i'~w'~ \land~ f~i'~w' < f~i~w) ~ \rangle
    \end{aligned}}}
    {\entailsL{\iter{k}{i}}{w}{\AU{\varphi}{\varphi'}}}}
  \end{mathpar}
}
\label{f:ticl:liveness-split}
\end{theorem}

\emph{Liveness split} breaks up a liveness proof to two parts, before and
after a user-defined intermediate point. This reduces a proof of loop
liveness to two ``smaller'' liveness proofs, connected together by an
intermediate relation ($\R_I$).  Those smaller liveness proofs have
smaller, simpler ranking functions ($f$). What is left afterwards is
the same liveness proof we started with, but starting from a better
position
($\R_I~i~w \to \entailsL{\iter{k}{i}}{w}{\AU{\varphi}{\varphi'}}$). One can continue
splitting liveness proofs in this way, by specifying convenient
intermediate relations $\R_I$ and ranking functions $f$. The notion of
modularity for liveness proofs extends beyond program structures (i.e:
loops), to logical structures, like the phases of the leader election
protocol.

The complete liveness proof for the leader election protocol is given
in Appendix~\ref{appendix:ticl:election}. Verification of the protocol
required 123 lines of Rocq definitions and 115 lines of proofs.

\newcommand{\electiontikz}[6]
{
  \scalebox{0.8}{
  \begin{tikzpicture}

    % Define the nodes in a ring
    \draw (0,0) node(A) {#4};
    \draw (1,1) node(B) {#5};
    \draw (2,0) node(C) {#6};

    % Draw curved arrows between nodes counter-clockwise
    \draw [->, bend left] (A) to node[midway, left]  {#1} (B);
    \draw [->, bend left] (B) to node[midway, right] {#2} (C);
    \draw [->, bend left] (C) to node[midway, below] {#3} (A);
  \end{tikzpicture}
}
}

\begin{figure}[t!]
  \setlength{\tabcolsep}{.6cm}

  \hspace*{-1cm}
  \begin{tabular}{m{1cm} m{1cm} m{1cm} m{1cm} m{1cm} m{1cm}}
    \electiontikz{$C_1$}{$C_2$}{$C_3$}{1}{2}{3}
      & \electiontikz{$C_3$}{$~$}{~}{1}{2}{3}
      & \electiontikz{~}{$C_3$}{~}{1}{2}{3}
      & \electiontikz{~}{~}{$E_3$}{1}{2}{\textbf{\textit{3}}}
      & \electiontikz{$E_3$}{$~$}{~}{\textbf{\textit{1}}}{2}{\textbf{\textit{3}}}
      & \electiontikz{~}{$E_3$}{~}{\textbf{\textit{1}}}{\textbf{\textit{2}}}{\textbf{\textit{3}}} \\
  \end{tabular}
  \caption{A unidirectional ring of three processes running the
    leader election protocol starting at PID$=1$.}

  \label{f:ticl:election-draw}
\end{figure}

\section{Discussion and related work}\label{s:ticl:discussion}
\Ticl is a temporal logic for mechanized, modular verification of
safety and liveness properties over effectful, nondeterministic, and
potentially nonterminating programs.  Its salient aspect is that it
can prove general temporal properties, over any programming language
denoting to coinductive
trees~\cite{xia2019interaction,chappe2023choice,YZZ22,itrees-spec,SH+23,ZHK+21,cpp19,lee2023fair},
with high-level lemmas and without the bureaucracy of (co\mbox{-})inductive
proofs. 

\subsection{Comparison with ITrees and CTrees}\label{s:discussion:trees}
\ICTrees{}{} are a computational model that is more expressive than
Interaction Trees (ITrees)~\cite{xia2019interaction} but less expressive than Choice Trees (CTrees)~\cite{chappe2023choice}.
We introduce this intermediate model of computation, instead of adopting
CTrees, because CTrees support two types of nondeterminism but we were only able to prove key lemmas of \ticl formulas for one of them.
In particular, CTrees support \emph{stepping choices} and \emph{delayed choices}.
In the context of labelled transition systems (LTS)---which give CTrees their operational semantics---stepping choices
correspond to \emph{$\tau$ transitions} (they model internal actions that do not change the observable behavior of the system),
while delayed choices do not correspond to a transition and defer to a transition in a child node.
Delayed choices are important for modeling the equational theory of certain models of concurrency, such as CCS~\cite{milner1980calculus}.
\ICTrees support stepping choices but not delayed choices\footnote{
  It remains an open question whether \ticl could support CTrees.
  The presence of delayed choice nodes (\tm{BrD}) in CTrees makes structural
  proofs of $t \bindc k$ (bind), where $t$ can make transitive, nondeterministic choices, much harder---specifically, it is not clear what is the inductive invariant on $t$ that characterizes the result of the entire bind.}.
%In contrast to ITrees and CTrees, where the goal is to complete proofs purely by
%equational reasoning~\cite{xia2019interaction,chappe2023choice}, \ticl's
%focus is on the transition relation of coinductive trees, used to
%define the \ticl entailment relation in Section~\ref{s:ticl:ctree-kripke}.

\subsection{Comparison with LTL, CTL, and TLA}\label{s:ticl:ctl-comparison}
The main difference between \ticl{} and temporal logics like
LTL~\cite{pnueli1977temporal}, CTL~\cite{emerson1982using},
and TLA~\cite{lamport1994temporal}
is that \ticl{} is designed with composition as a guiding principle.
As a result, \ticl{} needs to handle both finite programs with
postconditions, and infinite programs. In more detail, \Ticl's
treatment of termination is fundamentally different from LTL, CTL and
TLA, which assume a total Kripke transition relation (i.e:
$\forall~m,~\exists~m', R~m~m'$)~\cite{denicola1990action,emerson1986sometimes}.
\Ticl{} uses a non-total transition relation for \ICTree
structures (Section~\ref{s:ticl:ctree-kripke}), a variation on
finite trace LTL~\cite{de2013linear,artale2019you}. However, finite trace
logics do not support infinite traces and the \emph{always} operator,
and their support for postcondition specifications is limited---\ticl uses
the proof assistants metalanguage to describe complex postconditions.

\Ticl supports both finite and infinite properties, all CTL operators
($\varphi$ in Section~\ref{s:ticl:syntax}) and complex postconditions
($\psix$). By building on recent advances in monadic, coinductive
structures~\cite{capretta2005general,xia2019interaction,chappe2023choice},
\ticl proof composition follows the monadic composition
lemmas. Specifically, it allows $\Ret{v}$ to transition to a nullary
state, ensuring proper sequencing with continuation $k$ in one step,
as required by the monad laws (Figure\ref{f:ticl:ctree-algebra}).
Although there are well-known embeddings of CTL and TLA in the Rocq
proof assistant~\cite{doczkal2016completeness,sun2024anvil}, our
different approach to modularity and the large number of structural
proof lemmas we discovered as a result
(Figure~\ref{f:ticl:ticl-table}) indicates this is a still unexplored
area of research.

\subsection{Comparison with program logics}\label{s:ticl:deductive-comparison}
Comparing \ticl to existing program logics is straightforward, as
those are usually transparent on which property classes they
target. This is reflected in their choice of inductive or coinductive big-step semantics, which
ties the logic to \emph{eventally} or \emph{always} properties, with
no possibility to prove the other class in the future. While current
approaches excel within those boundaries, none offer a general,
compositional solution for proving arbitrary temporal properties like
\emph{always-eventually}.

{\bf Iris and Transfinite Iris: } Iris~\cite{jung2018iris} is a
concurrent-separation logic framework for Rocq that uses step-indexed
logical relations to prove safety properties of concurrent
programs. The recent extension \emph{Transfinite
  Iris}~\cite{spies2021transfinite} extends the step-indexing relation
from the naturals to ordinals, allowing total-correctness properties
to be proved by transfinite induction. A fundamental limitation of
step-indexing is that there is only one index; in the case of
``always-eventually'' properties, a hierarchy of induction and
transfinite induction proofs are required---this hierarchy is implicit
in the definition of $\vDash_{L,R}$ in \ticl
(Figure~\ref{f:ticl:ticl-entails}).  At the same time, \ticl, unlike
Iris, has no facilities for separation logic. One can imagine having
the ``best of both worlds'', combining the separation logic reasoning
of Iris and temporal reasoning of \ticl.

{\bf Fair operational semantics: } Lee et al.~\cite{lee2023fair}
recognize the limited support for liveness properties in mechanized
formal verification and propose an operational semantics for fairness
(FOS). FOS uses implicit counters for \emph{bad} events and defines
operational semantics that prove no infinite chain of \emph{bad}
events happens.  FOS provides comprehensive support for the specific
case of \emph{binary} fairness (\emph{good} vs. \emph{bad} events), but
limited support for general temporal specifications, like safety,
liveness and termination. As with Iris, it would be interesting to
combine that approach with \ticl.

{\bf Maude:} The Maude language and Temporal Rewriting Logic
(TLR)~\cite{meseguer2008temporal,meseguer1992conditional} recognize
the benefits of structural approaches (namely term rewriting) to
temporal logic verification. In \ticl we enable term rewriting with
\emph{up-to-tau equivalence} under a temporal context
(Section~\ref{s:ticl:ctree-core}).  However, Maude operates on the level of
models, not on the level of executable programs. This creates a verification gap
between the executable code and target properties. Finally
proof composition in Maude is not modular in the sense of Hoare Logic and \ticl.

{\bf Dijkstra monads: } Several works on Dijkstra monads target
partial-correctness properties in the style of weakest
preconditions~\cite{ahman2017dijkstra,maillard2019dijkstra,
  winterhalter2022partial,silver2021dijkstra}. Recent work targets
total-correctness properties like ``always''~\cite{silver2021dijkstra}
but not general temporal properties like liveness.

{\bf Synthesising ranking functions: } Yao et al.~\cite{yao2024mostly}
propose an automated synthesis procedure for ranking functions,
specialized to proving liveness properties in a class of distributed
systems.  Similar to model checking, the systems are described as
specifications not as implementations which is different from \ticl.
At the same time, automated synthesis of ranking functions is a
particularly attractive feature for \ticl, as they can be used with
\ticl lemmas like \mkw{IterAU$_{L,\mathbb{N}}$} (Figure~\ref{f:ticl:ticl-iter})
to get mostly automated, formal proofs of liveness.

\subsection{General Liveness Properties and Completeness}\label{s:ticl:liveness}
Our focus with \ticl is on providing a convenient and useful temporal
logic. However, its compositionality does not come without a cost:
just as with all the other (standard) temporal logics described above,
\ticl's temporal operators are not \textit{complete}.  Classic results
from automata theory~\cite{wolper83} show that there exist liveness
properties that cannot be expressed solely via temporal-logic
combinators of the kind supported by \ticl. To achieve completeness,
one would instead have to use alternative means of specifying the
desired liveness properties, such as with B{\"u}chi
automata~\cite{alpern1987recognizing}.  But reasoning about the
liveness properties expressible by such B{\"u}chi automata can, in the
limit, require using arbitrarily complex well-foundedness arguments.
We can therefore think of temporal logics, generally, as hiding that
complexity for the common case where the property of interest is
expressible in the logic.  In the case of \ticl, there are a few more
subtleties, however: first, its notion of definable observations,
which define the set of predicates for the logic, can be an
\textit{infinite} set (in contrast to much prior work that uses a
finite set of observations), and second, because it is embedded in the
Rocq framework, one could always fall back on raw coinductive proofs
about some liveness property.  It seems possible to extend \ticl with
an ``escape hatch'' mechanism that would let such proofs act as
``axioms'' from the point of view of \ticl's logic.  This would, in
theory, recover completeness at the expense of more manual user
effort.

\subsection{Conclusion}
In this work we ask: is it possible to write modular proofs about programs in a
general temporal logic akin to proofs in Hoare logic? We believe we
have answered affirmatively, and in the process
developed Temporal Interaction and Choice Logic (\ticl), a
specification language capable of expressing general liveness and
safety properties (we summarize \ticl in
Figure~\ref{f:ticl:ticl-syntax}). Along the way, we also designed an
extensive metatheory of structural lemmas
(Figures~\ref{f:ticl:ticl-table},~\ref{f:ticl:ticl-bind},~\ref{f:ticl:ticl-iter},~\ref{f:ticl:imp-lemmas})
that encapsulate complex (co-)inductive proofs to simple rule
application. We applied \ticl to several examples from
the T2 CTL benchmark suite~\cite{brockschmidt2016t2} and in three examples
inspired from computer systems as a way to demonstrate the metatheory
in action.

% Acknowledgements
\begin{acks}
This work was funded in part by NSF Grants CCF-2326576, CCF-2124184,
CNS-2107147, CNS-2321726, CNS-2247088. Opinions in this material are those of the authors and do not necessarily reflect the views of the NSF.

\end{acks}

\bibliographystyle{ACM-Reference-Format}
%\bibliography{conferences.bib,main.bib,refs.bib}
%%% -*-BibTeX-*-
%%% Do NOT edit. File created by BibTeX with style
%%% ACM-Reference-Format-Journals [18-Jan-2012].

%\section*{Appendix}\label{s:ticl:appendix}
\appendix
\section{Coinductive Proofs and Up-to Principles in Rocq}\label{appendix:ticl:coinduction}
In this appendix we focus on the low-level coinduction constructs in
\ticl used to define the \emph{forever} operators \AG{}, \EG{}. The
implementation details of coinductive structures and proofs in proof
assistants differs. We focus on the Rocq proof assistant, where the
infrastructure for coinductive proofs in provided by external
libraries~\cite{hur2013power,pous2016coinduction,zakowski2020equational}.
\Ticl relies on the \texttt{coinduction}
library by Damien Pous~\cite{pous2016coinduction} to define greatest fixpoints over
the complete lattice of Rocq propositions.

The primary construction offered by the library is a greatest fixpoint
operator ($\gfp~b:~X$) for any complete lattice $X$ and monotone
endofunction $b : X \to X$. Specifically, the library proves Rocq
propositions form a complete lattice, as do any
functions from an arbitrary type into a complete
lattice. Consequently, coinductive relations of arbitrary arity over
arbitrary types can be constructed using this combinator. In \ticl, we
target coinductive predicates over \ICTrees and worlds so we
work in the complete lattice $\ictree{E}{X} \to \World{E} \to \PROP$.

The \texttt{coinduction} library~\cite{pous2016coinduction} provides tactic support for
coinductive proofs based on Knaster-Tarski’s theorem: any
post-fixpoint is below the greatest fixpoint.  Given an endofunction
$b$, a (sound) enhanced coinduction principle, also known as an up-to
principle, involves an additional function $f : X \to X$ allowing one to
work with $b \circ f$ (the composition of $b$ with $f$) instead of
$b$: any post-fixpoint of $b \circ f$ is below the greatest fixpoint of
$f$. Practically, this gives the user access to a new proof
principle. Rather than needing to ``fall back'' precisely into their
coinduction hypothesis after ``stepping'' through $b$, they may first
apply $f$.

In Figure~\ref{f:ticl:ticl-upto} we give the up-to-principles for
coinduction proofs in \ticl. The $\upto{upto}(equiv)$ principle is used
to show $\upto{upto}(equiv) \leq \lambda~t.~\gfp(\tm{anc}~\varphi)~t$, meaning equivalent
trees (abstracting over the exact equivalence relation) satisfy the same
$\AG{\varphi}$ formula (similarly $\EG{\varphi}$). Note,
$\upto{upto}(equiv)$ is sufficiently general; any equivalence relation
\tm{equiv} satisfying the \mkw{ExEquiv} lemma in
Figure~\ref{f:ticl:kripke-lemmas} can be used in place of up-to-tau equivalence ($\sbisim$).

\begin{figure}[hb!]
  {\small
  \begin{alignat*}{2}
    \upto{upto}(equiv)~\R   & \triangleq &&~ \{t \sep \exists~t',~ \tm{equiv}~t~t' ~\land~ \R~t' \} \\
    \upto{bindAG}(\varphi,\Px)~\R & \triangleq &&~ \{(t\bindc k, w) \sep \entailsR{t}{w}{\AU{\varphi}{\AX{\done{\Px}}}} \\
    \hfill & \hfill && \hspace*{2.5cm}\land~(\forall~x,w,~\Px~x~w \to \R~(k~x)~w) \} \\
    \upto{bindEG}(\varphi,\Px)~\R & \triangleq &&~ \{(t\bindc k, w) \sep \entailsR{t}{w}{\EU{\varphi}{\EX{\done{\Px}}}} \\
    \hfill & \hfill && \hspace*{2.5cm}\land~(\forall~x,w,~\Px~x~w \to \R~(k~x)~w) \}
  \end{alignat*}
}
  \caption{Up-to-principles for coinductive \AG{}, \EG{} proofs.}
  \label{f:ticl:ticl-upto}
\end{figure}

Up-to-principles
$\upto{bindAG}(\varphi,~\Px),\,\upto{bindEG}(\varphi,~ \Px)$, parametrized by a
prefix formula $\varphi$ and a postcondition \typed{\Px}{X \to \World{E} \to
  \PROP} are used to prove the bind lemmas in
Figure~\ref{f:ticl:ticl-bind}. Specifically, by showing the bind principle
is under the greatest fixpoint
$\upto{bindAG}(\varphi,~ \Px) \leq \gfp(\tm{anc}~\varphi)$ we reduce a coinductive
proof \entailsL{\bind{t}{k}}{w}{\AG{\varphi}} to an \emph{inductive} proof
on the finite prefix \entailsR{t}{w}{\AU{\varphi}{\AX{\done{\Px}}}} and a
coinductive proof about its continuation $k$. The iteration lemmas
in Figure~\ref{f:ticl:ticl-iter} reduce to using the same bind up-to-principles.

\section{Secure concurrent shared memory proof}\label{appendix:ticl:mes}
This appendix provides the complete formalization of the
security-typed memory system introduced in
Section~\ref{s:ticl:ex-sec}. We present the instrumentation
handler ($h_{\Lbl}$), denotational semantics (\sembra{\Lbl}{}, \sembra{R}{}),
and proof lemmas (Figure~\ref{f:ticl:mes-lemmas}) for proving
confidentiality properties in concurrent memory access scenarios.

The goal for this example is to show confindentiality; if every read (\SRead{l_i}{x})
has an instruction label $l_i$, and accesses a memory label with
$l_m$, then it should always be true that $l_m \leq l_i$.

The two processes are in the bottom of Figure~\ref{f:ticl:mes-proof},
\tm{alice} has high-security access and \emph{writes} to \emph{odd}
numbered memory indices, while \tm{bob} who has low-security access
and \emph{reads} from \emph{even} memory indices.  The two processes
are written in the \meS{} language---with reads, writes and
conditionals--- while the interleaving scheduler is written in the
scheduler language \meS{} (Figure~\ref{f:ticl:mes-syntax})---with
infinite loops, nondeterministic choice, sequential composition and
calls to \meS{}. Having a different scheduler language from the
process language simplifies the example, however, both languages
denote to a common coinductive structure (\ictree{E_{\Lbl}}{}) in
Figure~\ref{f:ticl:mes-semantics}.

\begin{figure}[t!]
  \begin{adjustwidth}{0pt}{}
    {\small    
      %% MeS denotation
      \begin{typfunction}{h_{\Lbl}}{E_{\Lbl}  \leadsto \InstrM{\MS}{(\Lbl * \Lbl)}}
        & h_\Lbl~(\typed{\tm{Read}~l_i~x}{E_S})~(\typed{m}{\MS}) ~=~
        \begin{cases}
          \llog{(l_m, l_i)}~\tm{;\!;}~\Ret{(v, m)}, & \text{if } m[x] = \Some{l_m,v} \\
          \Ret{(0, m)}, & \text{otherwise}
        \end{cases} \hspace*{2cm} \\
        & h_\Lbl~(\typed{\tm{Write}~l_i~x~v}{E_S})~(\typed{m}{\MS}) ~=~ \Ret{(\ttt, (x \hookrightarrow v) \cup m)} \\
      \end{typfunction}
      \begin{typfunction}{\sembra{\Lbl}{\textcolor{gray}{\_}}}{\meS{A} \to \ictree{E_{\Lbl}}{A}}
        &\quad\sembra{\Lbl}{\SRead{l_i}{n}} ~=~ \trigger{(\tm{Read}~l_i~n)},
        \quad \sembra{\Lbl}{\SWrite{l_i}{n}{v}} ~=~ \trigger{(\tm{Write}~l_i~n~v)},\\
        &\quad\sembra{\Lbl}{\SIf{c}{t}{u}} ~=~
        \begin{cases}
          \sembra{\Lbl}{t}, &\text{if }c \\
          \sembra{\Lbl}{u}, &\text{otherwise}
        \end{cases}\\
        & \sembra{\Lbl}{\SBind{x}{k}} ~=~ \sembra{\Lbl}{x} \bindc (\lambda~a \Rightarrow \sembra{\Lbl}{k~a}),
        \quad \sembra{\Lbl}{\SRet{x}} ~=~ \Ret{x} \\
      \end{typfunction}
      %% MeR denotation
      \begin{typfunction}{\sembra{R}{\textcolor{gray}{\_}}}{\meR{A} \to \ictree{E_{\Lbl}}{A}}
        &\quad\sembra{R}{\RLoop{k}{x}} ~=~ \iter{(\lambda~x' \Rightarrow \sembra{\Lbl}{k~x'} \bindc (\lambda~v \Rightarrow \Ret{(\inl{v})}))}{x},
         \quad \sembra{R}{\RBr{l}{r}} ~=~ \sembra{R}{l} \oplus \sembra{R}{r}, \hspace*{1.5cm}\\
        &\quad\sembra{R}{\RCall{p}} ~=~ \sembra{\Lbl}{p},
         \quad\sembra{R}{\RRet{x}} ~=~ \Ret{x},
         \quad\sembra{R}{\RBind{x}{k}} ~=~ \sembra{R}{x} \bindc (\lambda~a \Rightarrow \sembra{R}{k~a})\\
      \end{typfunction}
      \vspace*{0.1cm}
      \typequation{\mentailsLR{(\typed{t}{\meS{A}})}{(\typed{m}{\MS})}{(\typed{l}{\Lbl * \Lbl})}{p}}{\PROP}
      {\entailsLR{\instrv{h_\Lbl}{\sembra{\Lbl}{p}}{m}}{\Obs{(\Log{l})}{\ttt}}{p}}
      \typequation{\rentailsLR{(\typed{r}{\meR{A}})}{(\typed{m}{\MS})}{(\typed{l}{\Lbl * \Lbl})}{p}}{\PROP}
      {\entailsLR{\instrv{h_\Lbl}{\sembra{R}{r}}{m}}{\Obs{(\Log{l})}{\ttt}}{p}}
    }
  \end{adjustwidth}
  \caption{Denotation of process and scheduler languages \meS{} and \meR{}, with a tagged heap. Observes memory
    labels and instruction labels to prove safety.}
  \label{f:ticl:mes-semantics}
\end{figure}

\def \MathparLineskip {\lineskip=0.2cm}
\begin{figure}[t!]
  {\footnotesize
    \begin{mathpar}
    \hspace*{-.5cm}
      {
        \mprset{vskip=.1cm}
        \mathscale{0.9}{
          \inferrule*[Right=Br$_R$AX]
          {
            \rentailsLR{a}{m}{l}{p}
            \\\\
            \rentailsLR{b}{m}{l}{p}
          }
          {
            \rentailsLR{\RBr{a}{b}}{m}{l}{\AX{p}}
          }
        }
    }
    \and
    \hspace*{-.5cm}
    {
      \mprset{vskip=.1cm}
      \mathscale{0.9}{
        \inferrule*[Right=Loop$_R$AG]{
          {
            \begin{aligned}
              \R&~i~m~l \to \forall~i,m,l,~\R~m~l \to \\
                & \rentailsL{\RLoop{k}{i}}{m}{l}{\varphi}~\land
                ~ \rentailsR{k~i}{m}{l}{\AX{(\AU{\varphi}{\AX{\done{\R}}})}}
            \end{aligned}
          }
        }
        {\rentailsL{\RLoop{k}{i}}{m}{l}{\AG{\varphi}}}
      }
    }
  \end{mathpar}
  }
  \caption{Representative \ticl lemmas for process and scheduler languages \meS{} and \meR{}.}
  \label{f:ticl:mes-lemmas}
\end{figure}

As \tm{bob} can potentially read every \emph{even} indexed location, we must
ensure the starting state has no high-security, even locations
to begin with; this is the $\tm{noleak}$ invariant on states and
indices (Figure~\ref{f:ticl:mes-proof}). The proof in
Figure.~\ref{f:ticl:mes-proof} starts by using the \emph{invariance} lemma
\mkw{Loop$_R$AG} (Figure~\ref{f:ticl:mes-lemmas}) with loop invariant
$\R$, then two proof obligations remain:
\begin{enumerate}
\item The \emph{loop} satisfies the safety property initially ($\obs{(l_m \leq l_i)}$).
\item The \emph{loop body} steps (outer \AX{}) then satisfies $\obs{(l_m \leq l_i)}$ until
  termination, at which point loop invariant $\R$ is satisfied at the end state.
\end{enumerate}

The invariance rule (\mkw{Loop$_R$AG}) abstracts the complex coinductive proof, proved
once and for all by lemma \mkw{IterAG}, adapted to language \meS{} in
a few lines of Rocq syntactic manipulations. The rest of the proof in
Figure~\ref{f:ticl:mes-proof} is straightforward. Proceed by examining
both cases of the nondeterministic choice
$(\RBr{(\tm{alice}~x~i)}{(\tm{bob}~i)})$ using the choice rule
\mkw{Br$_R$AX}. We consider both cases due to the universal
quantifier(\AX{}). The two subproofs proceed by case analysis on the
evenness of memory index $i$ and using the low-level theory of finite maps.

\begin{figure}[t!]
  \begin{adjustwidth}{0pt}{}
    \begin{mathpar}
    \inferrule*[Right=Loop$_R$AG,leftskip=1.5cm]{
        \mprset{vskip=.2cm}
      \inferrule*[Right=Br$_R$AX]{
        \mathscale{0.7}{
          \begin{aligned}
            &\forall~i,m,l_m,l_i,~\R~i~m~(l_m,l_i) \to \\
            &\rentailsR{\bm{alice}~x~i}{m}{(l_m,l_i)}{\AU{\obs{(l_m \leq l_i)}}{\AX{\done{\R}}}}\greencheck
          \end{aligned}
        }
        \and
        \mathscale{0.7}{
          \begin{aligned}
            &\forall~i,m,l_m,l_i,~\R~i~m~(l_m,l_i) \to \\
            &\rentailsR{\bm{bob}~i}{m}{(l_m,l_i)}{\AU{\obs{(l_m \leq l_i)}}{\AX{\done{\R}}}}\greencheck
          \end{aligned}
        }
      }
      {
        \mathscale{0.7}{
          \begin{aligned}
            \bm{\R}&~\_~m~(l_m,l_i)~\coloneq\\
              &l_m \leq l_i~\land~\forall~i, \bm{noleak}~i~m \\[0.8em]
            \R&~0~m~(l_m, l_i)~\greencheck \\
          \end{aligned}
        }
        \and
        \hspace*{-.5cm}
        \mathscale{0.7}{
          \begin{aligned}
            &\forall~i,m,l_m,l_i,~\R~i~m~(l_m,l_i) \to \\
            & \quad\begin{rentailsP}{.28}{L}{}
              {
                $\text{\textbf{loop}}~(\lambda~i\Rightarrow$
                \Indp \RBr{(\bm{alice}~x~i)}{(\bm{bob}~i)}\;
                \Indp \RRet{(i+1)}\;
                \Indm $)~i$
              }
              {m}
              {(l_m,l_i)}
              {\obs{(l_m \leq l_i)}}
              {9cm}
            \end{rentailsP}~\greencheck\\
            & \land~\rentailsR{\RBr{(\bm{alice}~x~i)}{(\bm{bob}~i)};~\RRet{(i+1)}}{m}{(l_m,l_i)}
              {\AX{(\AU{\obs{(l_m \leq l_i)}}{\AX{\done{\R}}})}}
          \end{aligned}
        }
      }
    }
    {
      {
        \begin{array}{l l}
        \begin{plet}{alice}{(\typed{x~i}{\mathbb{N}})}{
            \eIf{$\tm{even}~i$}{
              \SWrite{H}{(i+1)}{x}
            }{
              \SWrite{H}{i}{x}
            }
          }{0.7}{.25}
        \end{plet}
          &
            \begin{plet}{bob}{(\typed{i}{\mathbb{N}})}{
                \eIf{$\tm{even}~i$}{
                  \SRead{L}{i}
                }{
                  \SRead{L}{(i+1)}
                }
              }{0.7}{.25}
            \end{plet} \\[.8cm]
          \multicolumn{2}{l}{
          \mathscale{0.7}{
          \begin{aligned}
            \bm{noleak}&~(\typed{i}{\mathbb{N}})~(\typed{m}{\MS})~\coloneq \\
                       &\tm{even}~i \to \exists~v, m[i] = \Some{(L, v)}
          \end{aligned}
          }
          }
      \end{array}
    }
    \and
    \hspace{-.8cm}
      \mathscale{0.7}{
        \begin{aligned}
          & \forall~x,m,l_m,l_i,~l_m \leq l_i \to \forall~i,~\bm{noleak}~i~m \to \\
          &\begin{rentailsP}{.28}{L}{}
            {
              $\text{\textbf{loop}}~(\lambda~i\Rightarrow$
              \Indp \RBr{(\bm{alice}~x~i)}{(\bm{bob}~i)}\;
              \Indp \RRet{(i+1)}\;
              \Indm $)~0$
            }
            {m}
            {(l_m,l_i)}
            {\AG{\obs{(l_m \leq l_i)}}}
            {10cm}
          \end{rentailsP}
        \end{aligned}
      }
    }
  \end{mathpar}
\end{adjustwidth}
\caption{\Ticl safety proof for the concurrent secure heap. The goal is to show that
    \tm{read} instructions only access memory locations with a security level lower-or-equal than their own.}
\label{f:ticl:mes-proof}
\end{figure}

\section{Distributed Consensus proof}\label{appendix:ticl:election}
This appendix provides the complete formalization for the leader election
distributed consensus proof in Section~\ref{s:ticl:ex-election}.
A distributed, message-passing system is encoded using \ICTrees directly
and the liveness property ``eventually a consensus is reached''
is proved using the \ticl structural lemmas (Section~\ref{f:ticl:ticl-iter}).

\begin{figure}[t!]
  \begin{adjustwidth}{0pt}{}
    {\small
    \typequation{\Enet}{\Type \to \Type}{ \bbar \tm{Send} (\typed{id}{\Pid})(\typed{m}{\Msg}) \quad \bbar \tm{Recv} (\typed{id}{\Pid})}
    \begin{typfunction}{\hnet}{\Enet \leadsto \InstrM{\Mailboxes}{\Msg}}
      & \hnet~(\tm{Send}~id~msg)~(\typed{m}{\Mailboxes}) ~=~ \Ret{(\ttt, m[id~+~1~\%~n] \coloneq msg)} \hspace*{4cm} \\
      & \hnet~(\tm{Recv}~id)~(\typed{m}{\Mailboxes}) ~=~ \llog{(m[id])}~\tm{;\!;}~\Ret{(m[id], m)}
    \end{typfunction}
  }
  \end{adjustwidth}
  \caption{Message-passing events (\Enet) parametrized by the caller
    $id$ and their instrumentation ($\hnet$).}
  \label{f:ticl:election-syntax}
\end{figure}

Each process in the leader election protocol (Figure~\ref{f:ticl:election-proc}) is assigned a
process identifier ($\typed{pid}{\Pid}$). Message-passing events (\tm{send} and \tm{recv}) are defined in
Figure~\ref{f:ticl:election-syntax}. Their semantic meaning by an instrumentation handler ($\hnet$)---
messages are delivered in a unidirectional ring (\emph{uniring})
configuration in a clockwise manner, as shown in
Figure~\ref{f:ticl:election-draw}.

Scheduling follows the same pattern. Each process sends and receives
one of two kinds of messages: proposing a candidate PID $(C_i)$ and
announcing a leader ($E_i$). The formal definition of messages (\Msg)
and mailboxes (\Mailboxes) are in
Figure~\ref{f:ticl:election-vec}. Each process has exactly one message
in its mailbox. Messages and process identifiers (\Pid), are both indexed by
the number of processes in the system ($\typed{n}{\mathbb{N}}$).

\begin{figure}[t!]
  \begin{adjustwidth}{0pt}{}
    {\small
      \typequation{\Pid}{\Type}{\fin{n}}
      \typequation{\Msg}{\Type}{\bbar C~(\typed{p}{\Pid}) \quad \bbar E~(\typed{p}{\Pid})}
      \typequation{\Mailboxes}{\Type}{\tm{Vector}~n~\Msg}
      \quad \typed{(\typed{m}{\Mailboxes})[ \typed{p}{\Pid} ]}{\Msg}, \\
      \quad \typed{(\typed{m}{\Mailboxes})[ \typed{p}{\Pid} ] \coloneq (\typed{msg}{\Msg})}{\Mailboxes}
    }
  \end{adjustwidth}
  \caption{Process identifiers, messages and mailboxes with \emph{get} ($m[p]$) and \emph{set} ($m[p]\coloneq msg$) access.}
  \label{f:ticl:election-vec}
\end{figure}

The liveness proof (``eventually a consensus is reached'') in
Figure~\ref{f:ticl:election-proof} is proven using the \emph{liveness
  split} lemma (Theorem~\ref{f:ticl:liveness-split}).  \emph{Liveness
  split} divides a liveness proof into two sections, separated by a
user-specified intermediate point. This transforms a loop liveness
proof into two ``smaller'' liveness proofs, joined by an intermediate
relation $(\R_I)$. These reduced liveness proofs utilize more
straightforward, compact ranking functions ($f$). The result is the
original liveness proof, but starting from a later position
($\R_I~i~w \to \entailsL{\iter{k}{i}}{w}{\AU{\varphi}{\varphi'}}$). This splitting
process can be continued by defining appropriate intermediate
relations ($\R_I$) and ranking functions ($f$). This way, liveness
proof modularity extends to logical structures, such as the phases in
the leader election protocol (Section~\ref{s:ticl:ex-election}).

\def\largeDelim#1#2{\vcenter{\hbox{$\left#1 \rule{0pt}{#2} \right.$}}}

\begin{figure}[t!]
    \begin{mathpar}
      \hspace*{-1.6cm}
      \inferrule*[Right=BindAU$_L$]{
        \inferrule*[Right=SplitAU$_{L,\mathbb{N}}$]{
          \inferrule*[Right=iter]{
            \inferrule*[Right=BindAU$_L$]{
              \inferrule*[Right=iter]{
                \mathscale{0.7}{
                  \forall~w,~\notdone{w} \to 
                  \entailsL{\tm{instr}~\hnet~
                    \mbox{$\largeDelim{(}{.9cm}$
                      \begin{lstlisting}[style=customcoq,basicstyle=\small\ttfamily]^^J
                        proc $\ 1$; ^^J
                        iter (fun $\ i$ => ^^J
                        $\quad$ proc $\ i$; ^^J
                        $\quad$ Ret (inl (($i$ + 1) \% $n$)) ^^J
                        )$\ 2$
                      \end{lstlisting}$\largeDelim{)}{.9cm}$
                    }~ [E_3, C_3, C_3]}{w}{\AF{\obs{E_3}}}\greencheck
                }
              }{
                \mathscale{0.7}{
                  \forall~w,~\notdone{w} \to 
                  \entailsL{\tm{instr}~\hnet~
                    \mbox{$\largeDelim{(}{.9cm}$
                      \begin{lstlisting}[style=customcoq,basicstyle=\small\ttfamily]^^J
                        iter (fun $\ i$ => ^^J
                        $\quad$ proc $\ i$; ^^J
                        $\quad$ Ret (inl (($i$ + 1) \% $n$)) ^^J
                        )$\ 1$
                      \end{lstlisting}$\largeDelim{)}{.9cm}$
                    }~ [E_3, C_3, C_3]}{w}{\AF{\obs{E_3}}}
                }
              }
            }{
              \mathscale{0.7}{
                \forall~w,~\notdone{w} \to 
                \entailsL{\tm{instr}~\hnet~
                  \mbox{$\largeDelim{(}{.9cm}$
                    \begin{lstlisting}[style=customcoq,basicstyle=\small\ttfamily]^^J
                      proc $\ 3$; ^^J
                      iter (fun $\ i$ => ^^J
                      $\quad$ proc $\ i$; ^^J
                      $\quad$ Ret (inl (($i$ + 1) \% $n$)) ^^J
                      )$\ 1$
                    \end{lstlisting}$\largeDelim{)}{.9cm}$
                  }~ [C_3, C_3, C_3]}{w}{\AF{\obs{E_3}}}
              }
            }
          }
          {
            \mathscale{0.7}{
            \begin{aligned}          
              f&~id~ms~\coloneq \\[-.4em]
               & \bm{\tm{match}}~id, ms~\bm{\tm{with}}\\[-.8em]
               &\mbox{
                  \begin{lstlisting}[style=customcoq,basicstyle=\small\ttfamily]^^J
                    | 2, $[ C_3, C_3, C_2 ]$ => 1^^J
                    | 1, $[C_3, C_1, C_2]$ => 2^^J                      
                    | 3, $[ C_3, C_1, C_2 ]$ => 3^^J
                    | 2, $[ C_3, C_1, C_2 ]$ => 4^^J
                    | _, _ => 10^^J
                    end.^^J
                  \end{lstlisting}
                 } \\[.5em]
              \R&~id~ms~w~\coloneq \\[-.4em]
               & \bm{\tm{match}}~w~\bm{\tm{with}}\\[-.8em]
               &\mbox{
                 \begin{lstlisting}[style=customcoq,basicstyle=\small\ttfamily]^^J
                   | Pure => ^^J
                     $\quad$id = i $\ \land$ ms = $[C_3, C_1, C_2]$^^J
                   | Obs (Log $\ C_p$) tt => ^^J
                     $\quad$match (id, p) with^^J
                     $\quad$| (2, 3) => ms = $[C_3, C_3, C_2]$^^J
                     $\quad$| (3, 1) => ms = $[C_3, C_1, C_2]$^^J
                     $\quad$| (1, 2) => ms = $[C_3, C_1, C_2]$^^J
                     $\quad$| _ => $\bot$ ^^J
                     $\quad$end ^^J
                   | _ => $\bot$^^J
                   end^^J
                 \end{lstlisting}
                 } \\[1em]
              \R_I~& id~ms~w~\coloneq~\notdone{w} \\
                & \land~id = 3 \land ms = [C_3, C_3, C_3]
            \end{aligned}            
          }
          \and\hspace*{-.8cm}
          \mathscale{0.7}{
            \begin{aligned}
              \forall~i&~w~ms,~\R_I~i~w~ms \to \\[.5em]
              & \entailsL{\tm{instr}~\hnet~
                \mbox{$\largeDelim{(}{.9cm}$
                \begin{lstlisting}[style=customcoq,basicstyle=\small\ttfamily]^^J
                  iter (fun $\ i$ => ^^J
                  $\quad$ proc $\ i$; ^^J
                  $\quad$ Ret (inl (($i$ + 1) \% $n$)) ^^J
                  )$\ i$
                \end{lstlisting}$\largeDelim{)}{.9cm}$
                }~ ms}{w}{\AF{\obs{E_3}}}  \\[3em]
              \forall~i&~w~ms,~\R~i~w~ms \to \hspace*{.5cm}\greencheck \\[.5em]
              & \entailsRR{\tm{instr}~\hnet~
                \mbox{\bigg(
                \begin{lstlisting}[style=customcoq,basicstyle=\small\ttfamily]^^J
                  proc $\ i$; ^^J
                  Ret (inl (($i$ + 1) \% $n$)) ^^J
                \end{lstlisting}\bigg)
                }~ ms}{w}~\tm{AF}~\tm{AX}~\tm{done}(\tm{\lambda}~lr~w'~ms' \Rightarrow \\
              & \hspace*{3cm} ~\exists~i',~lr = \inl{i'} ~ \land~ \R_I~i'~w'~ms') \hspace*{.4cm}\mathscale{1.2}{\lor} \\[1em]
              & \entailsRR{\tm{instr}~\hnet~
                \mbox{\bigg(
                \begin{lstlisting}[style=customcoq,basicstyle=\small\ttfamily]^^J
                  proc $\ i$; ^^J
                  Ret (inl (($i$ + 1) \% $n$)) ^^J
                \end{lstlisting}\bigg)
                }~ ms}{w}~\tm{AF}~\tm{AX}~\tm{done}(\tm{\lambda}~lr~w'~ms' \Rightarrow \\
              & \hspace*{3cm} ~\exists~i',~lr = \inl{i'} ~ \land~ \R~i'~w'~ms' \land~ f~i'~w'~ms' < f~i~w~ms) 
            \end{aligned}            
          }
        }
      }            
      {
        \mathscale{0.7}{
        \begin{aligned}          
          & \forall~(\typed{i}{\Pid}),~\entailsL{\tm{instr}~\hnet~
            \mbox{$\largeDelim{(}{.9cm}$
            \begin{lstlisting}[style=customcoq,basicstyle=\small\ttfamily]^^J
              iter (fun $\ i$ => ^^J
              $\quad$ proc $\ i$; ^^J
              $\quad$ Ret (inl (($i$ + 1) \% $n$)) ^^J
              )$\ i$ 
            \end{lstlisting}$\largeDelim{)}{.9cm}$
            }~ [C_3, C_1, C_2]}{\Pure}{\AF{\obs{E_3}}}
        \end{aligned}
        }
      }
    }
    {
      \mathscale{0.7}{
        \begin{aligned}          
          & \entailsL{\tm{instr}~\hnet~
            \mbox{$\largeDelim{(}{1.1cm}$
            \begin{lstlisting}[style=customcoq,basicstyle=\small\ttfamily]^^J
              $\ i$ <- branch $\ n$; ^^J
              iter (fun $\ i$ => ^^J
              $\quad$ proc $\ i$; ^^J
              $\quad$ Ret (inl (($i$ + 1) \% $n$)) ^^J
              )$\ i$
            \end{lstlisting}$\largeDelim{)}{1.1cm}$
            }~ [C_3, C_1, C_2]}{\Pure}{\AF{\obs{E_3}}}
        \end{aligned}
      }
    }
  \end{mathpar}

\caption{\Ticl liveness proof for the distributed consensus example, is split in two
  subproofs using intermediate relation $\R_I$.}
\label{f:ticl:election-proof}
\end{figure}

The proof in Figure~\ref{f:ticl:election-proof} is for three processes ($n=3$), but the techniques used
generalize to any number. Start from the bottom, the goal liveness property is:

\begin{lemma}[Eventual leader consensus] \ \\
  \begin{center}
    \small
    \entailsL{\tm{instr}~\hnet~
      \mbox{$\largeDelim{(}{1.1cm}$
        \begin{lstlisting}[style=customcoq,basicstyle=\small\ttfamily]^^J
          $\ i$ <- branch $\ n$; ^^J
          iter (fun $\ i$ => ^^J
          $\quad$ proc $\ i$; ^^J
          $\quad$ Ret (inl (($i$ + 1) \% $n$)) ^^J
          )$\ i$
        \end{lstlisting}$\largeDelim{)}{1.1cm}$
      }~ [C_3, C_1, C_2]}{\Pure}{\AF{\obs{E_3}}}
  \end{center}
  \label{f:ticl:election-goal}
\end{lemma}

The shared state of this system captures the mailboxes starting with
$[C_3, C_1, C_2]$, meaning candidacy messages are waiting to be
received by their respective process. The ghost-state of the system
(see \hnet in Figure~\ref{f:ticl:election-syntax}) is the last
received message ($\llog{m[id]}$). Our goal specification
(Lemma~\ref{f:ticl:election-goal}) is to eventually observe $pid=3$
elected as the leader ($E_3$). When message $E_3$ is received at least
by one process, it will be infinitely propagated in a clockwise
manner until all processes know the leader.

Start by nondeterministically choosing a process to schedule
($i \gets \tm{branch}~n$) first.  Applying the sequence lemma
(\mkw{BindAU$_L$}) introduces $\typed{i}{\tm{PID}_3}$ to the proof
context.  We must prove that eventually a leader is elected,
regardless of which process the scheduler chooses first.
Figure~\ref{f:ticl:election-proof} uses liveness split
(\mkw{SplitAU_${L,\mathbb{N}}$}) with an intermediate relation ($\R_I$) to mark the
end of the candidate aggregation phase. Loop invariant ($\R$) shows
that, if we know which process ran last ($C_P$) and which process is
currently running ($id$) we can guess the state of mailboxes (either
$[C_3, C_3, C_2]$ or $[C_3, C_1, C_2]$ or $[C_3, C_1, C_2]$) by checking
Figure~\ref{f:ticl:election-draw}.

By case analysis (on $w, id, p$) we can show that regardless of the
nondeterministic choice of first PID ($i$), candidate aggregation
eventually ends in $\R_I$: The highest PID's candidacy message ($C_3$)
is in every mailbox and it its process is ready to be scheduled
($i=3$). Establishing this intermediate goal simplifies the ranking
function $f$ and invariant $\R$, which now only need to refer to
candidacy messages ($C_i$), not election announcement messages
($E_i$).

Now what's left is the remaining liveness proof, starting at a point ($\R_I$)
where the mailboxes have fully propagated the candidacy of $pid=3$ ($[C_3, C_3, C_3]$)
and also $i=3$ is scheduled to run. Using the equational theory of \ICTrees (Section~\ref{s:ticl:uptotau-proper})
we unfold one iteration of the loop ($\tm{proc}~3$). Consequently we use the sequencing lemma (\mkw{BindAU$_L$})
to evaluate the proccess with $pid=3$, updating the mailbox of its neighbor ($pid=1$). The state after
running $\tm{proc}~3$ is $[E_3, C_3, C_3]$ and the next process scheduled to run is $i=1$. When $\tm{proc}~1$
runs, it will receive the election announcement ($E_3$), satisfying the liveness property ($\obs{E_3}$) and
concluding the proof.

Consequently, from state $\R_I$, \proc{3} runs with $C_3$ in their
mailbox and announces themselves as the new leader ($E_3$). The next
process is then scheduled ($i=1$), receiving the leader announcement
($E_3$) and satisfying the goal ($\AF{\obs{E_3}}$).

%\section{CTree combinators and syntax}
%\input{appendix/ctree-notations}
\end{document}